\documentclass{aa} 

\usepackage{subfig}
\usepackage{graphicx}

\usepackage{movie15}

\usepackage{txfonts}
\usepackage{natbib}
\bibpunct{(}{)}{;}{a}{}{,} 

\graphicspath{{.}{images/}}

\usepackage{dblfloatfix}
\usepackage{array}
\usepackage{multirow}
\usepackage{hyperref}

\begin{document}
	\title{Towards consistent mapping of distant worlds:\\ secondary-eclipse scanning of the exoplanet \object{HD\,189733b}}


   \author{Julien de Wit\inst{1,2}
          \and
          Micha\"el Gillon\inst{3}
          \and
          Brice-Olivier Demory\inst{1}
          \and
          Sara Seager\inst{1,4}
          }

   \institute{Department of Earth, Atmospheric and Planetary Sciences, MIT, 77 Massachusetts Avenue, Cambridge, MA 02139, USA.\\
              \email{jdewit@mit.edu}
         \and
             Facult\'{e} des Sciences Appliqu\'{e}es, Universit\'{e} de Li\`{e}ge, Grande Traverse 12, 4000 Li\`ege, Belgium.
         \and
             Institut d'Astrophysique et de G\'{e}ophysique, Universit\'{e} de Li\`{e}ge, All\'{e}e du 6 Ao\^ut 17, 4000 Li\`ege, Belgium.
         \and
             Department of Physics and Kavli Institute for Astrophysics and Space Research, MIT, 77 Massachusetts Avenue, Cambridge, MA 02138, USA.             
             }

   \date{Received 16 February 2012 / Accepted 31 October 2012}


   \abstract
   {Mapping distant worlds is the next frontier for exoplanet infrared (IR) photometry studies. Ultimately, constraining spatial and temporal properties of an exoplanet atmosphere (e.g., its temperature) will provide further insight into its physics. For tidally-locked hot Jupiters that transit and are eclipsed by their host star, the first steps are now possible.}
   {Our aim is to constrain an exoplanet's \textbf{(1)} shape, \textbf{(2)} brightness distribution (BD) and \textbf{(3)} system parameters from its phase curve and eclipse measurements. In particular, we rely on the secondary-eclipse scanning which is obtained while an exoplanet is gradually masked by its host star.}
   {We use archived \textit{Spitzer}/IRAC 8-$\mu$m data of HD\,189733 (six transits, eight secondary eclipses, and a phase curve) in a global Markov Chain Monte Carlo (MCMC) procedure for mitigating systematics. We also include HD\,189733's out-of-transit radial velocity (RV) measurements to assess their incidence on the inferences obtained solely from the photometry.}
   {We find a 6\,$\sigma$ deviation from the expected occultation of a uniformly-bright disk. This deviation emerges mainly from a large-scale hot spot in HD\,189733b's atmosphere, not from HD\,189733b's shape. We indicate that the correlation of the exoplanet orbital eccentricity, $ e $, and BD (``uniform time offset'') does also depend on the stellar density, $ \rho_{\star} $, and the exoplanet impact parameter, $ b $ (``$ e $-$ b $-$\rho_{\star}$-BD correlation''). 
For HD\,189733b, we find that relaxing the eccentricity constraint and using more complex BDs lead to lower
stellar/planetary densities and a more localized and latitudinally-shifted hot spot. We, therefore, show that the light curve of an exoplanet does not constrain uniquely its brightness peak localization. Finally, we obtain an improved constraint on the upper limit of HD\,189733b's orbital eccentricity, $ e\leq 0.011$ ($95\%$ confidence), when including HD\,189733's RV measurements.}
   {Reanalysis of archived \object{HD\,189733}'s data constrains HD\,189733b's shape and BD at 8\,$\mu$m. Our study provides new insights into the analysis of exoplanet light curves and a proper framework for future eclipse-scanning observations. In particular, observations of the same exoplanet at different wavelengths could improve the constraints on HD\,189733's system parameters while ultimately yielding a large-scale time-dependent 3D map of HD\,189733b's atmosphere. Finally, we discuss the perspective of extending our method to observations in the visible (e.g., \textit{Kepler} data), in particular to better understand exoplanet albedos.}

   \keywords{eclipses -- planets and satellites : individual : HD\,189733b -- techniques : (photometric)}

   \titlerunning{Towards consistent constraints on transiting exoplanets}
   
   \maketitle

%

\section{Introduction}

\begin{figure*}

  \begin{center}
    \hspace{-0.cm}\includegraphics[trim = 80mm 80mm 90mm 75mm,clip,width=14cm,height=!]{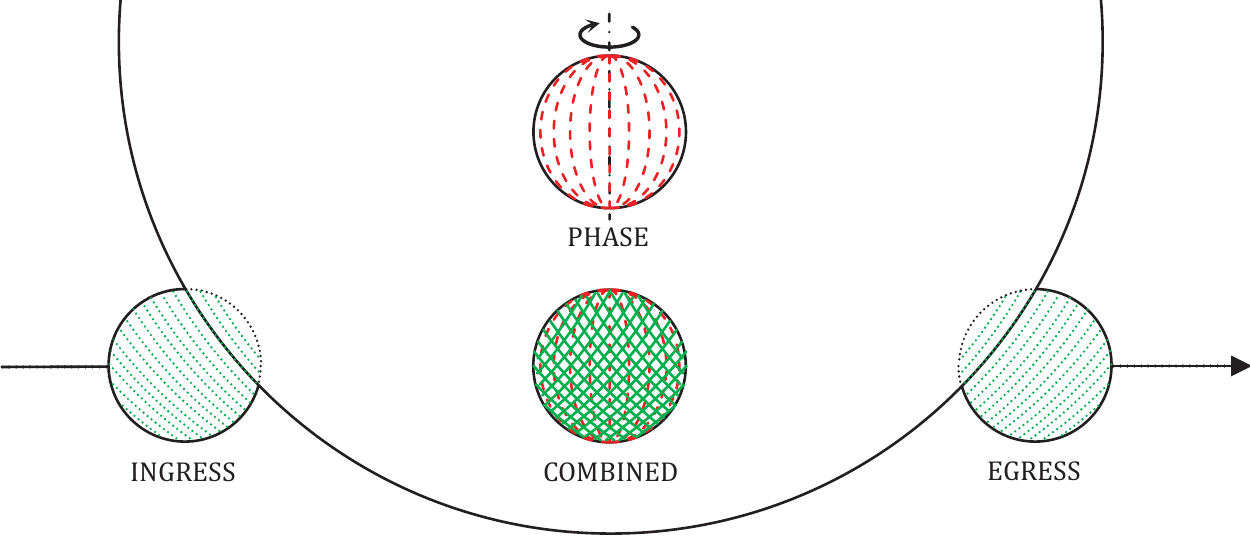}
  \end{center}
  \vspace{-0.5cm}
  \caption{Schematic description of the different scanning processes observable for an occulted exoplanet. The green dotted lines indicate the scanning processes during the exoplanet occultation ingress/egress. The red dashed line indicates the scanning process that results from the exoplanet rotation and produces the exoplanet phase curve---this scanning appears longitudinal for an observer as long as the exoplanet spin is close to the projection plane, e.g., for a transiting and synchronized exoplanet. The component labeled ``combined'' shows the specific grid generated by these three scanning processes.}
  \label{fig:scanning_processes}

    \end{figure*}

Among the hundreds of exoplanets detected to date\footnote{For an up-to-date list see the Extrasolar Planets Encyclopaedia: \url{http://exoplanet.eu }
}, the transiting ones can be most extensively characterized with current technology. 
Beyond derivation of the exoplanet orbital inclination and density \citep[e.g.,][]{Winn2010}, transiting exoplanets are key objects 
because their atmospheres are observationally accessible through transit-transmission and occultation-emission spectrophotometry \citep[see e.g.,][and references therein]{Seager2010a}. In addition, the phase-dependence of exoplanet IR flux  provides observational constraints on their atmospheric temperature distribution (see the component labeled ``\textit{phase}'' Fig.\,\ref{fig:scanning_processes}) and, consequently, further insight into their atmospheric structures. For instance, from phase-curve IR measurements of the exoplanet HD\,189733b, \citet{Knutson2007} derived the first longitudinal\footnote{An exoplanet phase curve contains information about the exoplanet hemisphere-integrated flux only.} brightness distribution (BD) of an exoplanet; it indicates an offset hot spot in agreement with the predictions for hot Jupiters with shifted atmospheric temperature distributions because of equatorial jets \citep[e.g.,][]{Showman2002}. 

Since then, significant theoretical developments have been achieved in modeling exoplanet atmospheres by combining hydrodynamic flow with thermal forcing \citep[e.g.,][]{Showman2009,Rauscher2010,Dobbs-Dixon2010,Rauscher2012} and/or with ohmic dissipation \citep[e.g.,][]{Batygin2011,Heng2012,Menou2012}. However, other possible forcing factors are yet unmodeled. For example, the magnetic star-planet interactions can also have a significant role in this matter---although, to our knowledge no studies discuss in detail their impact on the exoplanet atmospheric structure so far. Nevertheless,  magnetic interactions have to date been only observed at the stellar surface, in the form of chromospheric hot spots rotating synchronously with the companions \citep[e.g.,][]{Shkolnik2005,Lanza2009}.

The observation of specific spatial features within an exoplanet atmosphere, such as hot spots or cold vortices, is essential for constraining its structure and for gaining further insight into its physics. Eclipses have proved to be powerful tools for ``spatially resolving'' distant objects, including binary stars \citep[e.g.,][]{Warner1971} and accretion disks \citep[e.g.,][]{Horne1985}. Previous theoretical studies introduced the potential of eclipse scanning\footnote{Eclipse scanning is the process by which a body gradually masks another body.}(see Fig.\,\ref{fig:scanning_processes}) for exoplanets in order to disentangle atmospheric circulation regimes \citep[e.g.,][]{William2006,Rauscher2007}.

Ideally, the light curve of a transiting and occulted exoplanet with a non-zero impact parameter can enable a 2D surface brightness map of its day side. In fact, as represented in Fig.\,\ref{fig:scanning_processes}, such an exoplanet is scanned through several processes along its orbit. First, the exoplanet is gradually masked/unmasked by its host star during occultation ingress/egress. Secondly, the exoplanet rotation provides its phase-dependent hemisphere-integrated flux (i.e., its phase curve), which constrains its BD in longitudinal slices---as long as the exoplanet spin is close to the projection plane, e.g., for a transiting and synchronized exoplanet. In particular, the phase curve is modulated by the orbital period as long as the exoplanet is tidally locked. The three scanning processes (ingress, egress and phase curve) provide thus complementary pieces of information that could ultimately constrain the BD over a specific ``grid'' (e.g., see the component labeled ``\textit{combined}'' in Fig.\,\ref{fig:scanning_processes}). In this way, only a ``uniform time offset''\footnote{The uniform time offset measures the time lag between the observed secondary eclipse and that predicted by a planet with spatially uniform emission \citep[defined by][]{William2006}.} has been detected so far; \citet{Agol2010} showed that, assuming HD\,189733b's orbit to be circularized, its occultation was offset by 38$\pm$11 sec. This time offset is in agreement with the expected effect in occultation of the offset hot spot indicated by HD\,189733b's longitudinal map from \citet{Knutson2007}.

The subset of exoplanet thermal IR observations that aims to characterize a planetary BD is growing. Currently, the \textit{Spitzer Space Telescope} \citep{Werner2004} has observed thermal phase curves for a dozen different exoplanets  as well as the IR occultations of over thirty exoplanets. Among these, HD\,189733b \citep{Bouchy2005} is arguably the most favorable transiting exoplanet for detailed observational atmospheric studies; in particular, because its K-dwarf host is the closest star to Earth with a transiting hot Jupiter. This means the star is bright and the eclipses are relatively deep yielding favorable signal-to-noise ratio (SNR). As such, HD\,189733b represents a ``Rosetta Stone'' for the field of exoplanetology with one of the highest SNR secondary eclipses \citep{Deming2006,Charbonneau2008}, phase-curve observations \citep{Knutson2007,Knutson2009,Knutson2012} and, consequently, numerous atmospheric characterizations \citep[e.g.,][]{Grillmair2007,Pont2007,Tinetti2007,Redfield2008,Swain2008,Madhusudhan2009,D'esert2009,Deroo2010,Sing2011,Gibson2011,Huitson2012}. Although HD\,189733b's atmospheric models are in qualitative agreement with observations, important discrepancies remain between simulated and observed light curves as well as between emission spectra \citep[see e.g.,][Figs. 8 and 10]{Showman2009}. In addition, discrepancies exist between several published inferences---in particular molecular detections---which emphasize the impact of data reduction and analysis procedures \citep[e.g.,][]{D'esert2009,Gibson2011}. Hence, we undertake a global analysis of all HD\,189733's public photometry obtained with the \textit{Spitzer Space Telescope} for assessing the validity of published inferences (de Wit \& Gillon, in prep). 

In this paper, we present the first secondary-eclipse scanning of an exoplanet showing a deviation from the occultation of a uniformly-bright disk at the 6\,$\sigma$ level, which we obtain from the archived \textit{Spitzer}/IRAC 8-$\mu$m data of the star HD\,189733. In addition, we propose a new methodology of analysis for secondary-eclipse scanning to disentangle the possible contributing factors of such deviations. As a result, we perform a new step toward mapping distant worlds by constraining consistently (i.e., simultaneously) HD\,189733b's shape, BD at 8\,$\mu$m and system parameters.

At the time of submission, we learned about a similar study by \cite{Majeau2012}, hereafter M12, focusing on the derivation of HD\,189733b's 2D eclipse map, using the same data but different frameworks for data reduction and analysis. Our study differs from M12 in three main ways. First, we find a deviation from the occultation of a uniformly-bright disk at the 6\,$\sigma$ level in contrast to the $\lesssim$\,3.5\,$\sigma$ level deviation in the phase-folded light curves from \citet{Agol2010} (see their Fig.12) used in M12. Secondly, this deviation has multiple possible contributing factors (i.e., not only a non-uniform BD but also the exoplanet shape or biased orbital parameters). Our study provides a framework for constraining consistently these contributing factors. Thirdly, and related to the second point, we do not constrain \textit{a priori} the system parameters to the best-fit of a conventional analysis, nor the orbital eccentricity to zero; instead, we estimate the system parameters simultaneously with the BD. We compare the procedures and the results in Sect.\,\ref{sec:compare_2_M12}.

We begin with a summary description of the {\it Spitzer} 8\,$\mu$m data. In Sect.\,\ref{sec:data} we present our data reduction and conventional data analysis, i.e., for which we model HD\,189733b as a uniformly-bright disk. We present and assess the robustness of the structure detected in HD\,189733b's occultation in Sect.\,\ref{sec:results&discussI}. In Sect.\,\ref{sec:ecl}, we describe our new methodology for disentangling the possible contributing factors of such a structure. We present our results in Sect.\,\ref{sec:results}. We, then, discuss in Sect.\,\ref{sec:discussionII} the robustness of our results and the perspectives of our method. Finally, we conclude in Sect.\,\ref{sec:conclusion}.


\section{Data reduction \& conventional analysis}
\label{sec:data}

We present in this section the conventional data reduction and global data analysis performed for determining HD\,189733's system parameters (i.e., the orbital and physical parameters of the star and its companion) based on the \textit{Infrared Array Camera} \citep[IRAC:][]{Fazio2004} 8-$\mu$m eclipse photometry. We emphasize that we simultaneously analyze the whole data set in what we term a ``global analysis'', instead of combining each separately-analyzed eclipse events. The global analysis approach helps to mitigate the effects of noise by extracting simultaneously common information between multiple light curves. Therefore, a global procedure also enables detection of previously unnoticed effects in the data. We begin with a summary description of the data sets; then, we introduce the analysis method and the physical models used for the parameter determination.

\subsection{Data description and reduction}

The eight secondary eclipses and the six transits of HD\,189733b\footnote{Data available in the form of Basic Calibrated Data (BCD) on the \textit{Infrared Science Archive} : \url{http://
sha.ipac.caltech.edu/applications/Spitzer/SHA//}} used in this study are described (by
\textit{Astronomical Observation Requests}, hereafter AOR) in Table\,\ref{tab:AOR}.
The data were obtained from November 2005 to June 2008 with IRAC at 8\,$\mu$m. 
These are calibrated by the \textit{Spitzer} pipeline version S18.18.0. The new S18.18.0 version enables improvements in the quality 
of data reduction over the original published data sets that used older {\it Spitzer} pipeline versions\footnote{\url{http://irsa.ipac.caltech.edu/data/SPITZER/docs/irac/iracinstrumenthandbook/73 }}. Each AOR is composed of data sets; each of which corresponds to 64 individual subarray images of 32x32 pixels.

The data reduction consists in converting each AOR into a light curve; for that purpose, we follow a procedure \cite[see][hereafter G10, and references therein]{Gillon2010a} that
is performed individually for each AOR (de Wit \& Gillon, in prep). For each AOR, we first convert fluxes from \textit{Spitzer} units of specific intensity (MJy/sr) to photon counts; then, we perform aperture photometry on each image with the $\tt{IRAF}$\footnote{IRAF is distributed by the National Optical Astronomy Observatory, which is operated by the Association of Universities for Research in Astronomy, Inc., under cooperative agreement with the National Science Foundation.} $\tt{/DAOPHOT}$ software \citep{Stetson1987}. We estimate the PSF center by fitting a Gaussian profile to each image. We estimate the best aperture radius (see Table\,\ref{tab:AOR}) based on the instrument point-spread function (PSF\footnote{\url{http://ssc.spitzer.caltech.edu/irac/psf.html  }}), HD189733b's, HD189733's, \object{HD\,189733B}'s and the sky-background flux contributions. For each image, we correct the sky background by subtracting from the measured flux its mean contribution in an annulus extending from 10 to 16 pixels from the PSF center. Then, we discard:
\begin{itemize}
\item the first 30 minutes of each AOR for allowing the detector-sensitivity stabilization,
\item the few significant outliers to the bulk of the \textit{x-y} distribution of the PSF centers using a 10\,$\sigma$ median clipping,
\item and, for each subset of 64 subarray images, the few measurements with discrepant flux values, background and PSF center positions using a 10\,$\sigma$ median clipping.
\end{itemize}
Finally, for each AOR, the resulting light curve is the time series of the flux values averaged across each subset of 64 subarray images; while the photometric errors are the standard deviations on the averaged flux from each subset. 

\subsection{Photometry data analysis}

We describe now the procedure used for constraining HD\,189733's system parameters from the light curves obtained after data reduction. We present the analysis method and the models---eclipse and systematic models---used to fit the light curves. In addition, we describe the procedure used for taking into account the correlated noise for each light curve \citep[for details
on the effect of correlated noise see, e.g.,][]{Pont2006}. 

\subsubsection{Analysis method}
 
We use an adaptive Markov Chain Monte Carlo \citep[MCMC; see e.g.][]{Gregory2005,Ford2006} algorithm. MCMC is a Bayesian inference method based on stochastic simulations that sample the posterior probability distribution (PPD) of adjusted parameters for a given fitting model. We use here the implementation presented in detail in \citet{Gillon2009,Gillon2010b,Gillon2010a}. More specifically, this implementation uses Keplerian orbits and models the eclipse photometry using the model of \citet{Mandel2002}. In addition, the simulated eclipse photometry is multiplied by a baseline, different for each time-series, to take into account the systematics (see below).

We used in our first analysis the following jump parameters\footnote{Jump parameters are the model parameters that are randomly perturbed at each step of the MCMC method}; the planet/star area ratio $(R_p/R_{\star})^2$, the orbital period $P$, the transit duration (from the first to last contact) $W$, the time of minimum light $T_0$, $\sqrt{e}\cos\omega$, $\sqrt{e}\sin\omega$ and the impact parameter $b=a/R_{\star}\cos i$ (where $a$ is the exoplanet semi-major axis). We assumed a uniform prior distribution for all these jump parameters and draw at each step a random stellar mass, $M_{\star}$, based on the Gaussian prior: $M_{\star} = 0.84\pm0.06 M_{\odot}$ \citep{Southworth2010}. We estimate these parameters similarly to \citet{Agol2010}, setting $e = 0$ based on the small inferred value of $\lbrace e\cos \omega,e\sin \omega \rbrace$ and theoretical predictions advocating for HD\,189733b's orbital circularization \citep[e.g.,][]{Fabrycky2010}. 
We discuss the incidence of this assumption (hereafter COA for circularized-orbit assumption) in Sect.\,\ref{sec:results&discussI} and tackle it in detail in Sect.\,\ref{sec:results}.

\subsubsection{Eclipse models \& limb-darkening} 

We model the transit assuming a quadratic limb-darkening law for the star, and the secondary eclipse assuming the exoplanet to be a uniformly-bright disk. We draw the limb-darkening coefficients from the theoretical tables of  \citet{Claret2011}, $u_1 = 0.0473\pm0.0032$ and $u_2 = 0.0991\pm0.0036$, based on the spectroscopic parameters of HD\,189733 \citep[{$T_{eff} = 5050\pm50K$, $\log g = 4.61\pm0.03$ and $[\frac{Fe}{H}]-0.03\pm0.05$, see}][and references therein]{Southworth2010}. We add this \textit{a priori} knowledge as a Bayesian penalty to our merit function, using as additional jump parameters the combinations $c_1 = 2u_1+u_2 $ and $ c_2 = u_1-2u_2 $, as described in G10. The incidence of this coefficient choice does not affect our results (see Sect.\,\ref{sec:results&discussI}).

\subsubsection{Systematic correction models} 

IRAC instrumental systematic variations of the observed flux, such as the pixel-phase or the detector ramp, are well-documented \citep[e.g.,][and references therein]{D'esert2009}. At 8\,$\mu$m, Si:As-based detector showed a uniform intrapixel sensitivity (i.e., negligible pixel-phase effect) but temporal evolution of their pixels gain (i.e., detector ramp). Although \citet{Agol2010} advocate using a double-exponential for modeling the detector ramp, we find that the most adequate baseline model for the present AORs is the quadratic function of $\log(dt)$ introduced by \cite{Charbonneau2008}. We discuss this matter in Sect.~\ref{sec:results&discussI}.

 We also take into account possible low-frequency noise sources (e.g., instrumental and/or stellar) with a second-order time-dependent polynomial.

 The use of linear baseline models enables to determine their coefficients by linear least-squares minimization at each step of the MCMC. For this purpose, we employed the \textit{Singular Value Decomposition} (SVD) method \citep{Press92}.

\subsubsection{Correlated noise}

We take into account the correlated noise following a procedure similar to \citet{Winn2008} for obtaining reliable error bars on our parameters. For each light curve, we estimate the standard deviation of the best-fitting solution residuals for time bins ranging from 3.5 to 30 minutes in order to assess their deviation to the behavior of white noise with binning. For that purpose, the following factor $\beta_{red}$ is determined for each time bin:

\begin{eqnarray}
\beta_{red} & = & \frac{\sigma_N}{\sigma_1}\sqrt{\frac{N(M-1)}{M}}
      , \label{Bred}
\end{eqnarray}
where $N$ is the mean number of points in each bin, $M$ is the number of bins, and $\sigma_1$ and $\sigma_N$ are respectively the standard deviation of the unbinned and binned residuals. The largest value obtained with the different time bins is used to multiply the error bars of the measurements.

\section{Secondary eclipse: anomalous ingress/egress}
\label{sec:results&discussI}

\subsection{Significance}

\begin{figure*}

  \begin{center}
    {\hspace{-0.0cm}\includegraphics[trim = 30mm 80mm 50mm 91mm,clip,width=!,height=7cm]{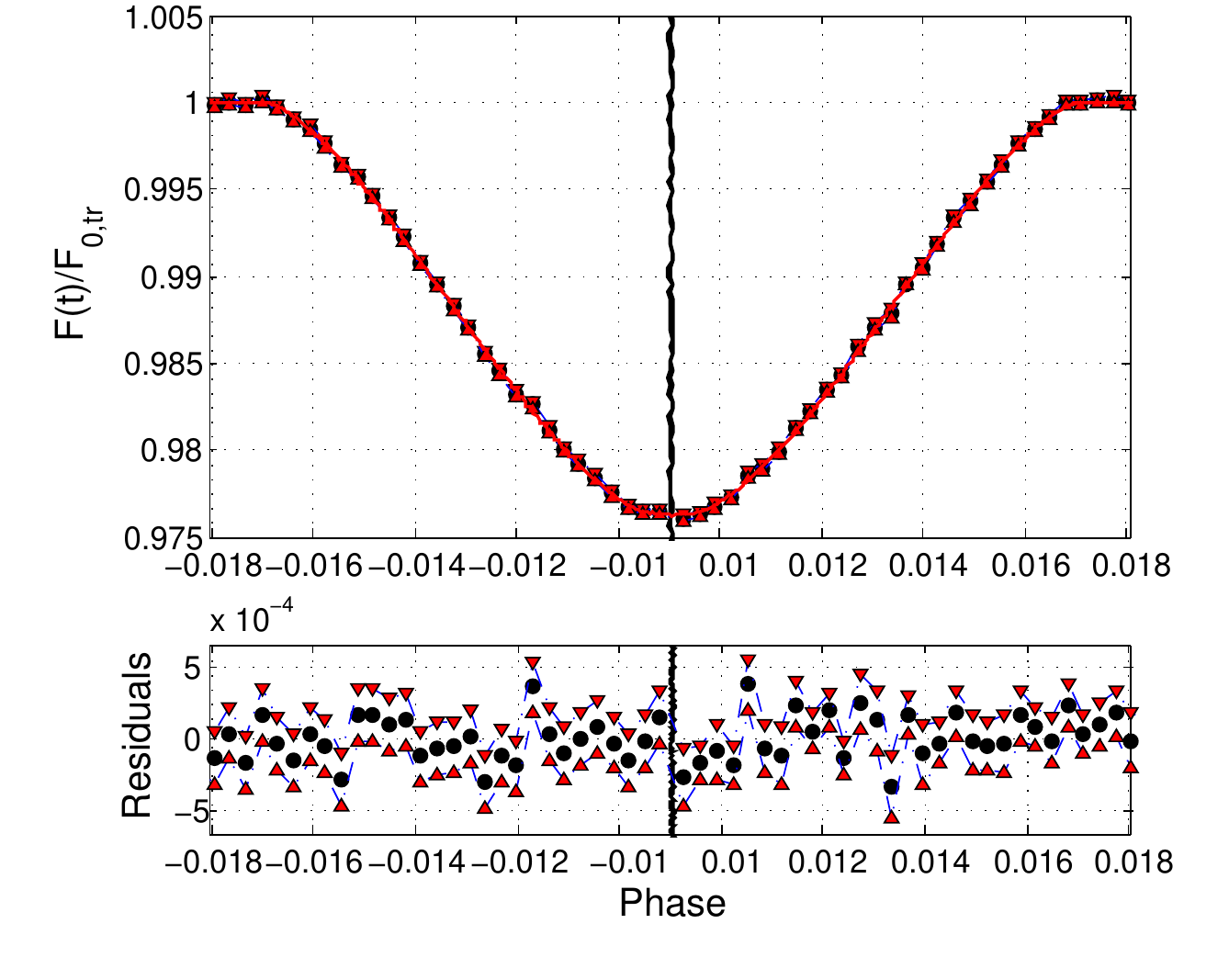}\hspace{-0.5cm}\includegraphics[trim = 30mm 80mm 50mm 91mm,clip,width=!,height=7cm]{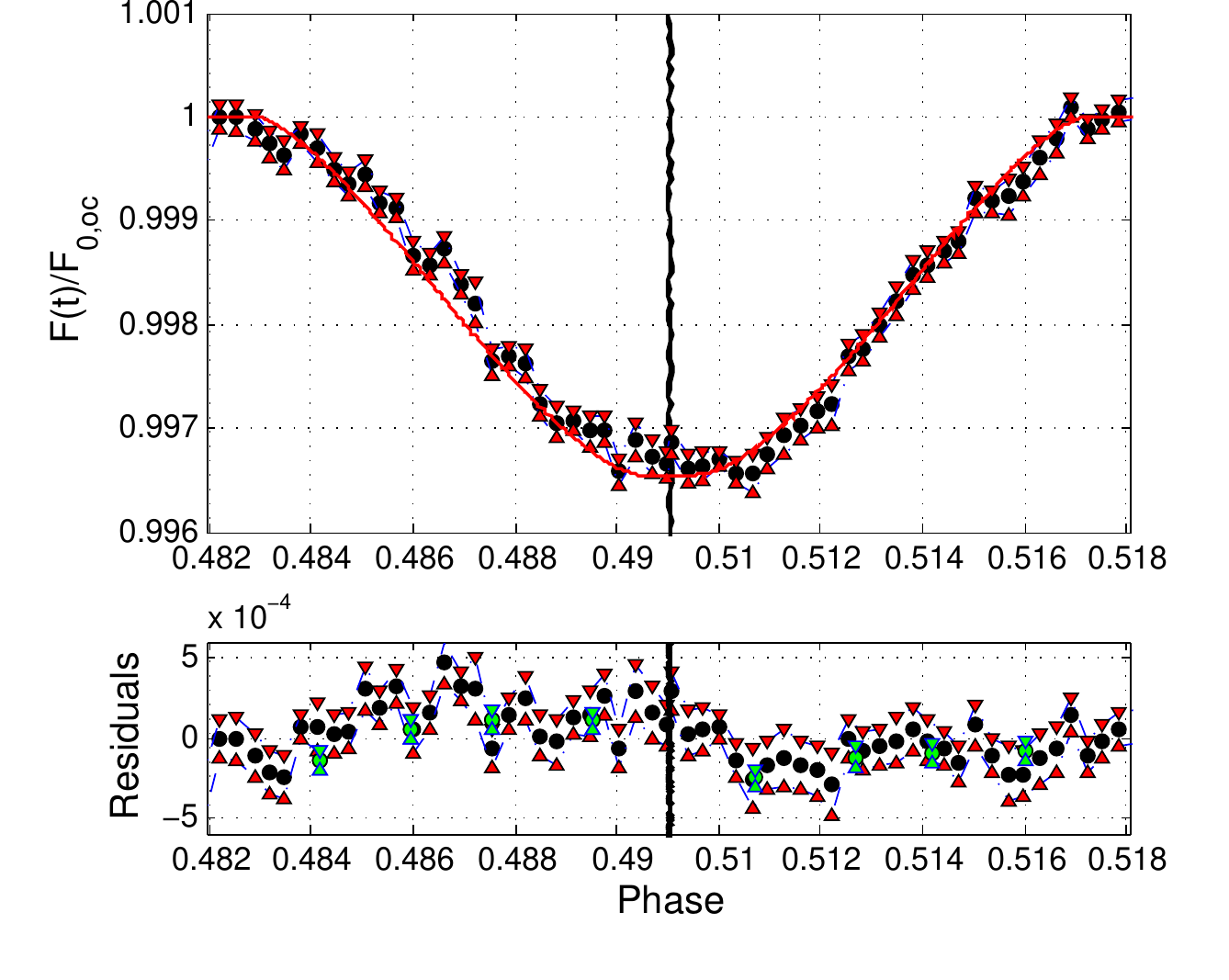}\hspace{-2.cm}}
  \end{center}
  \vspace{-0.5cm}
  \caption{IRAC 8-$\mu$m HD\,189733b's transit and occultation ingress/egress photometry binned per 1 minute and corrected for the systematics (black dots) with their 1\,$\sigma$ error bars (red triangles) and the best-fitting eclipse model superimposed (in red). The green dots present the residuals from \citet{Agol2010} (their Fig. 12), obtained using an average of the best-fit models from 7 individual eclipse analyses. \textbf{Left:} Phase-folded transits show no significant deviation to the transit of a disk during ingress/egress. \textbf{Right:} Phase-folded occultation ingress/egress deviate from the eclipse of a uniformly-bright disk, highlighting the secondary-eclipse scanning of HD\,189733b's dayside. \textbf{Top:} Phase-folded and corrected eclipse photometry. \textbf{Bottom:} Phase-folded residuals.}
  \label{fig:in_eg_structures}
\end{figure*}

We have detected an anomalous structure in the HD\,189733b occultation ingress/egress residuals (see Fig.\,\ref{fig:in_eg_structures}, bottom-right panel); which shows that the observations deviate from the occultation of uniformly-bright disk with a 6.2\,$\sigma$ significance\footnote{We determine the significance of this structure as
 $\sqrt{\sum_{i \in ingress} Y_i/\sigma_i - \sum_{i \in egress} Y_i/\sigma_i } $, where $Y_i$ and $\sigma_i$ are the flux measurement residual and its standard deviation at time $i$.}. 
 
 The IRAC 8-$\mu$m photometry of the eclipse ingress/egress corrected for the systematics and binned per 1 minute are shown in Fig.\,\ref{fig:in_eg_structures}, with the best-fitting eclipse model superimposed (solid red line). The error bars (red triangles) are rescaled by $\beta_{red}$ ($\sim$1.2) to take into account the correlated noise effects on our detection. In addition, we take advantage of the MCMC framework to account for the uncertainty induced by the systematic correction on the phase-folded light curves. For that purpose, we use the posterior distribution of the accepted baselines to estimate their median instead of using the best-fit model---which has no particular significance. Furthermore, we propagate the uncertainty of the systematic correction using the standard deviation of each bin from this posterior distribution; and increase their error bars accordingly (up to 20\%).
 
 We include in Fig.\,\ref{fig:in_eg_structures} the residuals from \citet{Agol2010} (their Fig. 12). These are obtained using an average of the best-fit models from 7 individual eclipse analyses, not from a global analysis. For further comparison, the structure detected in HD\,189733b's occultation leads to a uniform time offset of 37$\pm$6 sec ($\sim$6\,$\sigma$), in agreement with \citet{Agol2010} estimate of 38$\pm$11 sec ($\sim$3.5\,$\sigma$), light travel time deduced.

\subsection{Robustness}

We test the robustness of our results against various effects including the baseline models, the limb darkening, HD\,189733b's circularized-orbit assumption (COA) and the AORs. Although below we mainly discuss the robustness of the structure in occultation, the robustness of system parameter estimates is assessed simultaneously but tackled in detail in Sect.\,\ref{sec:uniform}.

As mentioned in Sect.\,\ref{sec:data}, \citet{Agol2010} advocate using a double-exponential for modeling the detector ramp; but we use the quadratic function of $\log(dt)$ introduced by \cite{Charbonneau2008}. The reason is that by taking advantage of our Bayesian framework we show that the quadratic function of $\log(dt)$ is the most adequate for correcting the present AORs. In particular, we use two different information criteria \citep[the BIC and the AIC see, e.g.,][]{Gelman2004} that prevent from overfitting based on the likelihood function and on a penalty term related to the number of parameters in the fitting model. (Note that the penalty term is larger in the BIC than in the AIC.) We obtain both a higher BIC ($\Delta$\,BIC $\sim$90) and a higher AIC ($\Delta$\,AIC $\sim$1.3) with the double-exponential; this means the additional parameters do not improve the fit enough, according to both criteria. The most adequate ramp model is thus the less complex quadratic function of $\log(dt)$. Nevertheless, we assess the robustness of our results to different baseline models including the double-exponential ramp, phase-pixel corrections and sinusoidal terms. These different MCMC simulations do not significantly affect the anomalous shape found in occultation ingress/egress to within  0.5\,$\sigma$. The reason is that the time scale of the baseline models are much larger than the structure detected in occultation ingress/egress; which, therefore, has no incidence on the baseline models, and \textit{vice versa}. 

We assess the incidence of the priors on the limb-darkening coefficients; for that purpose, we perform MCMC simulations with no priors on $u_1$ and $u_2$ (see Sect.\,\ref{sec:data}). Again, we observe no significant incidence to within 0.5\,$\sigma$. However, note that we outline in the next subsection the necessity of precise and independent constraints on the limb-darkening coefficient, even if these might appear to be well-constrained by high-SNR transit photometry. 

We assess the incidence of assuming HD\,189733b's orbital circularization. For that purpose, we perform MCMC simulations with a free eccentricity. We observe no significant incidence to within 0.5\,$\sigma$ for the jump parameters, except for the transit duration, the impact parameter, $\sqrt{e}\cos \omega$ and $\sqrt{e}\sin \omega$ (correlation detailed in Sect.\,\ref{sec:uniform}). In addition, we observe a net drop of the anomalous-shape significance which is explained by a compensation of the ``uniform time offset'', enabled when relaxing the constraint on $ e $ (see next subsection and Sect.\,\ref{sec:uniform}).

Finally, we also validate the independence of our results to inclusion or not of AOR subsets by analyzing different subsets of seven out of the eight eclipses. We observe relative significance decrease of $\sim$\,$ \sqrt{7/8} $; which are consistent with a significance drop due to a reduction of the sample. In particular, it shows that the anomalous shape in occultation ingress/egress is not due to one specific AOR. 

\subsection{Possible contributing factors}

The anomalous shape detected in occultation ingress/egress is in agreement with the expected signature of the offset thermal pattern indicated by HD\,189733b's phase curve from \citet{Knutson2007}. We discuss here the possible contributing factors---or origins---of this anomalous shape to propose further a consistent methodology for analysing secondary-eclipse scanning (see Sect.\,\ref{sec:ecl}).

An anomalous occultation ingress/egress possibly emerges from: \textbf{(1)} a non-circular projection of the exoplanet at conjunctions; \textbf{(2)} a biased eccentricity estimate (i.e., uniform time offset); and/or \textbf{(3)} a non-uniform brightness distribution (BD). We present in Fig.\,\ref{fig:shape_brightness} a schematic description of the anomalous occultation ingress/egress induced by a non-circular projection of the exoplanet at conjunctions (yellow) and a non-uniform BD (red). Both synthetic scenarios show specific deviations from the occultation photometry of uniformly-bright disk (black curve) in the occultation ingress/egress.

\textbf{(1)} We reject the non-circular-projection concept because the transit residuals show no anomalous structure (Fig.\,\ref{fig:in_eg_structures}, bottom-left panel). This is in agreement with current constraints on the HD\,189733b oblateness \citep[projected oblateness below 0.056, $95\%$ confidence,][]{Carter2010} and wind-driven shape \citep[expected to introduce a light curve deviation below 10\,ppm, see][]{Barne2009}. In particular, our transit residuals constrain HD\,189733b's oblateness ($95\%$ confidence) below 0.0267 in case of a projected obliquity of 45$^\circ$ and below 0.147 in case of a projected obliquity of 0$^\circ$---based on Fig.1 of \citet{Carter2010}. No significant deviation in transit ingress/egress means that HD\,189733b's shape-induced effects in occultation ingress/egress are negligible, because it is expected to be about one order of magnitude lower than in transit (i.e., effect ratio $\varpropto I_p/I_{\star}$, with $I_p$ and $I_{\star}$ respectively the exoplanet and the star mean intensities at 8\,$\mu $m). We therefore assume the exoplanet to be spherical, for a further analysis of the present data. 

We, therefore, emphasize that for constraining an exoplanet shape, one needs independent \textit{a priori} knowledge on the host star limb-darkening \citep[e.g.,][]{Claret2011} to avoid overfitting its possible signature in transit ingress/egress. 

We highlight that the phase curve also constrains a companion shape, similarly to the ellipsoidal light variations caused by a tidally-distorted star \citep[see][]{Russell1952,Kopal1959}. The contributions of the shape and the BD of an exoplanet to its phase curve may be disentangle because of their different periods \citep[see e.g.,][]{Faigler2011}. Indeed, for a synchronized exoplanet, the shape-induced modulation has a period of $ P/2 $---twice the same projection an orbital period, while the brightness-induced modulation as a period of $ P $. HD\,189733b's phase-curves show mainly a $ P $-modulation; what is in agreement with our previous constraint on HD\,189733b's shape.

\textbf{(2) and (3)} \citet{William2006} introduced the concept of uniform time offset to emphasize that a small variation of eccentricity can partially mimic the anomalous occultation ingress/egress of a non-uniformly-bright exoplanet. Therefore, an anomalous occultation ingress/egress can be partially compensated leading to biased estimates of $\sqrt{e}\cos\omega$ and $\sqrt{e}\sin\omega$. In fact, we show in Sect.\,\ref{sec:results} that not only $ e $ enables such partial compensations.

We present in Sect.\,\ref{sec:ecl} the methodology proposed to disentangle the factors possibly responsible for partial compensations.

\section{Analysis of HD\,189733b's scans}
\label{sec:ecl}

Our second analysis aims to investigate the contributing factors of the anomalous occultation ingress/egress we detect at the 6\,$\sigma$ level (see Sect.\,\ref{sec:results&discussI}). We require additional information to constrain the role of the remaining possible contributing factors. We, therefore, take advantage of the phase curve---in addition to the secondary-eclipse scanning---to constrain the exoplanet BD simultaneously with the system parameters.

We describe below the framework used for analysing HD\,189733b's secondary-eclipse scanning. We begin with an introduction of the data. Then we describe the analysis method and the models used for constraining HD\,189733b's BD at 8\,$\mu$m; we recall that HD\,189733b's shape has been constrained in Sect.\,\ref{sec:results&discussI}. 

\subsection{HD\,189733b's scans}

In this second analysis, we use the corrected and phase-folded light curves from Sect.\,\ref{sec:results&discussI} (Fig.\,\ref{fig:in_eg_structures}) and the phase curve from \citep{Knutson2007} corrected for stellar variability \citep[see][Fig. 11]{Agol2010}. As discussed in Sect.\,\ref{sec:results&discussI}, the transit and the phase curve enable to constrain respectively HD\,189733b's shape at conjunctions and its BD. Note that we discard the first third of the phase curve since it is strongly affected by systematics (i.e., detector stability).

To constrain the system parameters, we choose to use HD\,189733b's phase-folded transits instead of using the system parameter estimates from Sect.\,\ref{sec:data} (see Table\,\ref{tab:BFP}, column 2) for two reasons. First, these estimates are under the form of 1D Gaussian distribution while light curves provide a complex posterior probability distribution (PPD) over the whole parameter space. Secondly, these estimates are affected by HD\,189733b's COA (see implications in Sect.\,\ref{sec:results}). For those reasons, it is relevant to simultaneously analyse  HD\,189733b's scans and estimate the system parameters, to be consistent with our methodology of a ``global'' analysis and to avoid propagation of bias through inadequate priors, i.e., inadequate assumptions.

\subsection{Analysis method}

The procedure we develop for analysing HD\,189733b's scans uses a second, new, MCMC implementation. This implementation differs from the one introduced in Sect.\,\ref{sec:data} in that it models a non-uniformly-bright exoplanet. This implementation uses Keplerian orbits and models the transit photometry using the model from \citet{Mandel2002}. For that reason, it uses the jump parameters of the conventional-analysis method together with new jump parameters for the exoplanet brightness models (see below). However, note that we choose to use the stellar density, $\rho_{\star}$, instead of $W$ as jump parameter; because it relates directly to the orbital parameters \citep{Seager2003}. Furthermore, because we use the phase-folded light curves, the main constraint on the orbital period is missing. We, therefore, use a uniform prior on $P$, centered on the conventional-analysis estimate. In particular, we use a large prior (with an arbitrary symmetric extension of 10\,$\sigma_P$, where $\sigma_P$ is the estimated uncertainty on $P$, see Table\,\ref{tab:BFP}) to prevent our results to be affected by the assumption underlying this estimate. Note however that this has no incidence on our further results---the reason is that the orbital period is highly-constrained by the transit epochs and, therefore, is not affected by effects on the occultation.

\subsection{Non-uniformly-bright exoplanet: light curve models}

	We model the phase curve and the secondary eclipse by performing a numerical integration of the observed exoplanet flux. Model approximations include: ignoring the time variability of the target atmosphere \citep[in line with atmospheric models, e.g.,][]{Cooper2005}---especially because we use here phase-folded light curves; ignoring the planet limb darkening---expected to be weak for hot Jupiters, in particular at 8\,$\mu$m; and assuming HD\,189733b's rotational period to be synchronized with its orbital period---synchronization occurs over $\sim$\,$10^6$\,yr for hot Jupiters \citep[see e.g.,][]{Winn2010}. 
	
	From the exoplanet BD and the system orbits, we model the flux temporal evolution of the exoplanet by sampling its surface with a grid of $2N$ points in longitude ($\phi$) and $N+1$ points in latitude ($\theta$). We fix N to 100 to mitigate numerical effects up to $10^{-3}$ the secondary-eclipse depth, i.e. below 4\,ppm; in comparison, the data photometric precision is $\sim$130\,ppm. (We validate the independence of our inferences to higher grid resolutions.)
	
	Our brightness models (described in the next subsection) are composed of several modes/degrees, e.g., spherical harmonics. At each step of the MCMC, we \textbf{(1)} simulate the system orbits, \textbf{(2)} estimate the light curve corresponding to each of the brightness model modes and \textbf{(3)} determine the mode amplitude based on their simulated light curve by least-squares minimization using the SVD method. Practically, to employ consistently the SVD method within a stochastic framework, we use the covariance matrix to perturb the mode-amplitude estimates.
	
	We model here the transit as in our first analysis (see Sect.\,\ref{sec:data}) because we show that the projection of HD\,189733b at conjunctions does not deviate significantly from a disk (see Sect.\,\ref{sec:results&discussI}).

\subsection{Non-uniform brightness models}

To model the broad patterns of HD\,189733b's BD, we use two groups of models: the spherical harmonics and toy models. For the spherical harmonics, we use two additional jump parameters per degree for direction (i.e., longitude and latitude). The generalized formulation of the BD, $\Gamma_{SH,d}(\phi,\theta)$, can thus be written as;

\begin{eqnarray}	
      \Gamma_{SH,d}(\phi,\theta) & = & \sum_{l = 0}^d I_l Y_l^0(\phi-\Delta\phi_l,\theta-\Delta\theta_l),
       \label{SH}
\end{eqnarray}
where $Y_l^0(\phi-\Delta\phi_l,\theta-\Delta\theta_l)$ is the real spherical harmonic of degree $l$ and order $0$, directed to $\Delta\phi_l$ in longitude and $\Delta\theta_l$ in latitude. $I_l$ is the $l$th-mode amplitude that is estimated at each step of the MCMC using the perturbed SVD method. For instance, the additional parameters for a dipolar fitting model compared to the conventional-analysis method are: $\Delta\phi_1$ and $\Delta\theta_1$ as jump parameters and $I_1$ as a linear coefficient; $I_0$ is the amplitude of the uniformly-bright mode, included in the conventional-analysis method.

On the other hand, we use toy models that enable modeling a thermal pattern---a hot or cold spot---of various shape. Their expression can be written simply as;

\begin{eqnarray}	
      \Gamma_1(\phi,\theta) & = &  I_1 \phi_\circ^\alpha \exp[-\phi_\circ^\beta] \cos {}^\gamma\theta_\circ + I_0 , \label{BM1}\\
      \Gamma_2(\phi,\theta) & = & 
	\left\{{
   \begin{array}{c c}
    I_1\cos {}^\alpha \phi_\circ \cos {}^\gamma \theta_\circ + I_0 & \mbox{if } \phi_\circ\geq0 
  	\\
  	I_1\cos {}^\beta \phi_\circ \cos {}^\gamma\theta_\circ + I_0 & \mbox{if } \phi_\circ<0 
  \end{array}}
  	\right.
  	,\label{BM2}\\
      \Gamma_3(\phi,\theta) & = & 
	\left\{{
   \begin{array}{c c}
    I_1\exp[-(\frac{\phi_\circ}{\alpha})^2]\exp[-(\frac{\theta_\circ}{\gamma})^2] + I_0 & \mbox{if } \phi_\circ\geq0 
  	\\
  	I_1\exp[-(\frac{\phi_\circ}{\beta})^2]\exp[-(\frac{\theta_\circ}{\gamma})^2] + I_0 & \mbox{if } \phi_\circ<0
  \end{array}}
  	\right.
  	,\label{BM1}
\end{eqnarray}
where $\phi_\circ$ and $\theta_\circ$ are respectively the longitude and latitude relative to the position of the model extremum, i.e. $\phi_\circ = f(\phi,\Delta\phi,\alpha,\beta,\gamma)$ and $\theta_\circ = f(\theta,\Delta\theta,\alpha,\beta,\gamma)$. $\Delta\phi$ and $\Delta\theta$ are respectively the longitudinal and the latitudinal shift of the model peak from the substellar point. $\alpha,\beta$ and $\gamma$ parametrize the shape of the hot/cold spot. These toy models add to the conventional-analysis method five jump parameters ($\Delta\phi,\Delta\theta,\alpha,\beta,\gamma$) and a linear coefficient ($I_1$). 

Finally, we emphasize that the use of several brightness models is critical to assess the model-dependence of our results. We discuss this important matter in Sect.\,\ref{sec:discussionII}.

\section{Results}
\label{sec:results}

For HD\,189733's system, we find that relaxing the eccentricity constraint and using more complex brightness distributions (BDs) lead to lower
stellar/planetary densities and a more localized and latitudinally-shifted hot spot. In particular, we find that the more complex HD\,189733b's brightness model, the larger the eccentricity, the lower the densities, the larger the impact parameter and the more localized and latitudinally-shifted the hot spot estimated. This ``$ e $-$ b $-$\rho_{\star}$-BD correlation'' is of primary importance for data of sufficient quality. We present in this section our results for increasing model complexity to gain insight into the incidence of the model underlying assumptions---e.g., the COA and the uniformly-bright exoplanet assumption (hereafter UBEA). In particular, we relate their incidence to the $ e $-$ b $-$\rho_{\star}$-BD correlation. 

We gather the system parameter estimates for different fitting models in Table\,\ref{tab:BFP}; it shows the median values and the $68\%$ probability interval for our jump parameters. We compute our estimates based on the posterior probability distribution (PPD) of global MCMC simulations, i.e., not as a weighted mean of individual transit or eclipse analyses. Our conventional-analysis estimates are in good agreement with previous studies \citep[e.g.,][]{Winn2007,Triaud2009,Agol2010}. We discuss further the system parameter estimates obtained from our second analysis.

\subsection{Assuming HD\,189733b to be uniformly bright}
\label{sec:uniform}

 We show here the similar effects of the COA and the UBEA. Both assumptions prevent from exploring dimensions of the parameter space and, therefore, lead to more localized, and possibly biased, PPD.
 
\begin{figure*}
   \centering

  \begin{center}
    \hspace{-0.cm}\includegraphics[trim = 70mm 50mm 85mm 95mm,clip,width=14cm,height=!]{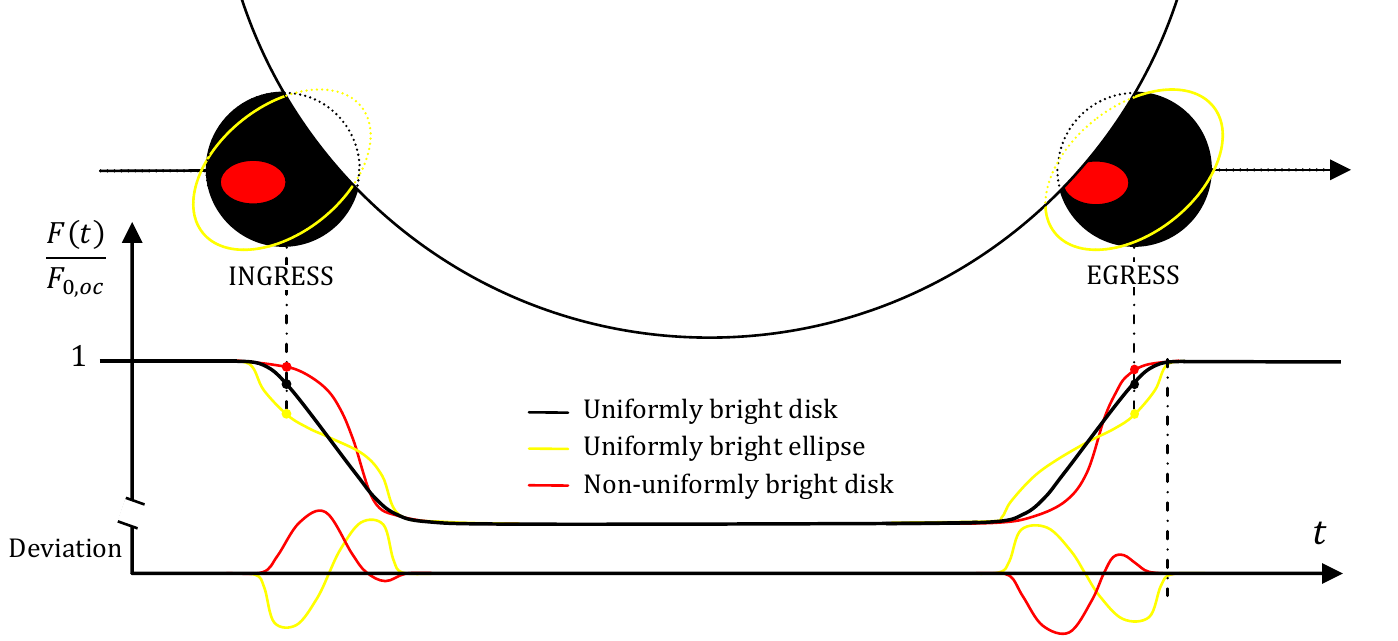}
  \end{center}
  \vspace{-0.0cm}
  \caption{Schematic description of the anomalous occultation ingress/egress induced by the shape or the brightness distribution of an exoplanet. The red curve indicates the occultation photometry for a non-uniformly-bright disk (hot spot in red). The yellow curve indicates the occultation photometry for an oblate exoplanet (yellow ellipse). Both synthetic scenarios
show specific deviations from the occultation photometry of uniformly-bright disk (black curve) in the occultation ingress/egress.}
  \label{fig:shape_brightness}

    \end{figure*}

\begin{figure*}
   \centering

  \subfloat[]{\hspace{-05mm}\label{fig:uniform_e_ddp}\includegraphics[trim = 35mm 85mm 35mm 85mm,clip,width=0.35\textwidth,height=!]{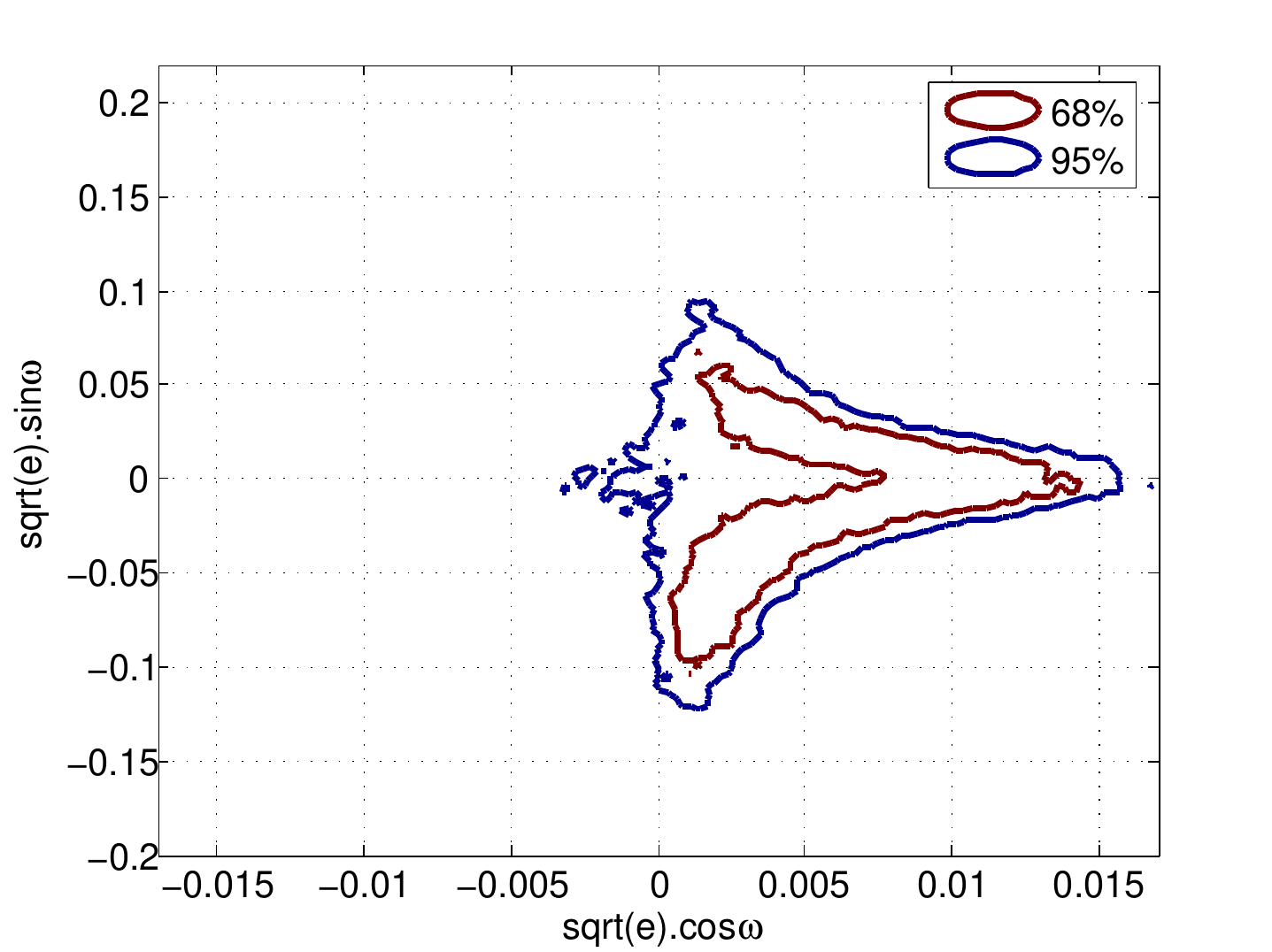}}
  ~ 
  \subfloat[]{\label{fig:uniform_brho_ddps}\includegraphics[trim = 35mm 85mm 35mm 85mm,clip,width=0.35\textwidth,height=!]{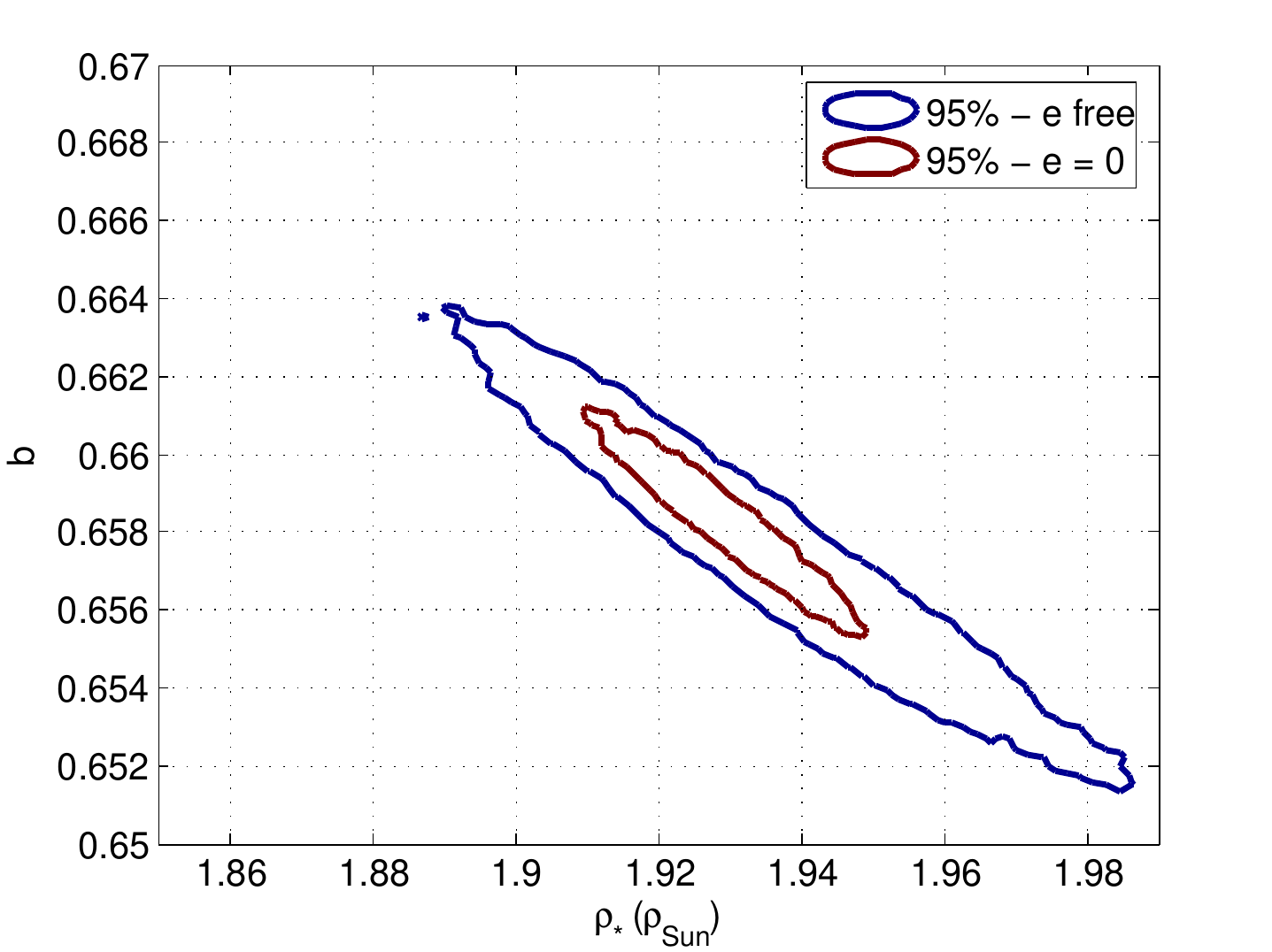}}
  ~ 
  \subfloat[]{\label{fig:Impact_syst_param}\includegraphics[trim = 35mm 85mm 35mm 85mm,clip,width=0.35\textwidth,height=!]{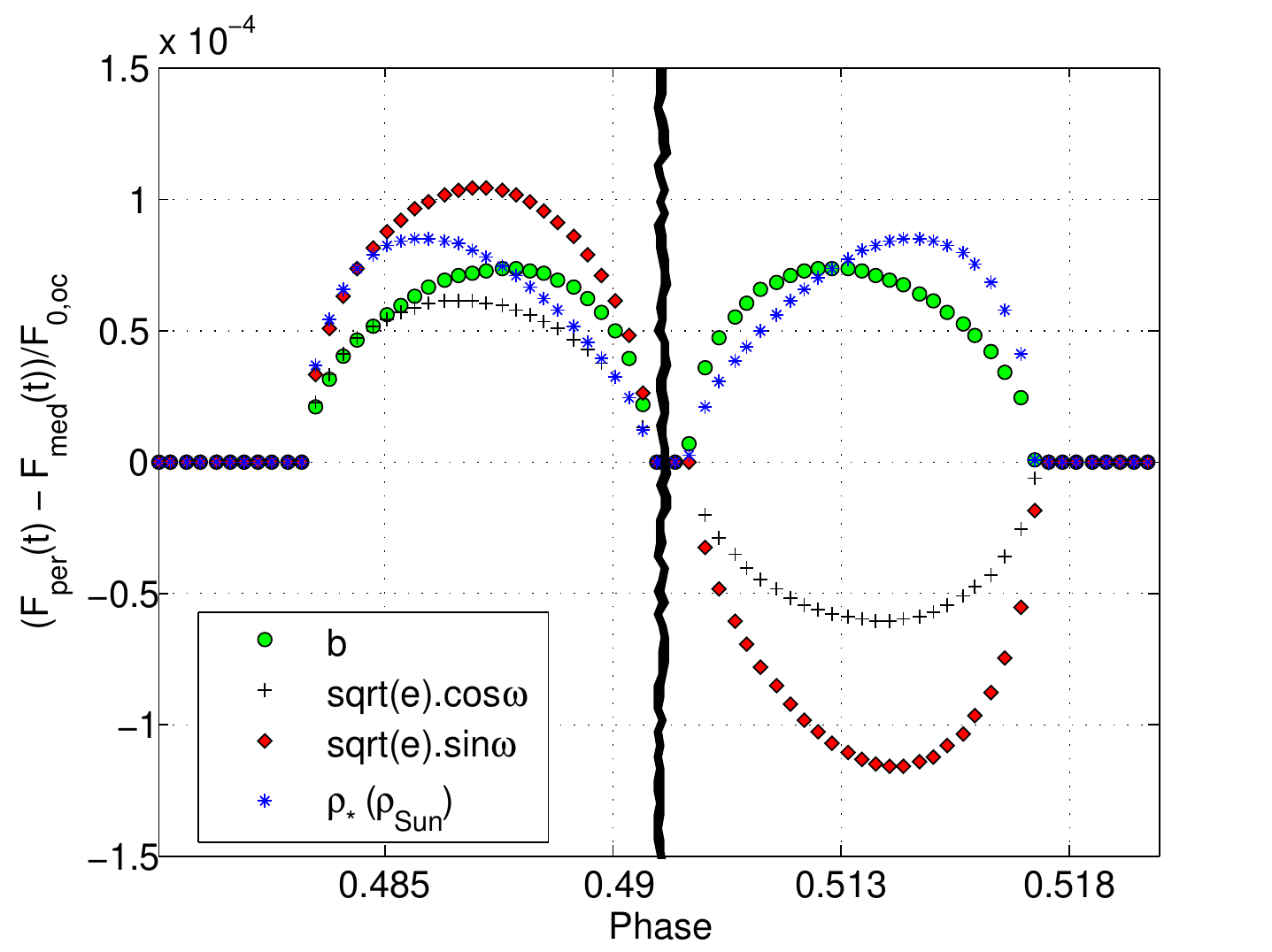}}

  \caption{Incidence of assuming an exoplanet to be uniformly bright and its orbit to be circularized. \protect\subref{fig:uniform_e_ddp} Marginal posterior probability distribution (PPD, $68\%$- and $95\%$-confidence intervals) of $\sqrt{e}\cos\omega$ and $\sqrt{e}\sin\omega$ that shows an unusual correlation. \protect\subref{fig:uniform_brho_ddps} Marginal PPD ($95\%$-confidence intervals) of $\rho_{\star}$ and $b$ that highlights the increase of adequate solutions enabled by the additional dimensions of the parameter space probed when relaxing the circularized orbit assumption. \protect\subref{fig:Impact_syst_param} Deviations in occultation ingress/egress from the median-fit model for the individual perturbations of $b$, $\sqrt{e}\cos\omega$, $\sqrt{e}\sin\omega$ and $\rho_{\star}$ by their estimated uncertainty (see Table\,\ref{tab:BFP}, column 2).
  It outlines that the system parameters enable compensation of an anomalous occultation that emerges from, e.g., a non-uniformly-bright exoplanet (see Fig.\,\ref{fig:shape_brightness}) leading to biased estimates of the system parameters.}
  \label{fig:correl_uniform}
\end{figure*}

 We first present an unusual correlation between $\sqrt{e}\cos \omega$ and $\sqrt{e}\sin \omega$ (see Fig.\,\ref{fig:uniform_e_ddp}).
It emerges from the partial compensation of the anomalous occultation ingress/egress enabled primarily by adequate combinations of $\sqrt{e}\cos \omega$ and $\sqrt{e}\sin \omega$, for conventional analyses. In particular, these parameters enable both to shift the occultation by
 \begin{eqnarray}	
      \Delta T_{0,oc} \approx \frac{2P}{\pi} e \cos\omega,
       \label{DeltaToc}
\end{eqnarray}
(i.e., uniform time offset) and to change its duration
 \begin{eqnarray}	
      \frac{W_{oc}}{W_{tr}} \approx 1 + 2 e \sin\omega,
       \label{Durations}
\end{eqnarray}
what is sufficient to partially compensate, e.g., the effect of a hot spot---compare the red and the black curves in Fig.\,\ref{fig:shape_brightness}. 
However, $\rho_{\star}$ and $b$ are also affected by the occultation shape and timing, as they constrain respectively $a/R_{\star}$ and $i$. We emphasize this point in Fig.\,\ref{fig:Impact_syst_param}. Fig.\,\ref{fig:Impact_syst_param} shows the deviations in occultation ingress/egress induced by individual perturbations of $b$, $\sqrt{e}\cos\omega$, $\sqrt{e}\sin\omega$ and $\rho_{\star}$ by their estimated uncertainty (see Table.\,\ref{tab:BFP}, column 3). This indicates that adequate combinations of these parameters mimic partially an anomalous occultation ingress/egress (see Fig.\,\ref{fig:in_eg_structures}, bottom-right panel, and Fig.\,\ref{fig:Impact_syst_param}). Therefore, the COA and the UBEA also affect the marginal PPD of $\lbrace\rho_{\star},b\rbrace$ by inhibiting the exploration of dimensions of the parameter space. We present in Fig.\,\ref{fig:uniform_brho_ddps} the extension of the $\lbrace\rho_{\star},b\rbrace$ marginal PPD that results from relaxing the COA. Similarly, we expect that the relaxation of the UBEA would significantly affect the system-parameter PPD---compare Figs.\,\ref{fig:shape_brightness} and \,\ref{fig:Impact_syst_param}. This motivates our global approach that constrains simultaneously the possible contributing factors of an anomalous occultation ingress/egress and, therefore, prevents the incidence of supplementary assumptions---e.g., the COA and the UBEA.

 \subsection{Assuming HD\,189733b to be non-uniformly bright}
 
\begin{figure*}
   \centering

  \begin{center}
    \hspace{-0.cm}\includegraphics[angle = -90, trim = 40mm 00mm 40mm 00mm,width=\textwidth,height=!]{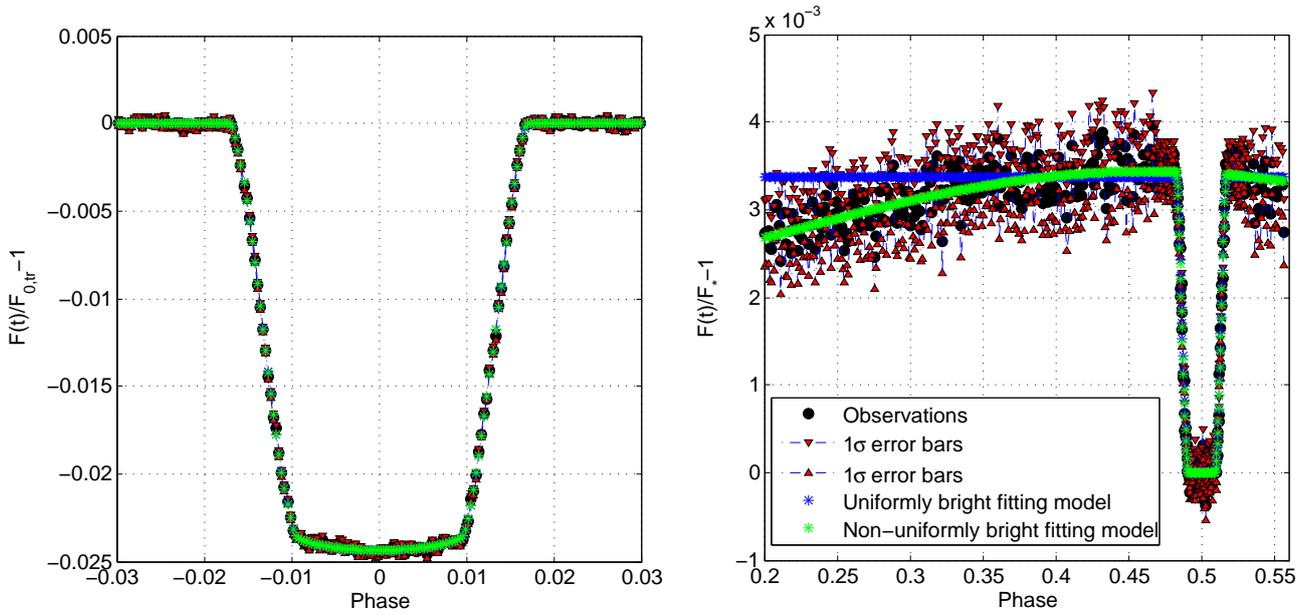}
  \end{center}
  \vspace{-0.5cm}
  \caption{Phase-folded IRAC 8-$\mu$m HD\,189733b's photometry binned per 1 minute and corrected for the systematics (black dots) with their 1\,$\sigma$ error bars (red triangles) and the best-fitting eclipse models superimposed. \textbf{Left:} Phase-folded transits. \textbf{Right:} Phase curve and phase-folded occultations that show the benefit of using non-uniform brightness model.}
  \label{fig:fits}

\end{figure*}

 We show in this subsection the incidence of relaxing the UBEA on the fit improvement and on the system-parameter PPD. We present in Fig.\,\ref{fig:fits} the phase-folded IRAC 8-$\mu$m photometry of HD\,189733b, corrected for the systematics with the best-fitting eclipse models for a uniformly (blue) and a non-uniformly (green) bright exoplanet superimposed. The best-fitting non-uniformly-bright eclipse model is shown for the $\Gamma_{SH,1}$ model---chosen arbitrarily, as the non-uniformly-bright models provide similar fits (see Table\,\ref{tab:BFP}). In particular, these models are significantly more adequate according to both the BIC and the AIC, see Table\,\ref{tab:BFP} (odds ratio: $\sim$\,$10^{36}$). 
 
 First, we introduce the results obtained using unipolar (i.e., with one spot on the planetary dayside) BDs to gain insight into the incidence of relaxing the UBEA. Then, we introduce the results obtained using multipolar BDs to assess the validity of trends observed in the unipolar-model results.

 \subsubsection{Unipolar brightness distribution}
 \label{sec:unipolar}
		
\begin{figure*}
  \centering
  
  \subfloat[]{\hspace{+00mm}\label{fig:101_e_ddp}\includegraphics[trim = 35mm 85mm 35mm 92mm,clip,width=0.5\textwidth,height=!]{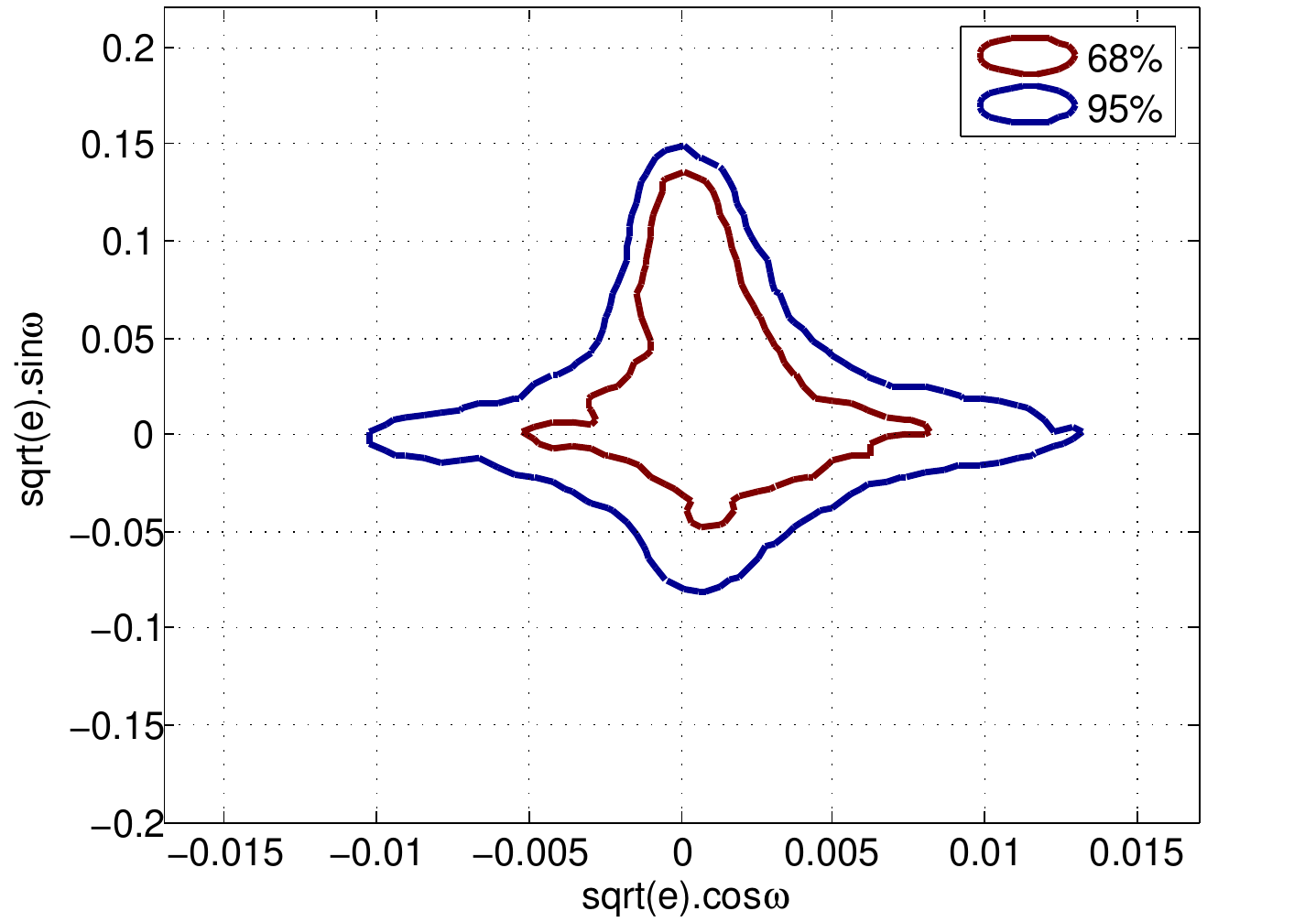}}
  ~ 
  \subfloat[]{\label{fig:3_e_ddp}\includegraphics[trim = 35mm 85mm 35mm 92mm,clip,width=0.5\textwidth,height=!]{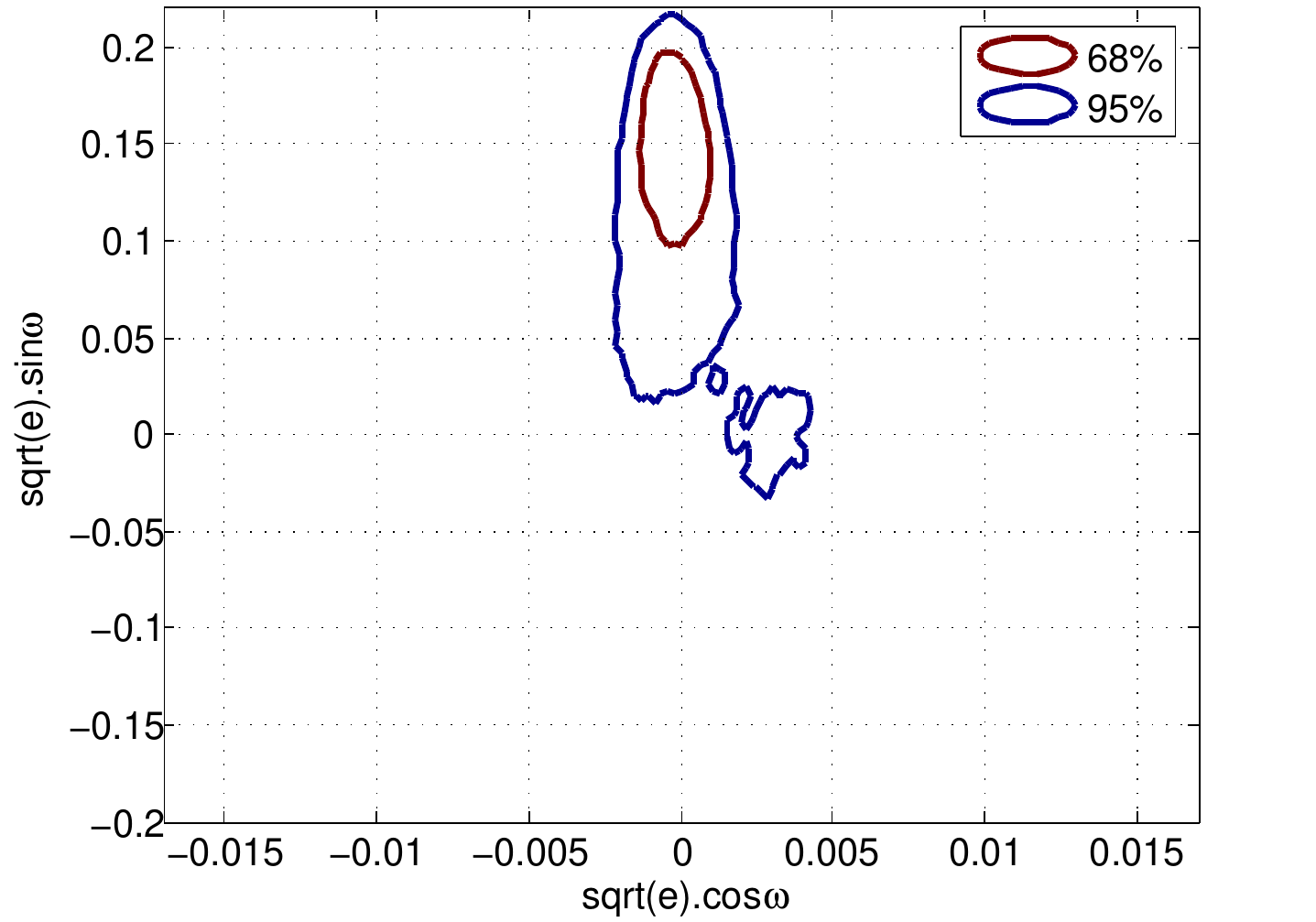}}

   \subfloat[]{\hspace{+00mm}\label{fig:101_erho_ddps}\includegraphics[trim = 35mm 85mm 35mm 92mm,clip,width=0.5\textwidth,height=!]{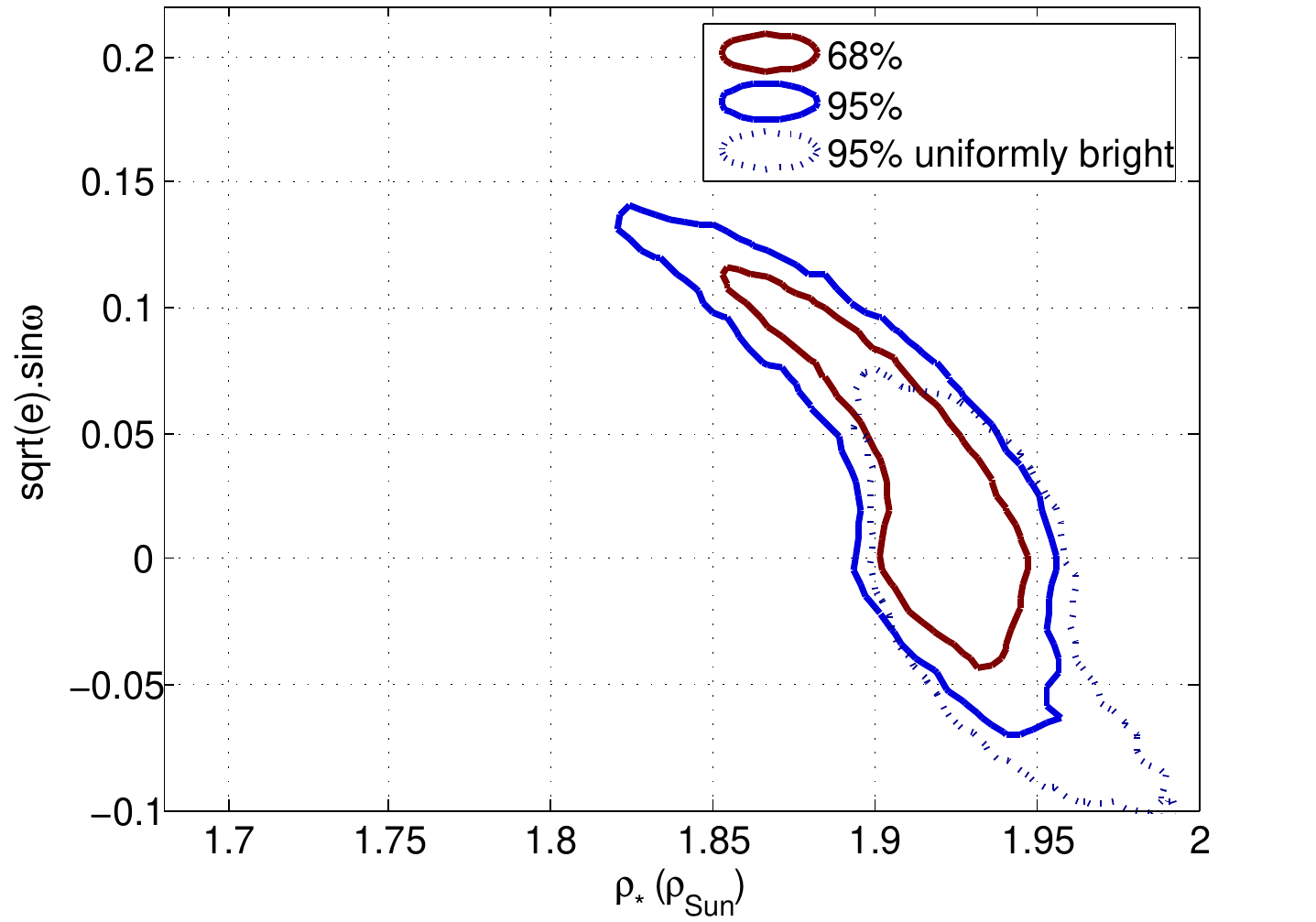}}
  ~  
  \subfloat[]{\label{fig:3_erho_ddps}\includegraphics[trim = 35mm 85mm 35mm 92mm,clip,width=0.5\textwidth,height=!]{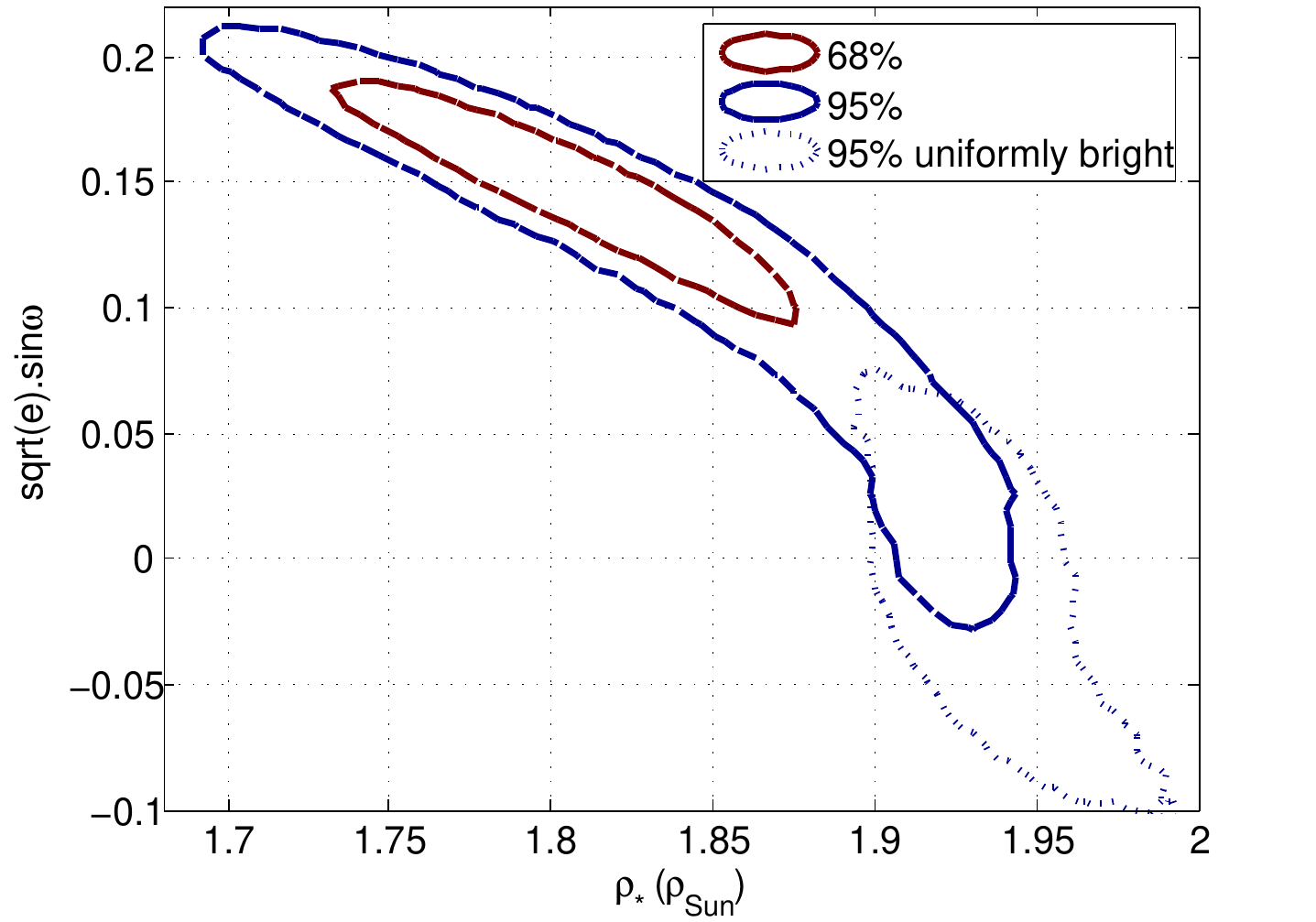}}

  \caption{Incidence of the brightness model complexity on the system-parameter posterior probability distribution (PPD), using unipolar models. \protect\subref{fig:101_e_ddp} \&  \protect\subref{fig:3_e_ddp} Marginal PPDs ($68\%$- and $95\%$-confidence intervals) of $\sqrt{e}\cos\omega$ and $\sqrt{e}\sin\omega$ for two unipolar brightness models, respectively, with a fixed and large structure ($\Gamma_{SH,1}$) and with free-confinement structure ($\Gamma_{2}$). A comparison with Fig.\,\ref{fig:uniform_e_ddp} shows that $\sqrt{e}\cos\omega$ is constrained closer to zero. The reason is that non-uniform brightness models enable the exploration of additional dimensions of the parameter space and, therefore, provide additional adequate combinations of the contributing factors to compensate an anomalous occultation. In particular, the uniform time offset is now mainly compensated by an non-uniformly BD, rather than by $\sqrt{e}\cos\omega$ as in conventional analysis. It shows also the evolution of the marginal PPD toward larger $\sqrt{e}\sin\omega$ when using more complex brightness models---which enable more localized brightness structure. \protect\subref{fig:101_erho_ddps} \&  \protect\subref{fig:3_erho_ddps} Marginal PPDs of $\rho_{\star}$ and $\sqrt{e}\sin\omega$ for, respectively, $\Gamma_{SH,1}$ and $\Gamma_{2}$. These show the impact of the brightness distribution on the retrieved system parameters (i.e., $ e $-$ b $-$\rho_{\star}$-BD correlation); in particular, it outlines the possible overestimation of $\rho_{\star}$ by 5\% (i.e., at 6\,$\sigma$ of the conventional estimate) when using extended brightness models, e.g., a uniform brightness distribution.}
  \label{fig:correl_mono}
\end{figure*}

	We first present the incidence of relaxing the UBEA on the system parameters. For that purpose, we show in Figs.\,\ref{fig:101_e_ddp} and \ref{fig:101_erho_ddps} respectively the marginal PPDs of $\lbrace\sqrt{e}\cos\omega,\sqrt{e}\sin\omega\rbrace$ and  $\lbrace\rho_{\star},\sqrt{e}\sin\omega\rbrace$ for the $\Gamma_{SH,1}$ model, and the ones for the $\Gamma_{2}$ model in Figs.\,\ref{fig:3_e_ddp} and \ref{fig:3_erho_ddps}. We superimpose in Figs.\,\ref{fig:101_erho_ddps} and \ref{fig:3_erho_ddps} the 95\%-confidence interval obtained for the uniformly-bright model to extend our previous observations regarding the incidence of relaxing supplementary assumptions, e.g., the COA (see Sect.\,\ref{sec:uniform}; Fig.\,\ref{fig:uniform_brho_ddps}). We observe the increases of $\sqrt{e}\sin\omega$ and $b$ and the decrease of $\rho_{\star}$ while $\sqrt{e}\cos\omega$ is constrained closer to zero (see Table\,\ref{tab:BFP}). 
	The reason is that the compensation of HD\,189773b's anomalous occultation is now also enabled by the non-uniform brightness models; which provide a better compensation than $e$ solely---with $\sqrt{e}\cos\omega$ (i.e., uniform time offset) for conventional analysis. Therefore, numerous combinations of $e$- and BD-based compensations are adequate. In other words, $e$ and the BD are correlated as highlighted by the PPD in Fig.\,\ref{fig:101_e_phi_ddp}. Finally, we note a progressive evolution of the system-parameter PPD with the brightness-model complexity (from uniform to $\Gamma_2$). We assess further the validity of these observations, using spherical harmonics of higher degree.

\begin{figure*}
   \centering

  \begin{center}
    \hspace{-0.cm}\includegraphics[angle = -90, trim = 70mm 10mm 70mm 10mm,clip,width=\textwidth,height=!]{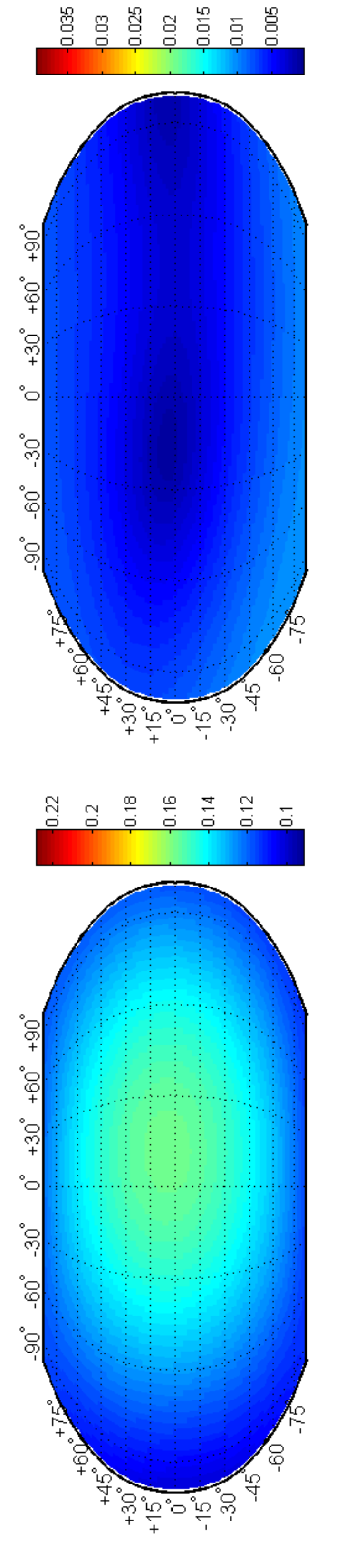}
  \end{center}
  \vspace{-0.5cm}
  \caption{Estimate of HD\,189733b's global brightness distribution in the IRAC 8-$\mu$m channel using the $\Gamma_{SH,1}$ brightness model. \textbf{Left:} Relative brightness distribution at HD\,189733b's dayside. \textbf{Right:} Dayside standard deviation. Because of its fixed and large structure, the $\Gamma_{SH,1}$ brightness model is well-constrained in amplitude (by the occultation depth) and in longitudinal localization (by the phase curve). However, it is less constrained in latitude (by the secondary-eclipse scanning) than more confined model (schematic description in Fig.\,\ref{fig:shape_brightness}), e.g., $\Gamma_{2}$ (see Fig.\,\ref{fig:3_brightness_ddps}). These model-induced constraints are observable on the dayside standard deviation; which is significantly lower than for more complex brightness models (see Figs.\,\ref{fig:3_brightness_ddps},\,\ref{fig:102_brightness_ddps} and \,\ref{fig:103_brightness_ddps}). In addition, the standard deviation distribution for the $\Gamma_{SH,1}$ model is related to its gradient (with a larger variation from the brightness peak localization along the latitude axis than along the longitude axis), because the brightness distributions accepted along the MCMC simulations differ from each other mainly in latitudinal orientation (see Fig.\,\ref{fig:ddps_101}).}
  \label{fig:101_brightness_ddps}

\end{figure*}

\begin{figure*}
   \centering

  \begin{center}
    \hspace{-0.cm}\includegraphics[angle = -90, trim = 70mm 10mm 70mm 10mm,clip,width=\textwidth,height=!]{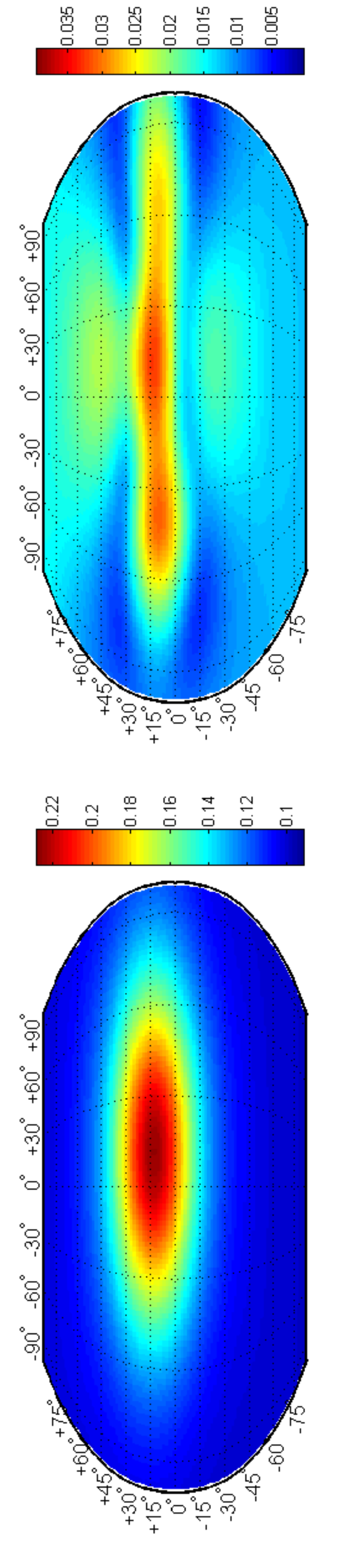}
  \end{center}
  \vspace{-0.5cm}
    \caption{Estimate of HD\,189733b's global brightness distribution in the IRAC 8-$\mu$m channel using the $\Gamma_{2}$ brightness model. \textbf{Left:} Relative brightness distribution at HD\,189733b's dayside. \textbf{Right:} Dayside standard deviation. Because of its increased complexity, the $\Gamma_{2}$ brightness model enable more localized structure that are less constrained in amplitude (by the occultation depth) and in longitudinal localization (by the phase curve) than large-and-fixed-structure model, e.g., the $\Gamma_{SH,1}$ model (see Fig.\,\ref{fig:101_brightness_ddps}). However, it is well-constrained in latitude by the secondary-eclipse scanning that is sensitive to confined brightness structure. These model-induced constraints are observable on the dayside standard deviation; which shows a maximum at the brightness peak localization and extended wings towards west and east along the planetary equator. The reason is that the brightness distributions accepted along the MCMC simulations mainly affect the former by their amplitude change and the latter by their structure change (see Fig.\,\ref{fig:ddps_3}).}
  \label{fig:3_brightness_ddps}

\end{figure*}

\begin{figure*}
  \centering
  
   \subfloat[]{\label{fig:101_e_phi_ddp}\includegraphics[trim = 35mm 85mm 35mm 92mm,clip,width=0.5\textwidth,height=!]{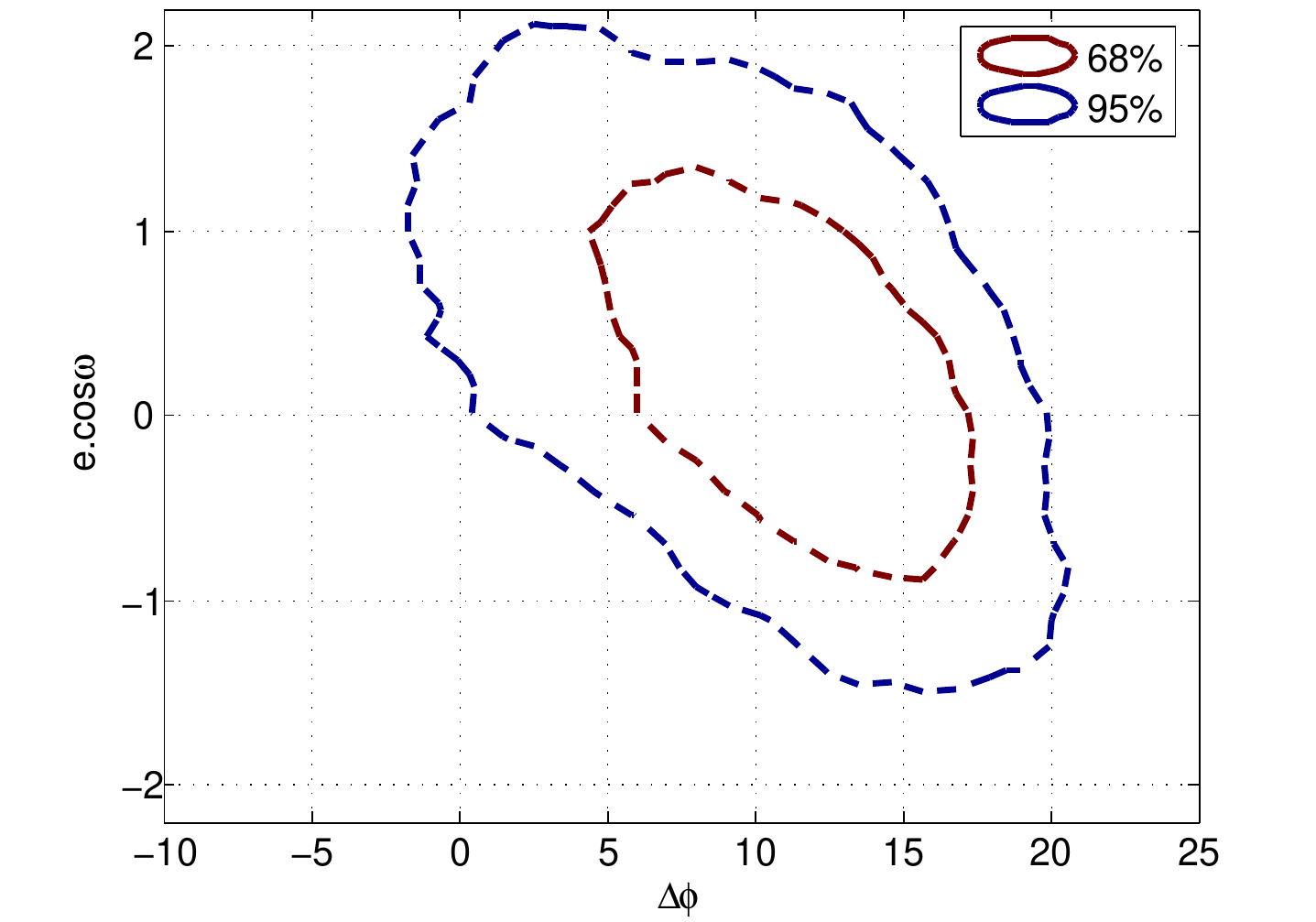}}	
  ~  
  \subfloat[]
  {\hspace{+00mm}\label{fig:101_peak_ddps}\includegraphics[trim = 35mm 85mm 35mm 92mm,clip,width=0.5\textwidth,height=!]{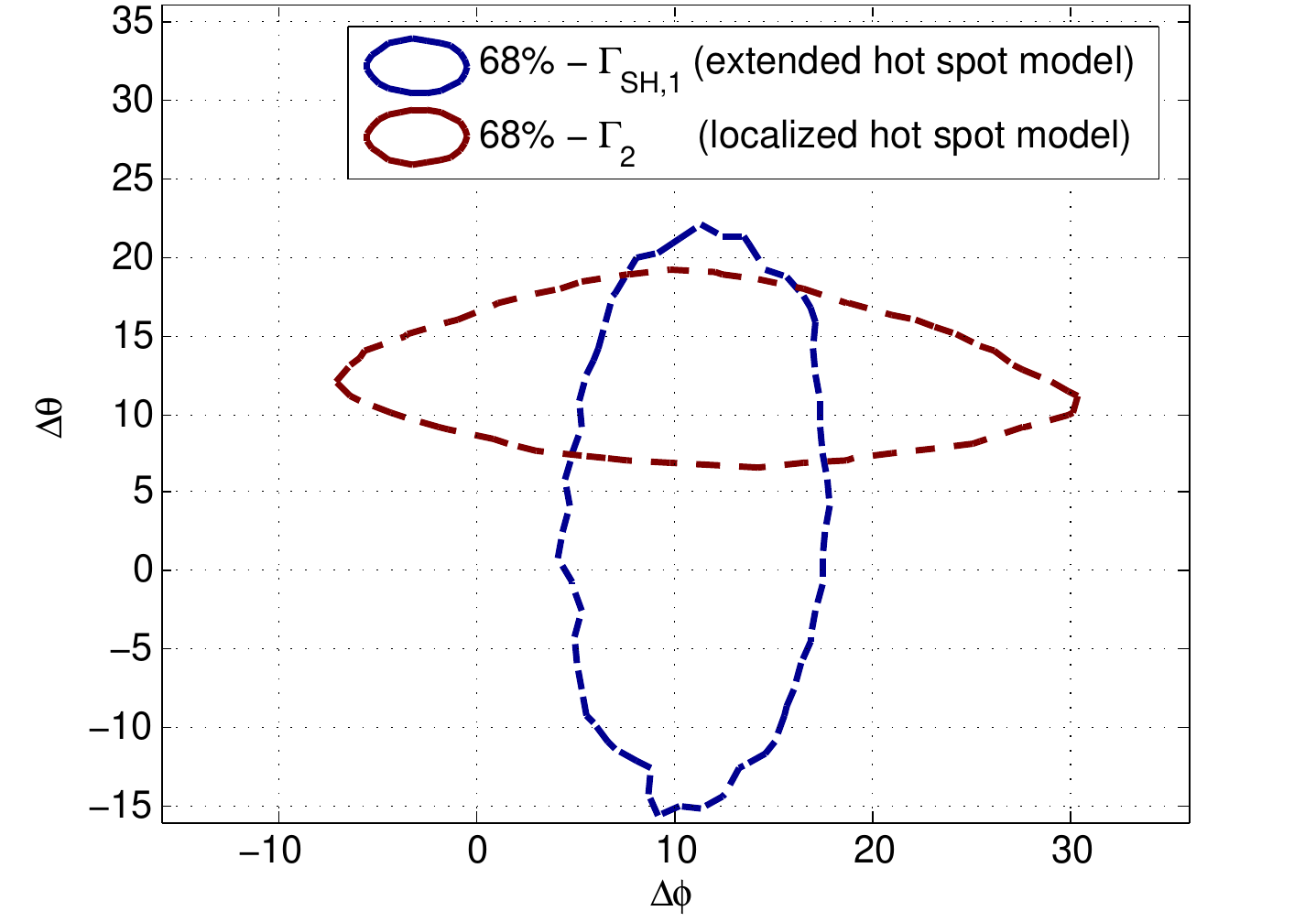}}
  
  \caption{Dependence and significance of the brightness peak localization. \protect\subref{fig:101_e_phi_ddp} Marginal PPD ($68\%$- and $95\%$-confidence intervals) of the brightness peak longitude, $\Delta\phi$, and $e\cos\omega$ for the $\Gamma_{SH,1}$ brightness model---the simplest non-uniform brightness model used in this study. This shows the correlation between the brightness model and the orbital eccentricity. This correlation emerges from enabling compensation of the anomalous occultation with a larger set of contributing factors, i.e., by including non-uniform brightness models. In particular, the uniform time offset is now mainly compensated by an non-uniformly-bright model, rather than by $e\cos\omega$ as in conventional analysis (see Fig.\,\ref{fig:shape_brightness}, Eq.\,\ref{DeltaToc} and Fig.\,\ref{fig:Impact_syst_param}).   \protect\subref{fig:101_peak_ddps} Marginal PPDs ($68\%$-confidence intervals) of the brightness peak localization for the $\Gamma_{SH,1}$ and $\Gamma_{2}$ brightness models. It shows that the brightness peak localization is model-dependent. For example, the longitudinal $\Gamma_{SH,1}$ peak localization is constrained by the phase curve because of its large and constant extension; while the free extension of the $\Gamma_2$ model relaxes this longitudinal constraint (see Sect.\,\ref{sec:results}). Therefore, the light curve of an exoplanet does not constrain uniquely its brightness peak localization; furthermore, the brightness peak localization is not an adequate parameter to characterize complex exoplanet brightness distributions.}
  \label{fig:peak_101}
\end{figure*}

 	We now turn to HD\,189733b's BD. We show the dayside estimates for the $\Gamma_{SH,1}$ and $\Gamma_{2}$ models with their corresponding uncertainties in Figs.\,\ref{fig:101_brightness_ddps} and \ref{fig:3_brightness_ddps} respectively; we focus on HD\,189733b's dayside as it is effectively constrained by the combination of the phase curve and the secondary-eclipse scanning. In particular, note that Figs.\,\ref{fig:101_brightness_ddps} and \ref{fig:3_brightness_ddps} present HD\,189733b's brightness relative to HD\,189733's hemisphere-averaged brightness in the IRAC 8-$\mu$m channel. In addition, the figures are time-averaged; our estimates aim to approach the global pattern of HD\,189733b's BD based on eight snapshots taken from November 2005 to June 2008.  Finally, these estimates correspond to the median and standard deviation of the map trials accepted along the MCMC simulations, similarly to our approach for the corrected and phase-folded light curves (see Sect.\,\ref{sec:results&discussI}).

 	Both models retrieve a spatial feature in HD\,189733b's BD; which corresponds to a hot spot. The $\Gamma_{SH,1}$ model retrieves a hot spot shifted to the east of the substellar point, see Fig.\,\ref{fig:101_brightness_ddps}. The $\Gamma_{2}$ model retrieves a hot spot shifted to the east of the substellar point but also away from the equator, see Fig.\,\ref{fig:3_brightness_ddps}. However, we cannot discuss the direction of this latitudinal shift due to a North-South ambiguity (E. Agol, private communication). 
 	
 	The BD estimates shown in Figs.\,\ref{fig:101_brightness_ddps} and \ref{fig:3_brightness_ddps} are significantly different both in pattern and in intensity. These differences are due to the estimate model-dependence; which motivates the use of different fitting models to enable a thorough discussion. For example, brightness models with non-constant structure (``complex'', i.e., in opposition to a dipole) are less constrained by a phase curve that is only dependent on the hemisphere-integrated brightness. To emphasize these model-induced constraints, we present in Fig.\,\ref{fig:ddps_video} animations showing compilations of dayside BDs accepted along the MCMC simulations for the $\Gamma_{SH,1}$ and $\Gamma_{2}$ models. These compilations show that \textbf{(1)} the amplitude and \textbf{(2)} the longitudinal localization for the $\Gamma_{SH,1}$-BD are more constrained than for the $\Gamma_{2}$ model (by the occultation depth and by the phase curve, respectively) because of its fixed and large structure. However, \textbf{(3)} the $\Gamma_{SH,1}$ model is less constrained in latitude (by the secondary-eclipse scanning) than the more complex $\Gamma_{2}$ model which enables more confined structures that induce larger deviations in occultation ingress/egress. For that reason, the brightness peak localization for the $\Gamma_{2}$ model is well-constrained in latitude (see Fig.\,\ref{fig:101_peak_ddps}), while for the $\Gamma_{SH,1}$ model it is well-constrained in longitude.
 	
 	This shows that the light curve of an exoplanet does not constrain uniquely its brightness peak localization without \textit{a priori} assumption (e.g., assuming a dipolar BD). Therefore, we will further refer to our brightness-distribution estimates instead of the brightness peak localization; which is not representative of complex BDs, in addition to being model-dependent. Nevertheless, we propose in Sect.\,\ref{sec:bpl} another unidimensional parameter to replace the brightness peak localization.

 	Finally, note that these model-induced constraints are also observable on the dayside standard deviation; which is significantly lower for $\Gamma_{SH,1}$ model than for more complex models. In particular, the standard deviation distribution for the $\Gamma_{SH,1}$ model is related to its gradient---with a larger variation from the brightness peak localization along the latitude axis than along the longitude axis, because the BDs accepted along the MCMC simulations differ from each other mainly in (latitudinal) orientation, see Fig.\,\ref{fig:ddps_101}. This is in contrast with the standard deviation distribution for the $\Gamma_{2}$ model that shows a maximum at the brightness peak localization and extended wings towards west and east along the equator; because the BDs accepted along the MCMC simulations mainly affect the former by their amplitude change and the latter by their structure change (see Fig.\,\ref{fig:ddps_3}).

 \subsubsection{Multipolar brightness distribution}
 
 We observe an evolution of our inferences (system-parameter PPD and BD) when increasing the complexity of our fitting model. To assess the validity of this observation, we present here the results obtained when using spherical harmonics up to the degrees 2 (quadrupole) and 3 (octupole).
 
 We present in Figs.\,\ref{fig:102_e_ddp} and \ref{fig:102_erho_ddps} respectively the marginal PPDs of $\lbrace\sqrt{e}\cos\omega,\sqrt{e}\sin\omega\rbrace$ and  $\lbrace\rho_{\star},\sqrt{e}\sin\omega\rbrace$ for the $\Gamma_{SH,2}$ model, and in Figs.\,\ref{fig:103_e_ddp} and \ref{fig:103_erho_ddps} for the $\Gamma_{SH,3}$ model. These PPDs appear as intermediate steps between the results  obtained with the $\Gamma_{SH,1}$ and $\Gamma_{2}$ models (see Fig.\,\ref{fig:correl_mono}). This, therefore, confirms the strong incidence of HD\,189733b's brightness model on the retrieved system parameters, because of the $ e $-$ b $-$\rho_{\star}$-BD correlation.
 
 We show the dayside brightness estimates for the $\Gamma_{SH,2}$ and $\Gamma_{SH,3}$ models with their uncertainty in Figs.\,\ref{fig:102_brightness_ddps} and \ref{fig:103_brightness_ddps}. As observed for the system-parameter PPDs, the BDs also appear as intermediate steps. The major evolutions are the shrinking of the structure retrieved and its shift away from the equator. 
 
 These progressive evolutions of both the system-parameter PPD and the retrieved brightness structure show that, for HD\,189733's system, relaxing the eccentricity constraint and using more complex BDs lead to lower
stellar/planetary densities and more localized and latitudinally-shifted hot spot. In particular, we find that the more complex HD\,189733b's brightness model, the larger the eccentricity, the lower the densities, the larger the impact parameter and the more localized and latitudinally-shifted the hot spot estimated. We discuss the significance of these suggestions in the next section.

\section{Discussion}
\label{sec:discussionII}

	\subsection{The most adequate model}
	
	We present in Sect.\,\ref{sec:results} the results obtained using several fitting-models for assessing the model-dependence of our inferences. As a result, this enables us to reveal the significant correlation between the system parameters and the BD of an exoplanet (i.e., ``$ e $-$ b $-$\rho_{\star}$-BD correlation''). In particular, we show a progressive evolution of the system-parameter PPD and the BD estimate when increasing the complexity of the fitting model. For that reason, we discuss below the relevance of our model-complexity increase, which may ultimately lead to overfitting the data.
	
	We take advantage of our Bayesian framework using the BIC and the AIC. Both information criteria are in favor of models that relax the COA and the UBEA. In particular, the $\Gamma_{SH,1}$ and $\Gamma_{SH,3}$ models are favored; the BIC insignificantly favors $\Gamma_{SH,3}$ (odds ratio $\sim$1.02) while the AIC significantly favors $\Gamma_{SH,3}$ (odds ratio $\sim$3.5). Because not decisive, the information criteria only suggest that the $\Gamma_{SH,3}$ model provides the most adequate constraints on
HD\,189733's system. (Note that theoretical studies favor the AIC, e.g., \citet{Burnham2002,Yang2005}.) 
			
	In particular, it suggests that HD\,189733b's hot spot is shifted both east of the substellar point and away from the equator and HD\,189733b's density has been overestimated by 3.6\%. Furthermore, it suggests that HD\,189733b's orbit is possibly not fully circularized ($ e = 0.015\pm^{0.09}_{0.012} $), although its eccentricity is consistent with zero. This emphasizes that the assumption of circularized orbit has to be continuously assessed, because of data of constantly increasing quality; including for old hot Jupiters that may show a hint of eccentricity \citep[e.g., CoRoT-16b, see][]{Ollivier2012}. Finally, for data-quality reason, the interpretation of HD\,189733b's BD has to focus on global trends: the presence of an asymmetrical hot spot.

	HD\,189733b's dayside presents a shifted hot spot. The eastward shift is in agreement with the literature: (1) with previous derivations, from HD\,189733b's phase curve \citep{Knutson2007} and an eclipse-timing constraint \citep{Agol2010}; and (2) atmospheric models suggesting a super-rotating equatorial jet \citep[e.g.,][]{Showman2009}. In opposition, the suggested shift away from the equator is new. The small-scale origin of this latitudinal asymmetry remains unconstrained because we use large-scale brightness models to be consistent with the data quality. For that reason, additional observations would be required to improve our understanding of HD\,189733b's atmosphere (see Sect.\,\ref{sec:perspectives}); in particular, its yet unmodeled interaction with HD\,189733 \citep[see e.g.,][]{Lecavelier2012} that could induce unexpected thermal patterns, e.g., asymmetric patterns in its BD. For example, magnetic star-planet interactions may lead to energy dissipation due to the stellar field penetration into the exoplanet envelope \citep[e.g.,][]{Laine2008} and to extensive energy injections into the auroral zones of the exoplanet from magnetic reconnections \citep[e.g.,][]{Ip2004}---similarly to the Jupiter-Io flux tube \citep[e.g.,][]{Bigg1964}. However, such magnetic reconnections have so far been only observed at the stellar surface, in the form of chromospheric hot spots rotating synchronously with the companions  \citep[e.g.,][]{Shkolnik2005,Lanza2009}.	
	
	\subsection{Adequacy of conventional analyses}
	\label{sec:impact1}

	We discuss here the possible limitation of conventional analyses for interpreting light curves of ``sufficient'' data quality. We highlight the $ e $-$ b $-$\rho_{\star}$-BD correlation and demonstrate in this context the significant impact on the system parameter estimates of assuming HD\,189733b to be uniformly bright---conceptually similar to neglecting the limb darkening of the host star for transit-photometry analysis. Nevertheless, the significance of this impact is related to the significance of HD\,189733b's secondary-eclipse scanning; which is enabled by the high-SNR \textit{Spitzer}/IRAC 8-$\mu $m photometry. This detection, therefore, motivates using more complex fitting models and outlines the limitation of conventional analyses because of the $ e $-$ b $-$\rho_{\star}$-BD correlation; which would have remained hidden if the photometric precision was less. In this context, a ``sufficient'' photometric precision requires resolving the occultation ingress/egress; therefore, it has to be about one order of magnitude less than the occultation depth, for a time bin about one order of magnitude less than the occultation ingress/egress duration. For less photometric precisions, modelling the eclipse using a uniformly-bright disk, therefore, is adequate, so are the conventional analyses.
	
	Finally, we briefly outline that the conventional assumption of a uniformly-bright exoplanet could also affect the inferred planetary interior models; because $ \rho_{\star} $ is possibly affected and, therefore, $ \rho_{p} $ is too (e.g., suggested 3.6\%-overestimation for HD\,189733b's).

	\subsection{Relevance of reconstruction methods}
	\label{sec:impact2}
	
	We discuss here the relevance of mapping methods based on direct reconstruction from an exoplanet ``slicing'' (e.g., the slice method introduced in M12). We emphasize that a 2D map is not directly accessible over a specific ``grid'' (see Fig.\,\ref{fig:scanning_processes}). The main reason is that such method requires the use of fixed system parameters, which are derived from a conventional analysis. Therefore, it leads to biased and overly-precise estimates by neglecting the correlation between the system parameters and the exoplanet BD---and shape, see Sect.\,\ref{sec:results}. Note that in addition, it also neglects the exoplanet rotation during its occultation ($\sim$\,$12^\circ$ for HD\,189733b) and does not take advantage of its phase curve (see M12). Finally, the degeneracy induced by the limited data ($ N $ slices by scanning processes while roughly $ N^2 $ cells over the map, see M12) implies the use of \textit{a priori} constraint on the BD.
	
	 \subsection{Reducing brightness distributions to 1D parameters}
 \label{sec:bpl}

We have shown in Sect.\,\ref{sec:unipolar} that  the light curve of an exoplanet does not constrain uniquely its brightness peak localization. We discuss here the reasons why we strongly advocate discussing the BD estimates and, if necessary, using with care the dayside barycenter as a representative parameter.

Discussing the significance of a hot spot with complex structure from a 2D map is difficult; but may be simplified using a unidimensional parameter. We show in Sect.\,\ref{sec:unipolar} that the brightness peak localization poorly represents complex BDs, in addition to being model-dependent. Therefore, we investigate using the dayside barycenter to replace the brightness peak localization. The reason is that the dayside brightness barycenter weights the BD according to the geometrical configuration at superior conjunction (i.e., it contains partial 2D information).

We show in Fig.\,\ref{fig:barycenter} the marginal PPDs ($68\%$-confidence intervals) of the brightness peak localization for the $\Gamma_{SH,1}$ and $\Gamma_{2}$ brightness models. A comparison with the marginal PPDs of the brightness peak localization (Fig.\,\ref{fig:101_peak_ddps}) shows the reduced model-dependence of the dayside barycenter. In particular, it shows a less-extended PPD for the $\Gamma_{2}$-model barycenter; because this extension for the brightness-peak-localization PPD emerges from the model wings---weighted by the dayside barycenter. In addition, it shows the shift and slight shrinking of the $\Gamma_{SH,1}$ PPDs that reflect the barycenter weighting according to the geometrical configuration at superior conjunction; map cells closer to the substellar point have more weight. This emphasizes the primary drawback of the dayside barycenter that is to attenuate the offset of BDs. This recalls that unidimensional parameters cannot stand adequately for complex BDs and, therefore, have to be used complementary to BD estimates.

	\subsection{Perspectives}	
	\label{sec:perspectives}
	
	We show that conventional analysis---which assume uniformly-bright exoplanets---may lead to biased estimates of the system parameters. We introduce below two of the possible applications of our method in this context: \textbf{(1)} IR multi-wavelength observations of HD\,189733b for improving the constraints on the system parameters and, ultimately, for yielding a time-dependent 3D map and \textbf{(2)} observations, in the visible, of targets with apparently high albedos.

	 \textbf{(1)} The atmospheric layers scanned during an occultation ingress/egress are wavelength-dependent. Therefore, additional high quality occultations at different wavelengths (i.e., optical depths) would contain the same information about the system parameters while different information about the BD. These would enable us to improve our constraints on the system parameters while gaining insights into the atmospheric physics of HD\,189733b from its thermal patterns at different depths.
	 
	 In particular, the origin of HD\,189733's hot spot could be revealed from observations at shorter wavelengths. These observations would refer to a higher temperature zone whose localization and extent could disentangle between its possible origins, e.g., radiative or magnetic.  
	 
	 In addition, the time variability of these spatial features could also be targeted. Indeed, as discussed in \citet{Rauscher2007}, \textit{JWST}/NIRspec performance should allow high significance detection of anomalous occultation ingress/egress based on a unique eclipse (e.g., for the grating centered at $4\mu m$). The time variability would then be assessed from the BDs derived for different occultations. In other words, the future exoplanetary-atmosphere investigations of spaced-based facilities like JWST \citep{Clampin2010} and EChO \citep{Tinetti2011} could ultimately lead to time-dependent 3D maps of distant worlds.

	 \textbf{(2)} Extension of our method to observations in the visible would also be valuable, both for ground- and space-based observations. For example, the \textit{Kepler} mission provides additional targets that would be relevant for similar analyses, thanks to its outstanding photometry quality. \textit{Kepler} provides broadband visible photometry and is, therefore, capable of observing the reflected light from exoplanet atmospheres. Recent studies found high geometric albedos for two hot Jupiters \citep{Demory2011,Kipping2011,Berdyugina2011}. In addition, \citet{Demory2011} suggest hazes/cloud and Rayleigh scattering as plausible explanations for such high albedos. Recently, \citet{Madhusudhan2012} provide a theoretical framework for interpreting geometrical albedos from phase curves, which is is indicative of the scattering and absorptive properties of the atmosphere. We emphasize that the albedo origins could make the exoplanet non-uniformly bright and modify its occultation ingress/egress, similarly to the effect of an hot spot. Therefore, our study could constrain independently the geometrical albedo by scanning an exoplanet dayside, what would be a complementary asset to \citet{Madhusudhan2012}'s framework. Furthermore, our study will be necessary to constrain consistently the system parameters from high-SNR light curves, see Sect.\,\ref{sec:impact1}.

\subsection{Comparison with \cite{Majeau2012}}
 \label{sec:compare_2_M12}
We obtain qualitatively similar results as M12: an offset hot spot. Nevertheless, significant differences exist between both studies. \textbf{(1)} The starting point of our second analysis is our detection of HD\,189733b's eclipse scanning at the 6$\sigma$ level (in contrast to their $\sim$3.5\,$\sigma$). \textbf{(2)} Eclipse scanning has multiple possible contributing factors (i.e., not only a non-uniform BD). Our study provides a framework for consistently constraining their contribution, e.g., HD\,189733b's shape. \textbf{(3)} In addition, and related to the second point, we do not constrain \textit{a priori} the system parameters to the best-fit of a conventional analysis nor the orbital eccentricity to zero; instead, we estimate the system parameters simultaneously with the BD. \textbf{(4)} We also investigate the model-dependence of our inferences, in contrast to M12 who focused on a dipolar brightness model (see related discussion in Sect.\,\ref{sec:unipolar}). In particular, we show their impact on the system-parameter PPD (see Sect.\,\ref{sec:results}). Finally, M12 proposed an additional method (the slice method) to generate an exoplanet map from its occultation ingress/egress only; we discuss the relevance of similar methods in Sect.\,\ref{sec:impact2}. 

As a consequence, M12 estimate the brightness peak localization with narrow error bars ($21.8\pm1.5^\circ$east and $3.1\pm9.4^\circ$ away from the equator) for a circularized HD\,189733b's orbit; while we show that the brightness peak localization, as well as the system-parameter PPD, are model-dependent because of the $ e $-$ b $-$\rho_{\star}$-BD correlation. In particular, we show that the light curve of an exoplanet does not constrain uniquely its brightness peak localization. Nevertheless, for a direct comparison to M12's estimate of the brightness peak localization for a dipolar brightness model, we estimate it to $11.5\pm4.3^\circ$east and $3.1\pm11.4^\circ$ away from the equator, see Sect.\,\ref{sec:discussionII}, Fig.\,\ref{fig:101_peak_ddps}.

\subsection{Complementary analysis}
 \label{sec:compl_analysis}
 
 We emphasize in our study the necessity of global approaches for fitting consistently the available data, in particular by highlighting the $ e $-$ b $-$\rho_{\star}$-BD correlation. Regarding this correlation, we discuss in this subsection the effect of the stellar activity (through the planetary transit analysis) and of radial velocity
(RV) measurements (i.e., for constraining $\sqrt{e}\sin\omega$).

\subsubsection{Effect of HD\,189733's activity}

HD\,189733 presents high-activity levels that may affect the transit parameters---incl., $ (R_p/R_\star)^2 $, due to occulted/unocculted star spots \citep[][]{Pont2007,Sing2011}. Therefore, treating coherently spots of active stars is necessary in the present context---global approach and $ e $-$ b $-$\rho_{\star}$-BD correlation. However, while important in the optical, the stellar activity may be negligible at 8\,$\mu $m, regarding the data quality. We assess this statement performing individual analysis of the 6 transits used in this study (see Table\,\ref{tab:AOR}). We present in Fig.\,\ref{fig:DdF} our individual transit-depth estimates. These estimates show no significant temporal variation \citep[in opposition to][who attributed these to stellar activity]{Agol2010}. Similarly, we observe no significant variation of the transit parameters from one individual analysis to another. In addition, we observe no pattern specific to the occultation of a star spot \citep[i.e., similar to, e.g., ``Features A and B'' in the Fig. 1 of][]{Pont2007}. Therefore, we consider that our time-averaged  inferences (see Sect.\,\ref{sec:results}) are not biased by HD\,189733's activity.

\subsubsection{Complementary input of HD\,189733's RV measurements}

RV measurements may help constraining $\sqrt{e}\sin\omega$ and, therefore, providing a complementary insight into the $ e $-$ b $-$\rho_{\star}$-BD correlation. Therefore, we analysed HD\,189733's out-of-transit RV
data \citep{Winn2006,Boisse2009} while using priors on
independent system parameters; $ (R_p/R_\star)^2 $, $ P $, $ W $, and $ b $. This enable us to assess the constraint derived solely from the RV data on the eccentricity. We present in Fig.\,\ref{fig:RV_inputs} our overall Keplerian fit and the marginal PPD of the parameters $\sqrt{e}\cos\omega$ and $\sqrt{e}\sin\omega$. This shows that the RV data does not constrain HD\,189733b's eccentricity further than the \textit{Spitzer}/IRAC 8-$\mu$m photometry for low complexity brightness models (see Figs.\,\ref{fig:uniform_e_ddp}, \ref{fig:101_e_ddp} and \ref{fig:102_e_ddp}). However HD\,189733's RV measurements may affect our inferences for more complex models that favor a localized hot spot and larger eccentricity (see Figs.\,\ref{fig:103_e_ddp} and \ref{fig:3_e_ddp}), by rejecting the solutions involving $\sqrt{e}\sin\omega$ $ \gtrsim $ 0.15. Therefore, we implement the analysis of RV data in our global method.

	We observe no change of the system parameters PPDs for simple brightness models (see Table\,\ref{tab:BFP_RV}), as expected from the comparison of Fig.\,\ref{fig:ddp_e_unif} and Figs.\,\ref{fig:101_e_ddp} and \ref{fig:102_e_ddp}. However, we observe the incidence of rejecting the solutions involving $\sqrt{e}\sin\omega$ $ \gtrsim $ 0.15---in the context of the $ e $-$ b $-$\rho_{\star}$-BD correlation---for complex brightness models ($\Gamma_{SH,3}$ and $\Gamma_2$). In particular, the evolution of the parameter estimates---toward lower densities, larger impact parameter and a more localized and latitudinally-shifted hot spot---providing larger eccentricity are attenuated. We present the incidence on the system-parameter PPD in Fig.\,\ref{fig:correl_multi_RV}. The $\lbrace\sqrt{e}\cos\omega,\sqrt{e}\sin\omega\rbrace$-PPDs (Figs.\,\ref{fig:103_e_ddp_RV} and \ref{fig:3_e_ddp_RV}) conceptually correspond to the marginalized product of the PPDs estimated using solely the photometry (see Figs.\,\ref{fig:103_e_ddp} and \,\ref{fig:3_e_ddp}) and using solely the RV data (Fig.\,\ref{fig:RVfit}). These highlight the redistribution of the probability density that results from the rejection of solutions involving $\sqrt{e}\sin\omega$ $ \gtrsim $ 0.15 by HD\,189733's RV data. As a consequence of this probability redistribution in $\lbrace\sqrt{e}\cos\omega,\sqrt{e}\sin\omega\rbrace$, the set of acceptable combinations to compensate HD\,189733b's anomalous occultation is affected too. In other words, the $ b $, $\rho_{\star}$ and BD estimates are affected. On the one hand, the system parameter estimates are consistent with those obtained for less complex brightness models (see Table\,\ref{tab:BFP_RV} and compare Figs.\,\ref{fig:103_erho_ddps_RV} and \ref{fig:3_erho_ddps_RV} to Figs.\,\ref{fig:101_erho_ddps} and \ref{fig:102_erho_ddps}). On the other hand, HD\,189733b's dayside brightness estimates present less confined patterns (compare Figs.\,\ref{fig:103_brightness_ddps} and \ref{fig:3_brightness_ddps} with Figs.\,\ref{fig:103_brightness_ddps_RV} and \ref{fig:3_brightness_ddps_RV}, respectively). This emphasizes the necessity of global approaches to consistently probe the highly-correlated parameter space of exoplanetary data.
	
	Finally, we present our improved constraint on the upper limit of HD\,189733b's orbital eccentricity, $ e\leq 0.011$ ($95\%$  confidence), based on our global analysis of the \textit{Spitzer}/IRAC 8-$\mu$m photometry and the RV measurements of HD\,189733.

\section{Summary}
\label{sec:conclusion}

We have performed two analyses of the HD\,189733b public eclipse data obtained at 8\,$\mu $m with \textit{Spitzer}/IRAC. The first, conventional analysis provides HD\,189733's system parameters and the corrected and phase-folded light curves; which deviate from the occultation of a uniformly-bright disk at the 6\,$\sigma$ level. The second analysis investigates the possible contributing factors of this deviation following a new, general methodology.

We demonstrate that this deviation emerges mainly from a large-scale brightness structure in HD\,18733b's atmosphere; we reject HD\,189733b's shape as a possible contributing factor based on the transit residuals. It indicates a hot spot within the atmospheric layers probed at 8\,$\mu $m as well as a hemispheric asymmetry. In addition, we show a correlation between the system parameters and HD\,189733b's BD---what we term the ``$ e $-$ b $-$\rho_{\star}$-BD correlation''. This enables us to outline that for data of sufficient quality the conventional assumption of a uniformly-bright exoplanet leads to an underestimated uncertainty on (and possibly biased estimates of) system parameters, in particular on $ \rho_{\star} $ and $\rho_{p} $. Notably, we find that the more complex HD\,189733b's brightness model, the larger the eccentricity, the lower the densities, the larger the impact parameter and the more localized and latitudinally-shifted the hot spot estimated. In this context, we show that the light curve of an exoplanet does not constrain uniquely its brightness peak localization. Therefore, we propose to use the dayside barycenter---when a unidimensional parameter is necessary---because it contains partial two-dimensional information from weighting the BD according to the geometrical configuration at superior conjunction.

Because of the highlighted $ e $-$ b $-$\rho_{\star}$-BD correlation, we assess the incidence of HD\,189733's RV measurements on our inferences obtained solely from \textit{Spitzer}/IRAC 8-$\mu $m photometry. We observe redistributions of the probability density for the PPDs obtained with complex brightness models, which result from the rejection by HD\,189733's RV data of solutions involving $\sqrt{e}\sin\omega$ $ \gtrsim $ 0.15, favored by the photometry in case of a localized hot spot---i.e., for complex brightness models. This emphasizes the necessity of global approaches to consistently probe the highly-correlated parameter space of exoplanetary data.

As a final result, we show that, for present data quality, the system parameter estimates for complex brightness model are similar to those obtained for a dipolar brightness models. In particular, we present our constraint on the upper limit of HD\,189733b's orbital eccentricity, $ e\leq 0.011$ ($95\%$  confidence).

	We discuss the perspectives of applying our methods to observations in different spectral bands than in the IRAC 8-$\mu$m channel. In particular, observations of HD\,189733b at other wavelengths could improve the constraints on the system parameters while ultimately yielding a large-scale 3D map---because different optical depths would be probed under the same orbital configuration. In addition, applications to observations in the visible (ground- and space-based, e.g., Kepler data) could also be valuable for targets non-uniformly bright because of albedo (e.g., cloud/hazes or Rayleigh scattering). We emphasize that the albedo origins could make the exoplanet non-uniformly bright and modify its occultation ingress/egress, similarly to the effect of an hot spot. Therefore, our study could constrain independently the geometrical albedo by scanning an exoplanet dayside, what would be a complementary asset to \citet{Madhusudhan2012}'s framework. We discuss the relevance of mapping methods based on direct reconstruction from an exoplanet ``slicing''. We emphasize the necessity of independent constraints on the host star to consistently constrain the exoplanet shape, in particular on its limb-darkening coefficient---although we emphasize that the phase curve also constrains the exoplanet shape. Finally, we suggest that the future exoplanetary-atmosphere investigations of spaced-based facilities like JWST and EChO could ultimately lead to time-dependent 3D maps of distant worlds.


\begin{acknowledgements}
      This study is based on observations made with the \textit{Spitzer Space Telescope},
which is operated by the Jet Propulsion Laboratory, California Institute of Technology, under contract to NASA. We thank the people who made and keep the \textit{Spitzer Space Telescope} such a wonderful scientific adventure. We thank H. Knutson for sharing HD\,189733b's phase curve data. We are grateful to A. Zsom, V. Stamenkovic, A. Triaud and E. Agol for helpful discussions and to the anonymous referee for her/his comments that strengthen this manuscript. JdW acknowledges the hospitality of the Institut d'Astrophysique et de G\'{e}ophysique, Universit\'{e} de Li\`{e}ge where portions of this study were completed and thanks A. Lanotte and S. Sohy for their help. JdW acknowledges support from the Grayce B. Kerr Foundation and the WBI (Wallonie-Bruxelles International) in the form of fellowships. Finally, JdW also acknowledges support provided by the Odissea Prize initiated by the Belgian Senate.
\end{acknowledgements}

\bibliographystyle{bibtex/aa} 

\begin{thebibliography}{75}
\expandafter\ifx\csname natexlab\endcsname\relax\def\natexlab#1{#1}\fi

\bibitem[{{Agol} {et~al.}(2010){Agol}, {Cowan}, {Knutson}, {Deming}, {Steffen},
  {Henry}, \& {Charbonneau}}]{Agol2010}
{Agol}, E., {Cowan}, N.~B., {Knutson}, H.~A., {et~al.} 2010, \apjl, 721, 1861

\bibitem[{{Barnes} {et~al.}(2009){Barnes}, {Cooper}, {Showman}, \&
  {Hubbard}}]{Barne2009}
{Barnes}, J.~W., {Cooper}, C.~S., {Showman}, A.~P., \& {Hubbard}, W.~B. 2009,
  \apj, 706, 877

\bibitem[{{Batygin} {et~al.}(2011){Batygin}, {Stevenson}, \&
  {Bodenheimer}}]{Batygin2011}
{Batygin}, K., {Stevenson}, D.~J., \& {Bodenheimer}, P.~H. 2011, \apj, 738, 1

\bibitem[{{Berdyugina} {et~al.}(2011){Berdyugina}, {Berdyugin}, {Fluri}, \&
  {Piirola}}]{Berdyugina2011}
{Berdyugina}, S.~V., {Berdyugin}, A.~V., {Fluri}, D.~M., \& {Piirola}, V. 2011,
  \apjl, 728, L6

\bibitem[{{Boisse} {et~al.}(2009){Boisse}, {Moutou}, {Vidal-Madjar}, {Bouchy},
  {Pont}, {H{\'e}brard}, {Bonfils}, {Croll}, {Delfosse}, {Desort}, {Forveille},
  {Lagrange}, {Loeillet}, {Lovis}, {Matthews}, {Mayor}, {Pepe}, {Perrier},
  {Queloz}, {Rowe}, {Santos}, {S{\'e}gransan}, \& {Udry}}]{Boisse2009}
{Boisse}, I., {Moutou}, C., {Vidal-Madjar}, A., {et~al.} 2009, \aap, 495, 959

\bibitem[{{Bouchy} {et~al.}(2005){Bouchy}, {Udry}, {Mayor}, {Moutou}, {Pont},
  {Iribarne}, {da Silva}, {Ilovaisky}, {Queloz}, {Santos}, {S{\'e}gransan}, \&
  {Zucker}}]{Bouchy2005}
{Bouchy}, F., {Udry}, S., {Mayor}, M., {et~al.} 2005, \aap, 444, L15

\bibitem[{Burnham \& Anderson(2002)}]{Burnham2002}
Burnham, K. \& Anderson, D. 2002, {Model Selection and Multimodel Inference: A
  Practical Information-theoretic Approach} (Springer)

\bibitem[{{Carter} \& {Winn}(2010)}]{Carter2010}
{Carter}, J.~A. \& {Winn}, J.~N. 2010, \apj, 709, 1219

\bibitem[{{Charbonneau} {et~al.}(2008){Charbonneau}, {Knutson}, {Barman},
  {Allen}, {Mayor}, {Megeath}, {Queloz}, \& {Udry}}]{Charbonneau2008}
{Charbonneau}, D., {Knutson}, H.~A., {Barman}, T., {et~al.} 2008, \apjl, 686,
  1341

\bibitem[{{Clampin}(2010)}]{Clampin2010}
{Clampin}, M. 2010, in Proceedings of the conference In the Spirit of Lyot 2010

\bibitem[{{Claret} \& {Bloemen}(2011)}]{Claret2011}
{Claret}, A. \& {Bloemen}, S. 2011, \aap, 529, A75

\bibitem[{{Cooper} \& {Showman}(2005)}]{Cooper2005}
{Cooper}, C.~S. \& {Showman}, A.~P. 2005, \apjl, 629, L45

\bibitem[{{Deming} {et~al.}(2006){Deming}, {Harrington}, {Seager}, \&
  {Richardson}}]{Deming2006}
{Deming}, D., {Harrington}, J., {Seager}, S., \& {Richardson}, L.~J. 2006,
  \apjl, 644, 560

\bibitem[{{Demory} {et~al.}(2011){Demory}, {Seager}, {Madhusudhan}, {Kjeldsen},
  {Christensen-Dalsgaard}, {Gillon}, {Rowe}, {Welsh}, {Adams}, {Dupree},
  {McCarthy}, {Kulesa}, {Borucki}, \& {Koch}}]{Demory2011}
{Demory}, B.-O., {Seager}, S., {Madhusudhan}, N., {et~al.} 2011, \apjl, 735,
  L12

\bibitem[{{Deroo} {et~al.}(2010){Deroo}, {Swain}, \& {Vasisht}}]{Deroo2010}
{Deroo}, P., {Swain}, M.~R., \& {Vasisht}, G. 2010, ArXiv e-prints

\bibitem[{{D{\'e}sert} {et~al.}(2009){D{\'e}sert}, {Lecavelier des Etangs},
  {H{\'e}brard}, {Sing}, {Ehrenreich}, {Ferlet}, \&
  {Vidal-Madjar}}]{D'esert2009}
{D{\'e}sert}, J.-M., {Lecavelier des Etangs}, A., {H{\'e}brard}, G., {et~al.}
  2009, \apjl, 699, 478

\bibitem[{{Dobbs-Dixon} {et~al.}(2010){Dobbs-Dixon}, {Cumming}, \&
  {Lin}}]{Dobbs-Dixon2010}
{Dobbs-Dixon}, I., {Cumming}, A., \& {Lin}, D.~N.~C. 2010, \apj, 710, 1395

\bibitem[{{Fabrycky}(2010)}]{Fabrycky2010}
{Fabrycky}, D.~C. 2010, {Non-Keplerian Dynamics of Exoplanets}, ed. {Seager,
  S.}, 217--238

\bibitem[{{Faigler} \& {Mazeh}(2011)}]{Faigler2011}
{Faigler}, S. \& {Mazeh}, T. 2011, \mnras, 415, 3921

\bibitem[{{Fazio} {et~al.}(2004){Fazio}, {Hora}, {Allen}, {Ashby}, {Barmby},
  {Deutsch}, {Huang}, {Kleiner}, {Marengo}, {Megeath}, {Melnick}, {Pahre},
  {Patten}, {Polizotti}, {Smith}, {Taylor}, {Wang}, {Willner}, {Hoffmann},
  {Pipher}, {Forrest}, {McMurty}, {McCreight}, {McKelvey}, {McMurray}, {Koch},
  {Moseley}, {Arendt}, {Mentzell}, {Marx}, {Losch}, {Mayman}, {Eichhorn},
  {Krebs}, {Jhabvala}, {Gezari}, {Fixsen}, {Flores}, {Shakoorzadeh}, {Jungo},
  {Hakun}, {Workman}, {Karpati}, {Kichak}, {Whitley}, {Mann}, {Tollestrup},
  {Eisenhardt}, {Stern}, {Gorjian}, {Bhattacharya}, {Carey}, {Nelson},
  {Glaccum}, {Lacy}, {Lowrance}, {Laine}, {Reach}, {Stauffer}, {Surace},
  {Wilson}, {Wright}, {Hoffman}, {Domingo}, \& {Cohen}}]{Fazio2004}
{Fazio}, G.~G., {Hora}, J.~L., {Allen}, L.~E., {et~al.} 2004, \apjs, 154, 10

\bibitem[{{Ford}(2006)}]{Ford2006}
{Ford}, E.~B. 2006, \apj, 642, 505

\bibitem[{Gelman {et~al.}(2004)Gelman, Carlin, Stern, \& Rubin}]{Gelman2004}
Gelman, A., Carlin, J., Stern, H., \& Rubin, D. 2004, Bayesian Data Analysis
  (Chapman \& Hall/CRC)

\bibitem[{{Gibson} {et~al.}(2011){Gibson}, {Pont}, \& {Aigrain}}]{Gibson2011}
{Gibson}, N.~P., {Pont}, F., \& {Aigrain}, S. 2011, \mnras, 411, 2199

\bibitem[{{Gillon} {et~al.}(2009){Gillon}, {Demory}, {Triaud}, {Barman},
  {Hebb}, {Montalb{\'a}n}, {Maxted}, {Queloz}, {Deleuil}, \&
  {Magain}}]{Gillon2009}
{Gillon}, M., {Demory}, B.-O., {Triaud}, A.~H.~M.~J., {et~al.} 2009, \aap, 506,
  359

\bibitem[{{Gillon} {et~al.}(2010{\natexlab{a}}){Gillon}, {Hatzes}, {Csizmadia},
  {Fridlund}, {Deleuil}, {Aigrain}, {Alonso}, {Auvergne}, {Baglin}, {Barge},
  {Barnes}, {Bonomo}, {Bord{\'e}}, {Bouchy}, {Bruntt}, {Cabrera}, {Carone},
  {Carpano}, {Cochran}, {Deeg}, {Dvorak}, {Endl}, {Erikson}, {Ferraz-Mello},
  {Gandolfi}, {Gazzano}, {Guenther}, {Guillot}, {Havel}, {H{\'e}brard},
  {Jorda}, {L{\'e}ger}, {Llebaria}, {Lammer}, {Lovis}, {Mayor}, {Mazeh},
  {Montalb{\'a}n}, {Moutou}, {Ofir}, {Ollivier}, {P{\"a}tzold}, {Pepe},
  {Queloz}, {Rauer}, {Rouan}, {Samuel}, {Santerne}, {Schneider}, {Tingley},
  {Udry}, {Weingrill}, \& {Wuchterl}}]{Gillon2010b}
{Gillon}, M., {Hatzes}, A., {Csizmadia}, S., {et~al.} 2010{\natexlab{a}}, \aap,
  520, A97

\bibitem[{{Gillon} {et~al.}(2010{\natexlab{b}}){Gillon}, {Lanotte}, {Barman},
  {Miller}, {Demory}, {Deleuil}, {Montalb{\'a}n}, {Bouchy}, {Collier Cameron},
  {Deeg}, {Fortney}, {Fridlund}, {Harrington}, {Magain}, {Moutou}, {Queloz},
  {Rauer}, {Rouan}, \& {Schneider}}]{Gillon2010a}
{Gillon}, M., {Lanotte}, A.~A., {Barman}, T., {et~al.} 2010{\natexlab{b}},
  \aap, 511, A3

\bibitem[{{Gregory}(2005)}]{Gregory2005}
{Gregory}, P.~C. 2005, {Bayesian Logical Data Analysis for the Physical
  Sciences: A Comparative Approach with ``Mathematica'' Support}, ed. {Gregory,
  P.~C.} (Cambridge University Press)

\bibitem[{{Grillmair} {et~al.}(2007){Grillmair}, {Charbonneau}, {Burrows},
  {Armus}, {Stauffer}, {Meadows}, {Van Cleve}, \& {Levine}}]{Grillmair2007}
{Grillmair}, C.~J., {Charbonneau}, D., {Burrows}, A., {et~al.} 2007, \apjl,
  658, L115

\bibitem[{{Heng}(2012)}]{Heng2012}
{Heng}, K. 2012, \apjl, 748, L17

\bibitem[{{Horne}(1985)}]{Horne1985}
{Horne}, K. 1985, \mnras, 213, 129

\bibitem[{{Huitson} {et~al.}(2012){Huitson}, {Sing}, {Vidal-Madjar},
  {Ballester}, {Lecavelier des Etangs}, {D{\'e}sert}, \& {Pont}}]{Huitson2012}
{Huitson}, C.~M., {Sing}, D.~K., {Vidal-Madjar}, A., {et~al.} 2012, \mnras,
  422, 2477

\bibitem[{{Ip} {et~al.}(2004){Ip}, {Kopp}, \& {Hu}}]{Ip2004}
{Ip}, W.-H., {Kopp}, A., \& {Hu}, J.-H. 2004, \apjl, 602, L53

\bibitem[{{Kipping} \& {Bakos}(2011)}]{Kipping2011}
{Kipping}, D. \& {Bakos}, G. 2011, \apj, 730, 50

\bibitem[{{Knutson} {et~al.}(2007){Knutson}, {Charbonneau}, {Allen}, {Fortney},
  {Agol}, {Cowan}, {Showman}, {Cooper}, \& {Megeath}}]{Knutson2007}
{Knutson}, H.~A., {Charbonneau}, D., {Allen}, L.~E., {et~al.} 2007, \nat, 447,
  183

\bibitem[{{Knutson} {et~al.}(2009){Knutson}, {Charbonneau}, {Cowan}, {Fortney},
  {Showman}, {Agol}, {Henry}, {Everett}, \& {Allen}}]{Knutson2009}
{Knutson}, H.~A., {Charbonneau}, D., {Cowan}, N.~B., {et~al.} 2009, \apj, 690,
  822

\bibitem[{{Knutson} {et~al.}(2012){Knutson}, {Lewis}, {Fortney}, {Burrows},
  {Showman}, {Cowan}, {Agol}, {Aigrain}, {Charbonneau}, {Deming}, {D{\'e}sert},
  {Henry}, {Langton}, \& {Laughlin}}]{Knutson2012}
{Knutson}, H.~A., {Lewis}, N., {Fortney}, J.~J., {et~al.} 2012, \apj, 754, 22

\bibitem[{{Kopal}(1959)}]{Kopal1959}
{Kopal}, Z. 1959, {Close binary systems}

\bibitem[{{Laine} {et~al.}(2008){Laine}, {Lin}, \& {Dong}}]{Laine2008}
{Laine}, R.~O., {Lin}, D.~N.~C., \& {Dong}, S. 2008, \apj, 685, 521

\bibitem[{{Lanza}(2009)}]{Lanza2009}
{Lanza}, A.~F. 2009, \aap, 505, 339

\bibitem[{{Lecavelier des Etangs} {et~al.}(2012){Lecavelier des Etangs},
  {Bourrier}, {Wheatley}, {Dupuy}, {Ehrenreich}, {Vidal-Madjar}, {H{\'e}brard},
  {Ballester}, {D{\'e}sert}, {Ferlet}, \& {Sing}}]{Lecavelier2012}
{Lecavelier des Etangs}, A., {Bourrier}, V., {Wheatley}, P.~J., {et~al.} 2012,
  \aap, 543, L4

\bibitem[{{Madhusudhan} \& {Burrows}(2012)}]{Madhusudhan2012}
{Madhusudhan}, N. \& {Burrows}, A. 2012, \apj, 747, 25

\bibitem[{{Madhusudhan} \& {Seager}(2009)}]{Madhusudhan2009}
{Madhusudhan}, N. \& {Seager}, S. 2009, \apj, 707, 24

\bibitem[{{Majeau} {et~al.}(2012){Majeau}, {Agol}, \& {Cowan}}]{Majeau2012}
{Majeau}, C., {Agol}, E., \& {Cowan}, N.~B. 2012, \apjl, 747, L20

\bibitem[{{Mandel} \& {Agol}(2002)}]{Mandel2002}
{Mandel}, K. \& {Agol}, E. 2002, \apjl, 580, L171

\bibitem[{{Menou}(2012)}]{Menou2012}
{Menou}, K. 2012, \apj, 745, 138

\bibitem[{{Moses} {et~al.}(2011){Moses}, {Visscher}, {Fortney}, {Showman},
  {Lewis}, {Griffith}, {Klippenstein}, {Shabram}, {Friedson}, {Marley}, \&
  {Freedman}}]{Moses2011}
{Moses}, J.~I., {Visscher}, C., {Fortney}, J.~J., {et~al.} 2011, \apj, 737, 15

\bibitem[{{Ollivier} {et~al.}(2012){Ollivier}, {Gillon}, {Santerne},
  {Wuchterl}, {Havel}, {Bruntt}, {Bord{\'e}}, {Pasternacki}, {Endl},
  {Gandolfi}, {Aigrain}, {Almenara}, {Alonso}, {Auvergne}, {Baglin}, {Barge},
  {Bonomo}, {Bouchy}, {Cabrera}, {Carone}, {Carpano}, {Cavarroc}, {Cochran},
  {Csizmadia}, {Deeg}, {Deleuil}, {Diaz}, {Dvorak}, {Erikson}, {Ferraz-Mello},
  {Fridlund}, {Gazzano}, {Grziwa}, {Guenther}, {Guillot}, {Guterman}, {Hatzes},
  {H{\'e}brard}, {Lammer}, {L{\'e}ger}, {Lovis}, {MacQueen}, {Mayor}, {Mazeh},
  {Moutou}, {Ofir}, {P{\"a}tzold}, {Queloz}, {Rauer}, {Rouan}, {Samuel},
  {Schneider}, {Tadeu dos Santos}, {Tal-Or}, {Tingley}, \&
  {Weingrill}}]{Ollivier2012}
{Ollivier}, M., {Gillon}, M., {Santerne}, A., {et~al.} 2012, \aap, 541, A149

\bibitem[{{Pont} {et~al.}(2007){Pont}, {Gilliland}, {Moutou}, {Charbonneau},
  {Bouchy}, {Brown}, {Mayor}, {Queloz}, {Santos}, \& {Udry}}]{Pont2007}
{Pont}, F., {Gilliland}, R.~L., {Moutou}, C., {et~al.} 2007, \aap, 476, 1347

\bibitem[{{Pont} {et~al.}(2006){Pont}, {Zucker}, \& {Queloz}}]{Pont2006}
{Pont}, F., {Zucker}, S., \& {Queloz}, D. 2006, \mnras, 373, 231

\bibitem[{{Press} {et~al.}(1992){Press}, {Flannery}, {Teukolsky}, \&
  {Vetterling}}]{Press92}
{Press}, W.~H., {Flannery}, B.~P., {Teukolsky}, S.~A., \& {Vetterling}, W.~T.
  1992, Numerical Recipes in FORTRAN: The Art of Scientific Computing, 2nd ed.
  (Cambridge University Press), 51--63

\bibitem[{{Rauscher} \& {Menou}(2010)}]{Rauscher2010}
{Rauscher}, E. \& {Menou}, K. 2010, \apj, 714, 1334

\bibitem[{{Rauscher} \& {Menou}(2012)}]{Rauscher2012}
{Rauscher}, E. \& {Menou}, K. 2012, \apj, 750, 96

\bibitem[{{Rauscher} {et~al.}(2007){Rauscher}, {Menou}, {Seager}, {Deming},
  {Cho}, \& {Hansen}}]{Rauscher2007}
{Rauscher}, E., {Menou}, K., {Seager}, S., {et~al.} 2007, \apj, 664, 1199

\bibitem[{{Redfield} {et~al.}(2008){Redfield}, {Endl}, {Cochran}, \&
  {Koesterke}}]{Redfield2008}
{Redfield}, S., {Endl}, M., {Cochran}, W.~D., \& {Koesterke}, L. 2008, \apjl,
  673, L87

\bibitem[{{Russell} \& {Merrill}(1952)}]{Russell1952}
{Russell}, H.~N. \& {Merrill}, J.~E. 1952, {The determination of the elements
  of eclipsing binaries}

\bibitem[{{Seager} \& {Deming}(2010)}]{Seager2010a}
{Seager}, S. \& {Deming}, D. 2010, \araa, 48, 631

\bibitem[{{Seager} \& {Mall{\'e}n-Ornelas}(2003)}]{Seager2003}
{Seager}, S. \& {Mall{\'e}n-Ornelas}, G. 2003, \apj, 585, 1038

\bibitem[{{Shkolnik} {et~al.}(2005){Shkolnik}, {Walker}, {Bohlender}, {Gu}, \&
  {K{\"u}rster}}]{Shkolnik2005}
{Shkolnik}, E., {Walker}, G.~A.~H., {Bohlender}, D.~A., {Gu}, P.-G., \&
  {K{\"u}rster}, M. 2005, \apj, 622, 1075

\bibitem[{{Showman} {et~al.}(2009){Showman}, {Fortney}, {Lian}, {Marley},
  {Freedman}, {Knutson}, \& {Charbonneau}}]{Showman2009}
{Showman}, A.~P., {Fortney}, J.~J., {Lian}, Y., {et~al.} 2009, \apj, 699, 564

\bibitem[{{Showman} \& {Guillot}(2002)}]{Showman2002}
{Showman}, A.~P. \& {Guillot}, T. 2002, \aap, 385, 166

\bibitem[{{Sing} {et~al.}(2011){Sing}, {Pont}, {Aigrain}, {Charbonneau},
  {D{\'e}sert}, {Gibson}, {Gilliland}, {Hayek}, {Henry}, {Knutson}, {Lecavelier
  Des Etangs}, {Mazeh}, \& {Shporer}}]{Sing2011}
{Sing}, D.~K., {Pont}, F., {Aigrain}, S., {et~al.} 2011, \mnras, 416, 1443

\bibitem[{{Southworth}(2010)}]{Southworth2010}
{Southworth}, J. 2010, \mnras, 408, 1689

\bibitem[{{Stetson}(1987)}]{Stetson1987}
{Stetson}, P.~B. 1987, \pasp, 99, 191

\bibitem[{{Swain} {et~al.}(2008){Swain}, {Vasisht}, \& {Tinetti}}]{Swain2008}
{Swain}, M.~R., {Vasisht}, G., \& {Tinetti}, G. 2008, \nat, 452, 329

\bibitem[{{Tinetti} {et~al.}(2011){Tinetti}, {Beaulieu}, {Henning}, {Meyer},
  {Micela}, {Ribas}, {Stam}, {Swain}, {Krause}, {Ollivier}, {Pace}, {Swinyard},
  {Aylward}, {van Boekel}, {Coradini}, {Encrenaz}, {Snellen},
  {Zapatero-Osorio}, {Bouwman}, {Y-K.~Cho}, {Coud{\'e} du Foresto}, {Guillot},
  {Lopez-Morales}, {Mueller-Wodarg}, {Palle}, {Selsis}, {Sozzetti}, {Ade},
  {Achilleos}, {Adriani}, {Agnor}, {Afonso}, {Allende Prieto}, {Bakos},
  {Barber}, {Barlow}, {Bernath}, {Bezard}, {Bord{\'e}}, {Brown}, {Cassan},
  {Cavarroc}, {Ciaravella}, {Cockell}, {Coustenis}, {Danielski}, {Decin}, {De
  Kok}, {Demangeon}, {Deroo}, {Doel}, {Drossart}, {Fletcher}, {Focardi},
  {Forget}, {Fossey}, {Fouqu{\'e}}, {Frith}, {Galand}, {Gaulme}, {Gonz{\'a}lez
  Hern{\'a}ndez}, {Grasset}, {Grassi}, {Grenfell}, {Griffin}, {Griffith},
  {Gr{\"o}zinger}, {Guedel}, {Guio}, {Hainaut}, {Hargreaves}, {Hauschildt},
  {Heng}, {Heyrovsky}, {Hueso}, {Irwin}, {Kaltenegger}, {Kervella}, {Kipping},
  {Koskinen}, {Kov{\'a}cs}, {La Barbera}, {Lammer}, {Lellouch}, {Leto}, {Lopez
  Morales}, {Lopez Valverde}, {Lopez-Puertas}, {Lovis}, {Maggio}, {Maillard},
  {Maldonado Prado}, {Marquette}, {Martin-Torres}, {Maxted}, {Miller},
  {Molinari}, {Montes}, {Moro-Martin}, {Moses}, {Mousis}, {Nguyen Tuong},
  {Nelson}, {Orton}, {Pantin}, {Pascale}, {Pezzuto}, {Pinfield}, {Poretti},
  {Prinja}, {Prisinzano}, {Rees}, {Reiners}, {Samuel}, {Sanchez-Lavega}, {Sanz
  Forcada}, {Sasselov}, {Savini}, {Sicardy}, {Smith}, {Stixrude}, {Strazzulla},
  {Tennyson}, {Tessenyi}, {Vasisht}, {Vinatier}, {Viti}, {Waldmann}, {White},
  {Widemann}, {Wordsworth}, {Yelle}, {Yung}, \& {Yurchenko}}]{Tinetti2011}
{Tinetti}, G., {Beaulieu}, J.~P., {Henning}, T., {et~al.} 2011, ArXiv e-prints

\bibitem[{{Tinetti} {et~al.}(2007){Tinetti}, {Vidal-Madjar}, {Liang},
  {Beaulieu}, {Yung}, {Carey}, {Barber}, {Tennyson}, {Ribas}, {Allard},
  {Ballester}, {Sing}, \& {Selsis}}]{Tinetti2007}
{Tinetti}, G., {Vidal-Madjar}, A., {Liang}, M.-C., {et~al.} 2007, \nat, 448,
  169

\bibitem[{{Triaud} {et~al.}(2009){Triaud}, {Queloz}, {Bouchy}, {Moutou},
  {Collier Cameron}, {Claret}, {Barge}, {Benz}, {Deleuil}, {Guillot},
  {H{\'e}brard}, {Lecavelier Des {\'E}tangs}, {Lovis}, {Mayor}, {Pepe}, \&
  {Udry}}]{Triaud2009}
{Triaud}, A.~H.~M.~J., {Queloz}, D., {Bouchy}, F., {et~al.} 2009, \aap, 506,
  377

\bibitem[{{Warner} {et~al.}(1971){Warner}, {Robinson}, \&
  {Nather}}]{Warner1971}
{Warner}, B., {Robinson}, E.~L., \& {Nather}, R.~E. 1971, \mnras, 154, 455

\bibitem[{{Werner} {et~al.}(2004){Werner}, {Roellig}, {Low}, {Rieke}, {Rieke},
  {Hoffmann}, {Young}, {Houck}, {Brandl}, {Fazio}, {Hora}, {Gehrz}, {Helou},
  {Soifer}, {Stauffer}, {Keene}, {Eisenhardt}, {Gallagher}, {Gautier}, {Irace},
  {Lawrence}, {Simmons}, {Van Cleve}, {Jura}, {Wright}, \&
  {Cruikshank}}]{Werner2004}
{Werner}, M.~W., {Roellig}, T.~L., {Low}, F.~J., {et~al.} 2004, \apjs, 154, 1

\bibitem[{{Williams} {et~al.}(2006){Williams}, {Charbonneau}, {Cooper},
  {Showman}, \& {Fortney}}]{William2006}
{Williams}, P.~K.~G., {Charbonneau}, D., {Cooper}, C.~S., {Showman}, A.~P., \&
  {Fortney}, J.~J. 2006, \apj, 649, 1020

\bibitem[{{Winn}(2010)}]{Winn2010}
{Winn}, J.~N. 2010, {Exoplanet Transits and Occultations}, ed. {Seager, S.},
  55--77

\bibitem[{{Winn} {et~al.}(2007){Winn}, {Holman}, {Henry}, {Roussanova}, {Enya},
  {Yoshii}, {Shporer}, {Mazeh}, {Johnson}, {Narita}, \& {Suto}}]{Winn2007}
{Winn}, J.~N., {Holman}, M.~J., {Henry}, G.~W., {et~al.} 2007, \aj, 133, 1828

\bibitem[{{Winn} {et~al.}(2008){Winn}, {Holman}, {Torres}, {McCullough},
  {Johns-Krull}, {Latham}, {Shporer}, {Mazeh}, {Garcia-Melendo}, {Foote},
  {Esquerdo}, \& {Everett}}]{Winn2008}
{Winn}, J.~N., {Holman}, M.~J., {Torres}, G., {et~al.} 2008, \apj, 683, 1076

\bibitem[{{Winn} {et~al.}(2006){Winn}, {Johnson}, {Marcy}, {Butler}, {Vogt},
  {Henry}, {Roussanova}, {Holman}, {Enya}, {Narita}, {Suto}, \&
  {Turner}}]{Winn2006}
{Winn}, J.~N., {Johnson}, J.~A., {Marcy}, G.~W., {et~al.} 2006, \apjl, 653, L69

\bibitem[{{Yang}(2005)}]{Yang2005}
{Yang}, Y. 2005, Biometrika, 92, 937

\end{thebibliography}

\section{Tables \& Figures}

\begin{table*}
	\caption{AOR's description}
	\label{tab:AOR}
	\centering	
	\setlength{\extrarowheight}{2pt}
	\begin{tabular}{cccc|cc}
		\hline\hline
		\textbf{AORKEY}($\bigoplus^{\mathrm{a}}$) & \textbf{PI}  & \textbf{Publication Ref.} & \textbf{Data sets}$^{\mathrm{b}}$ (64x) & \textbf{Exposure time [s]}  & \textbf{Aperture [px]}\\
		\hline
			$16343552$(O)	&	D. Charbonneau	&	\cite{Charbonneau2008}	&	$1359$ & $0.1$ & $3.6$ \\ 
			$20673792$(P)	&	D. Charbonneau	&	\cite{Knutson2007}	&	$1319$ & $0.4$ & $4.8$ \\ 
			$22808832$(O)	& & & & & \tabularnewline
			$22809088$(O)	& & & & & \tabularnewline
			$22809344$(O)	& & & & & \tabularnewline
			$22810112$(O)	& & & & & \tabularnewline
			$24537600$(O)	& & & & & \tabularnewline
			$27603456$(O)	&\multirow{-6}*{E. Agol} & \multirow{-6}*{\cite{Agol2010}} & \multirow{-6}*{$690$} & \multirow{-6}*{$0.4$}& \multirow{-6}*{$4.8$} \tabularnewline \hline
			$22807296$(T)	& & & & & \tabularnewline
			$22807552$(T)	& & & & & \tabularnewline
			$22807808$(T)	& & & & & \tabularnewline
			$24537856$(T)	& & & & & \tabularnewline
			$27603712$(T)	& & & & & \tabularnewline
			$27773440$(T)    & \multirow{-6}*{E. Agol} & \multirow{-6}*{\cite{Agol2010}} & \multirow{-6}*{$690$} & \multirow{-6}*{$0.4$}& \multirow{-6}*{$4.8$} \tabularnewline 
		\end{tabular}	 
\begin{list}{}{}
\item[$^{\mathrm{a}}$] {AORKEY target: T, O or P respectively transit, occultation or phase curve.}
\item[$^{\mathrm{b}}$] {Present AOR are composed of data sets, each data set corresponds to 64 individual subarray images of 32x32 px.}
\end{list}

\end{table*}

\begin{table}
\caption{Fit properties for different fitting models of HD\,189733's photometry in the \textit{Spitzer}/IRAC 8-$\mu $m channel \label{tab:BFP}}
	\centering
	\setlength{\extrarowheight}{3pt}
	\setlength{\tabcolsep}{8pt}

	\begin{tabular}{c c c c c c c}
	
	\hline\hline
	\multirow{2}{*}{\textbf{Parameters (units)}} & \multicolumn{2}{c}{\textbf{Uniform brightness}} & \multicolumn{2}{c}{\textbf{Unipolar brightness}} & \multicolumn{2}{c}{\textbf{Multipolar brightness}}\\
	  & $e = 0$ & $e$ free & $\Gamma_{SH,1}$ & $\Gamma_{2}$ & $\Gamma_{SH,2}$ & $\Gamma_{SH,3}$\\
	\hline

			$b (R_\star)$	&	$0.6576\pm^{0.0021}_{0.0021}$ & $0.6579\pm^{0.0021}_{0.0024}$ & $0.6598\pm^{0.0038}_{0.0024}$ & $0.6719\pm^{0.0063}_{0.0072}$ & $0.6609\pm^{0.0059}_{0.0031}$ & $0.6683\pm^{0.0071}_{0.0074}$
			  \\
			$\sqrt{e}\cos\omega$	& - &  $0.0043\pm^{0.0054}_{0.0027}$ & $0.0007\pm^{0.0032}_{0.0019}$ & $-0.0002\pm^{0.0008}_{0.0006}$ & $0.0001\pm^{0.0019}_{0.0018}$ & $0.0004\pm^{0.0013}_{0.0010}$
			 \\
			$\sqrt{e}\sin\omega$	& - & $-0.008\pm^{0.032}_{0.045}$ &  $0.016\pm^{0.066}_{0.034}$  &  $0.142\pm^{0.029}_{0.046}$  &  $0.046\pm^{0.065}_{0.055}$  & $0.121\pm^{0.036}_{0.064}$
			\\
			$\rho_\star (\rho_\odot)$	& $1.916\pm^{0.015}_{0.016}$ &			$1.932\pm^{0.018}_{0.015}$ &			$1.918\pm^{0.018}_{0.031}$ &			$1.816\pm^{0.058}_{0.048}$ &			$1.909\pm^{0.023}_{0.051}$ &			$1.845\pm^{0.061}_{0.058}$   
			\\
			\hline
			$\Delta$\,BIC / $\Delta$\,AIC &	$0 / 0$ &	$0.9 / -0.4$ & $-170.1 / -173.3$ & $-167.2 / -172.3$ & $-167.4 / -172.0$ & $-170.0 / -175.8$  \\
			\hline
			\hline
			\multicolumn{7}{c}{$(R_p/R_\star)^2 = 0.024068\pm^{0.000049}_{0.000049}$ ; $P (days) = 2.2185744\pm^{0.0000003}_{0.0000003}$ ; $T_0 (hjd-2453980) = 8.803352\pm^{0.000058}_{0.000061}$ and $F_p/F_\star|_{8\mu m} = 0.0034117\pm^{0.000037}_{0.000037}$}

	\end{tabular}
	
\end{table}

\clearpage

\clearpage

\clearpage

\clearpage

\onlfig{10}{
 
 \begin{figure*}[ht]
  \centering
  
   \subfloat[]{\label{fig:ddps_101}\includemovie[poster,toolbar,palindrome = false]{.5\textwidth}{.2825\textwidth}{images/ddps_101.mp4}}	
  ~  
  \subfloat[]
  {\hspace{+00mm}\label{fig:ddps_3}\includemovie[poster,toolbar,palindrome = false]{.5\textwidth}{.2825\textwidth}{images/ddps_3.mp4}}
  
  \caption{Insight into the brightness distribution estimates. Animations showing compilations of HD\,189733b's dayside brightness distributions accepted along the MCMC simulations for the $\Gamma_{SH,1}$ and $\Gamma_{2}$ models, \protect\subref{fig:ddps_101} and \protect\subref{fig:ddps_3} respectively. These animations show that \textbf{(1)} the amplitude and \textbf{(2)} the longitudinal localization for the $\Gamma_{SH,1}$ brightness model is more constrained than for the $\Gamma_{2}$ model (by the occultation depth and by the phase curve, respectively) because of its fixed and large structure. However, \textbf{(3)} the $\Gamma_{SH,1}$ model is less constrained in latitude (by the secondary-eclipse scanning) than the more complex $\Gamma_{2}$ model which enable more confined structures that induce larger deviation in occultation ingress/egress (schematic description in Fig.\,\ref{fig:shape_brightness}).}
  \label{fig:ddps_video}
\end{figure*}
}

\begin{figure*}
  \centering
  
  \subfloat[]{\hspace{+00mm}\label{fig:102_e_ddp}\includegraphics[trim = 35mm 85mm 35mm 92mm,clip,width=0.5\textwidth,height=!]{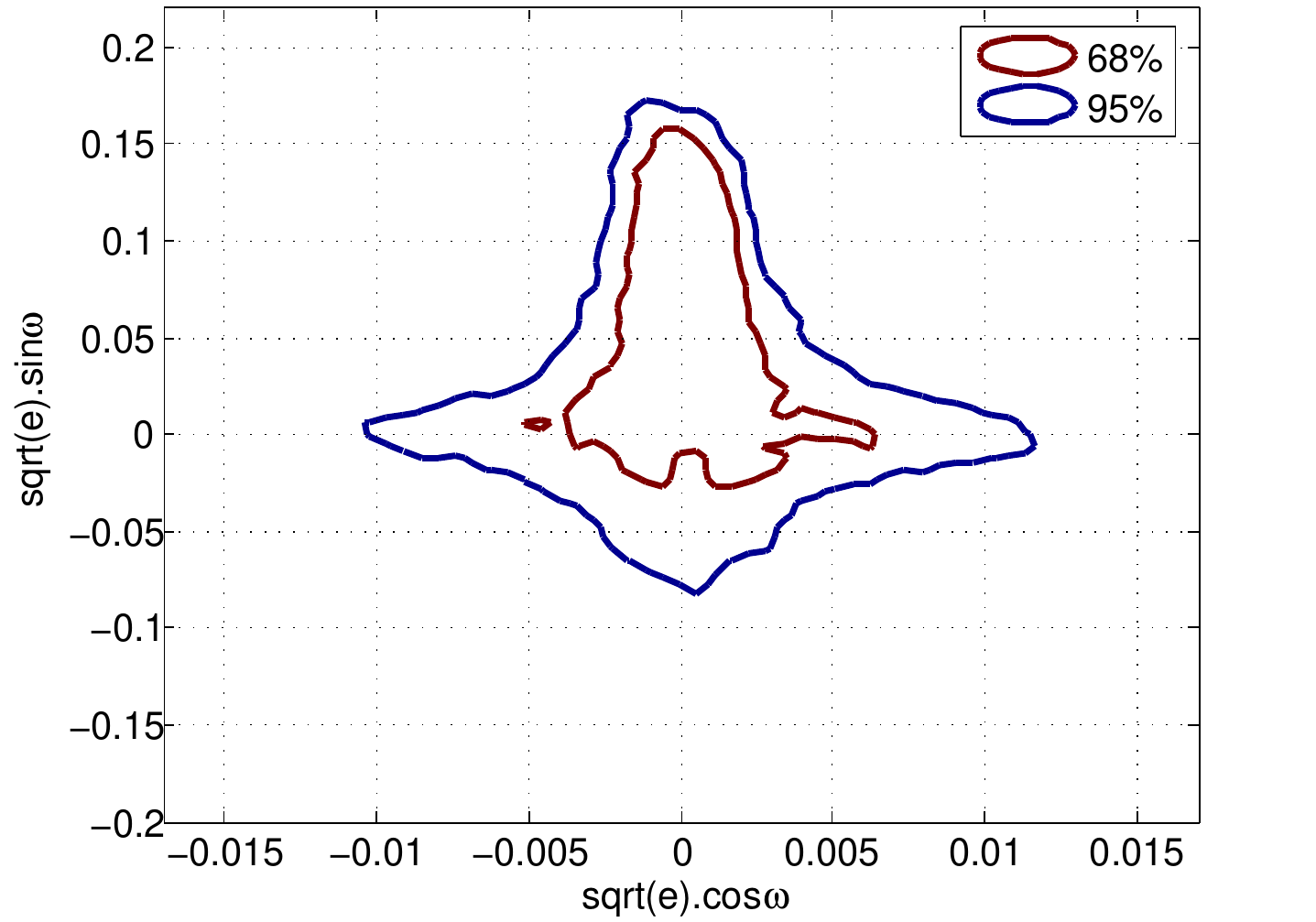}}
  ~ 
  \subfloat[]{\label{fig:103_e_ddp}\includegraphics[trim = 35mm 85mm 35mm 92mm,clip,width=0.5\textwidth,height=!]{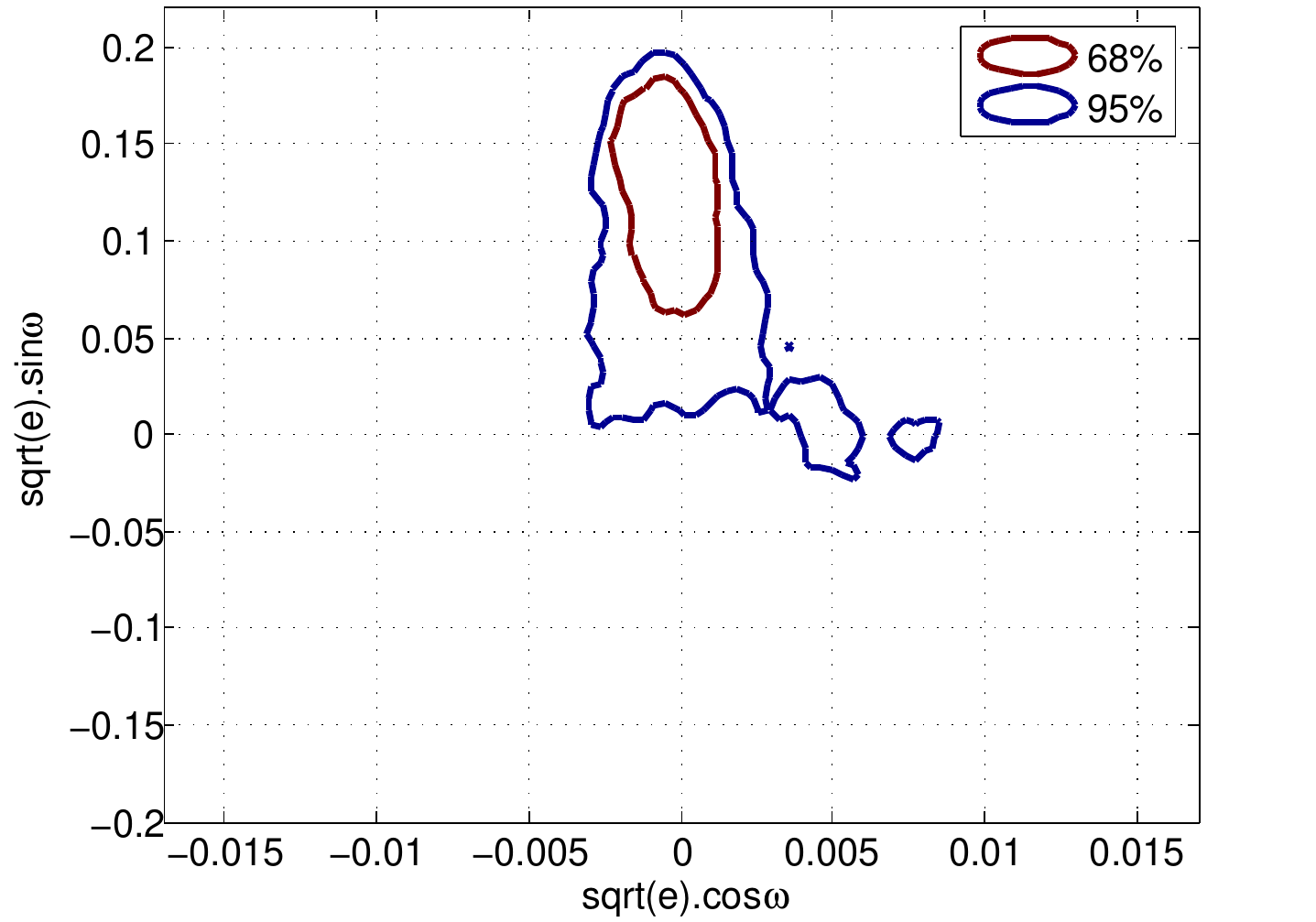}}

   \subfloat[]{\hspace{+00mm}\label{fig:102_erho_ddps}\includegraphics[trim = 35mm 85mm 35mm 92mm,clip,width=0.5\textwidth,height=!]{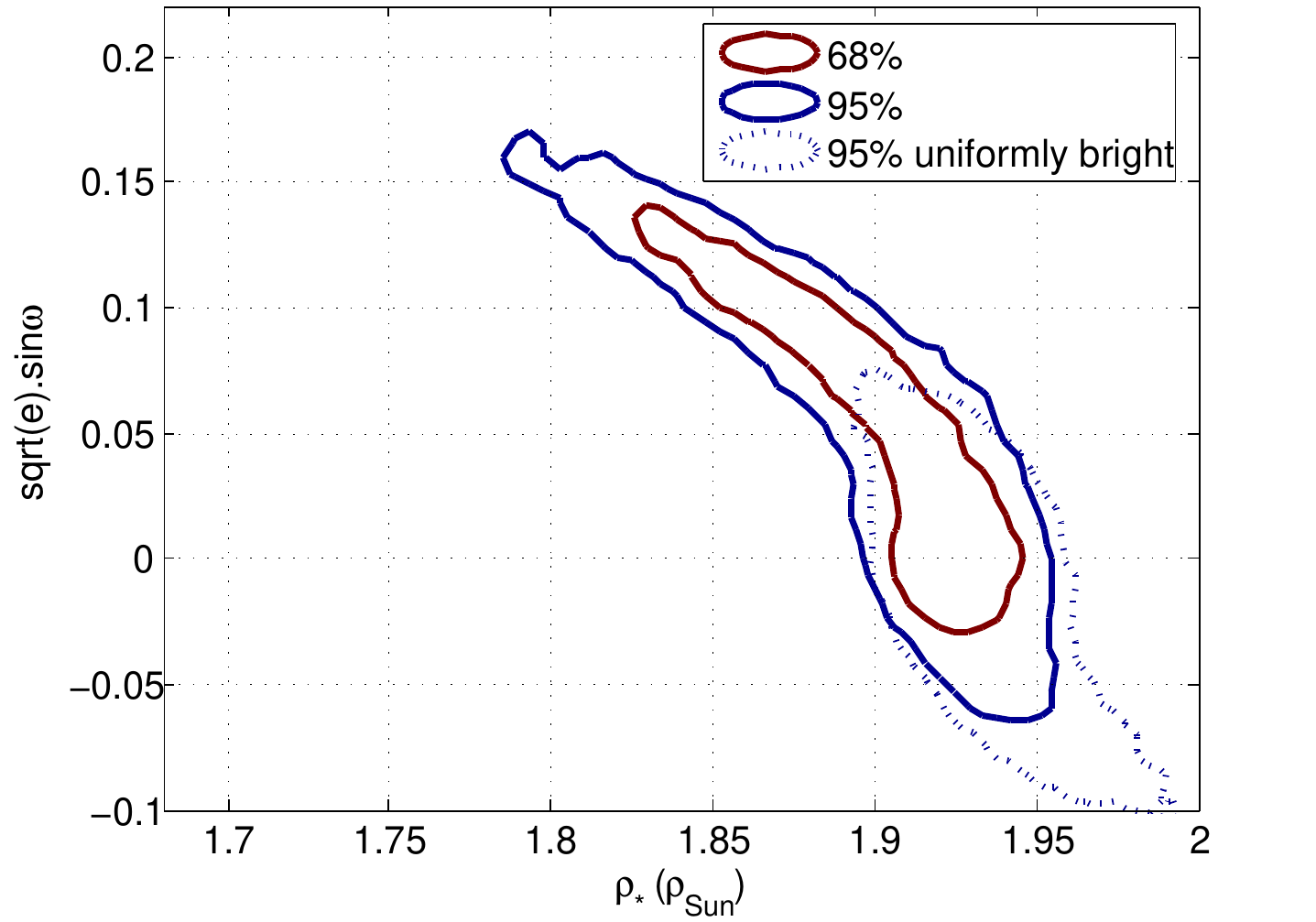}}
  ~  
  \subfloat[]{\label{fig:103_erho_ddps}\includegraphics[trim = 35mm 85mm 35mm 92mm,clip,width=0.5\textwidth,height=!]{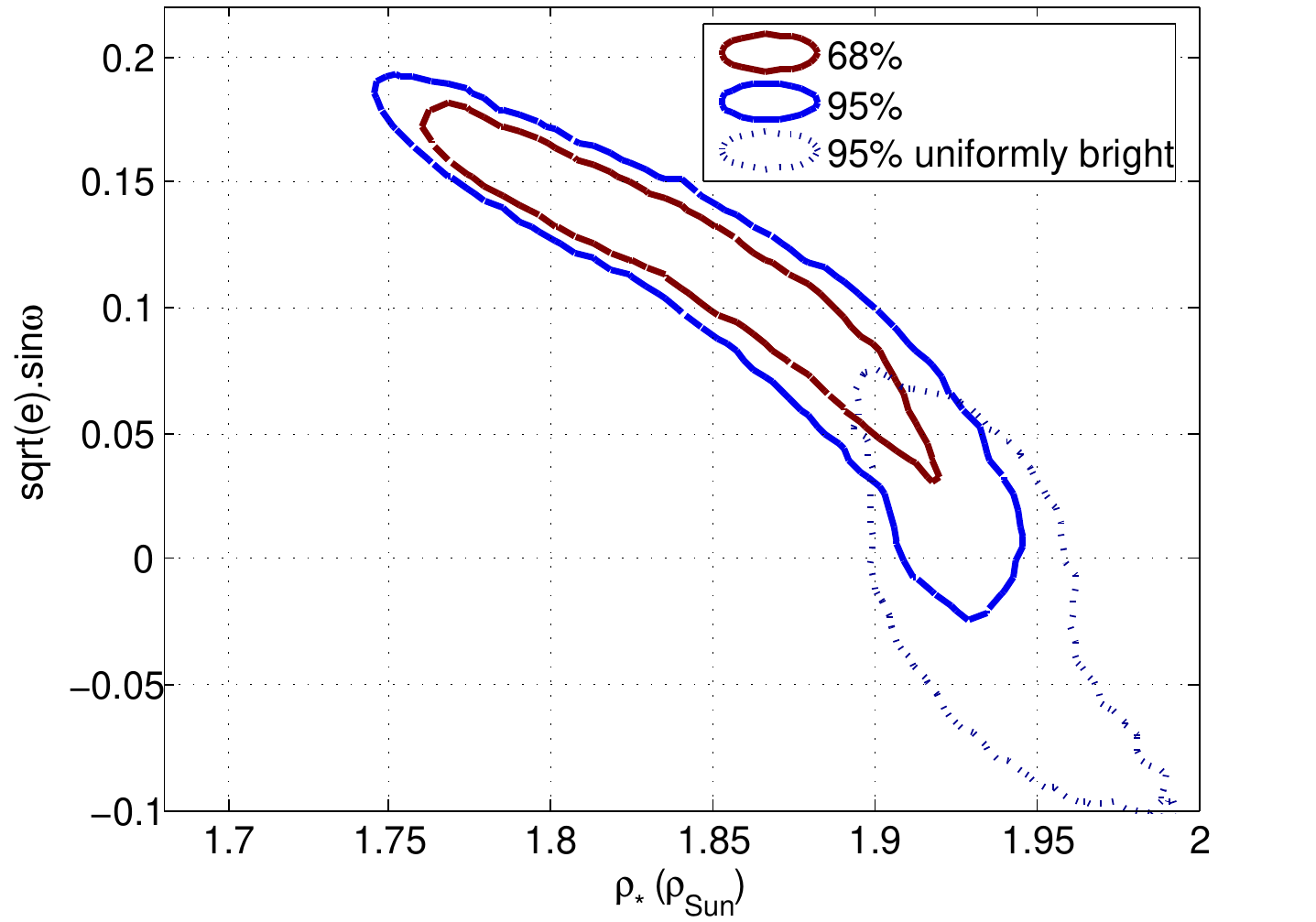}}

  \caption{Incidence of the brightness model extension on the system parameter posterior probability distribution (PPD), using multipolar models. \protect\subref{fig:102_e_ddp} \&  \protect\subref{fig:103_e_ddp} Marginal PPDs ($68\%$- and $95\%$-confidence intervals) of $\sqrt{e}\cos\omega$ and $\sqrt{e}\sin\omega$ for the quadrupolar and octupolar brightness models, respectively. \protect\subref{fig:102_erho_ddps} \&  \protect\subref{fig:103_erho_ddps} Marginal PPDs of $\rho_{\star}$ and $\sqrt{e}\sin\omega$ for the quadripolar and octopolar brightness models, respectively. These confirm the trend observed in Sect.\,\ref{sec:uniform} (see Fig.\,\ref{fig:correl_mono}) towards larger $\sqrt{e}\sin\omega$ and lower stellar density when increasing the complexity of the brightness distribution, i.e., when enabling more localized structures.}
  \label{fig:correl_multi}
\end{figure*}

\clearpage

\begin{figure*}
   \centering
   
      \subfloat[]{\hspace{+00mm}\label{fig:102_brightness_ddps}\includegraphics[angle = -90, trim = 70mm 10mm 70mm 10mm,clip,width=\textwidth,height=!]{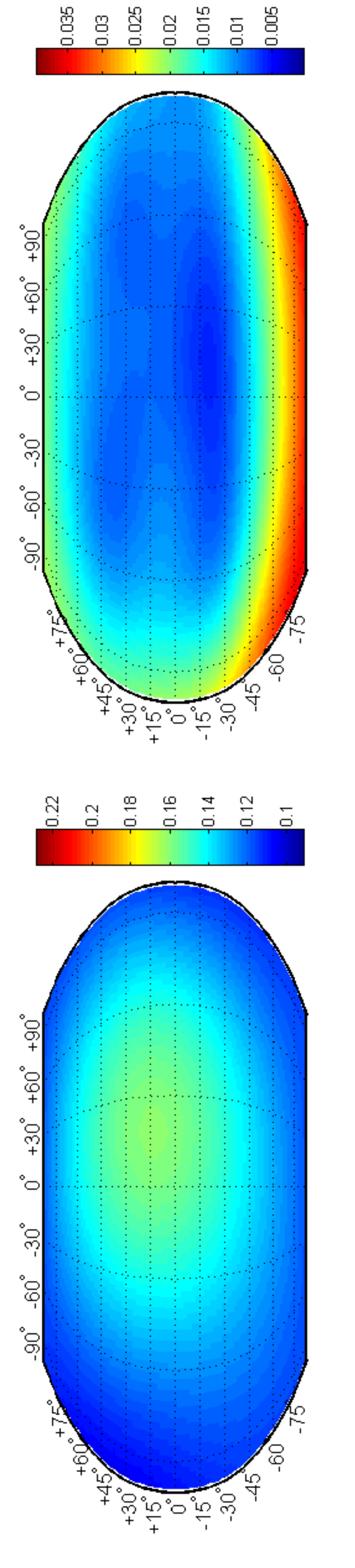}}
      
            \subfloat[]{\hspace{+00mm}\label{fig:103_brightness_ddps}\includegraphics[angle = -90, trim = 70mm 10mm 70mm 10mm,clip,width=\textwidth,height=!]{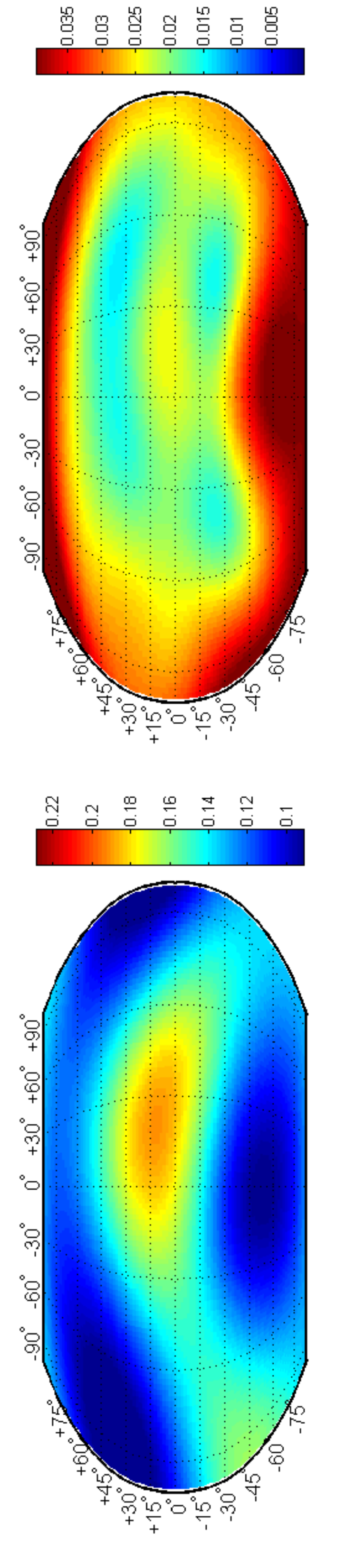}}

  \vspace{+0.5cm}
    \caption{Estimate of HD\,189733b's global brightness distribution in the IRAC 8-$\mu$m channel using multipolar brightness models. \textbf{Left:} Relative brightness distribution at HD\,189733b's dayside. \textbf{Right:} Dayside standard deviation. \protect\subref{fig:102_brightness_ddps} Estimate using the $\Gamma_{SH,2}$ brightness model. \protect\subref{fig:103_brightness_ddps} Estimate using the $\Gamma_{SH,3}$ brightness model. These confirm the trend toward a more localized and latitudinally-shifted hot spot when increasing the brightness-model complexity. Therefore, together with Fig.\,\ref{fig:correl_multi} these outline that the more complex HD\,189733b's brightness model, the larger the eccentricity, the lower the densities, the larger the impact parameter and the more localized and latitudinally-shifted the hot spot estimated.}
  \label{fig:complex_brightness}

\end{figure*}

\begin{figure*}
  \centering
   \begin{center}
    \hspace{-0.cm}\includegraphics[trim = 35mm 85mm 35mm 92mm,clip,width=0.5\textwidth,height=!]{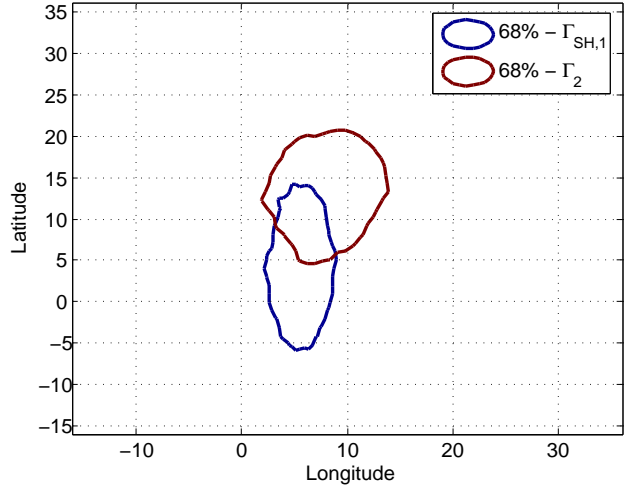}
  \end{center}
  \vspace{-0.5cm}  
  \caption{Reducing brightness distributions to unidimensional parameters. Marginal PPDs ($68\%$-confidence intervals) of the dayside barycenter brightness peak localization for the $\Gamma_{SH,1}$ and $\Gamma_{2}$ brightness models. A comparison with the marginal PPDs of the brightness peak localization (Fig.\,\ref{fig:101_peak_ddps}) shows the reduced model-dependence of the dayside barycenter. In particular, it shows a less-extended PPD for the $\Gamma_{2}$-model barycenter; because this extension for the brightness-peak-localization PPD emerges from the model wings, weighted by the dayside barycenter. In addition, it shows the shift and slight shrinking of the $\Gamma_{SH,1}$ PPD that reflect the barycenter weighting based on the geometrical configuration at superior conjunction; map cells closer to the substellar point have more weight.}
  \label{fig:barycenter}
\end{figure*}

\clearpage

\begin{figure*}
   \centering

  \begin{center}
    \hspace{-0.cm}\includegraphics[trim = 20mm 75mm 15mm 75mm,clip,width=0.5\textwidth,height=!]{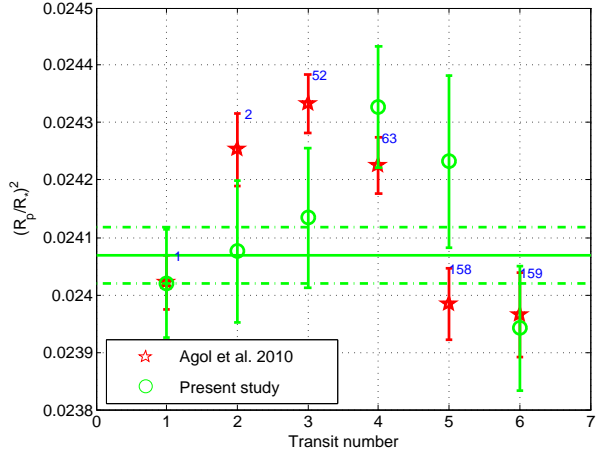}
  \end{center}
  \vspace{-0.5cm}
  \caption{Incidence of HD\,189733's variability on individual transit-depth estimates in the \textit{Spitzer}/IRAC 8-$\mu$m channel. Transit-depth estimated for six transits in individual (markers) and global (solid line, the dashed lines refer to the 1\,$\sigma$ error bar) analyses. Our estimates (green) show no significant transit-depth variation with the relative eclipse phase (blue numbers) in opposition to \citet{Agol2010} (red). Similarly, we observe no significant variation of the transit parameters from one individual analysis to another. In addition, we observe no pattern specific to the occultation of a star spot \citep[i.e., similar to, e.g., ``Features A and B'' in the Fig. 1 of][]{Pont2007}. Therefore, we consider that our time-averaged  inferences (see Sect.\,\ref{sec:results}) are not biased by HD\,189733's activity.}
  \label{fig:DdF}

\end{figure*}

\begin{figure*}
  \centering
  
  \subfloat[]{\hspace{+00mm}\label{fig:RVfit}\includegraphics[trim = 15mm 67mm 10mm 69mm,clip,width=!,height=70mm]{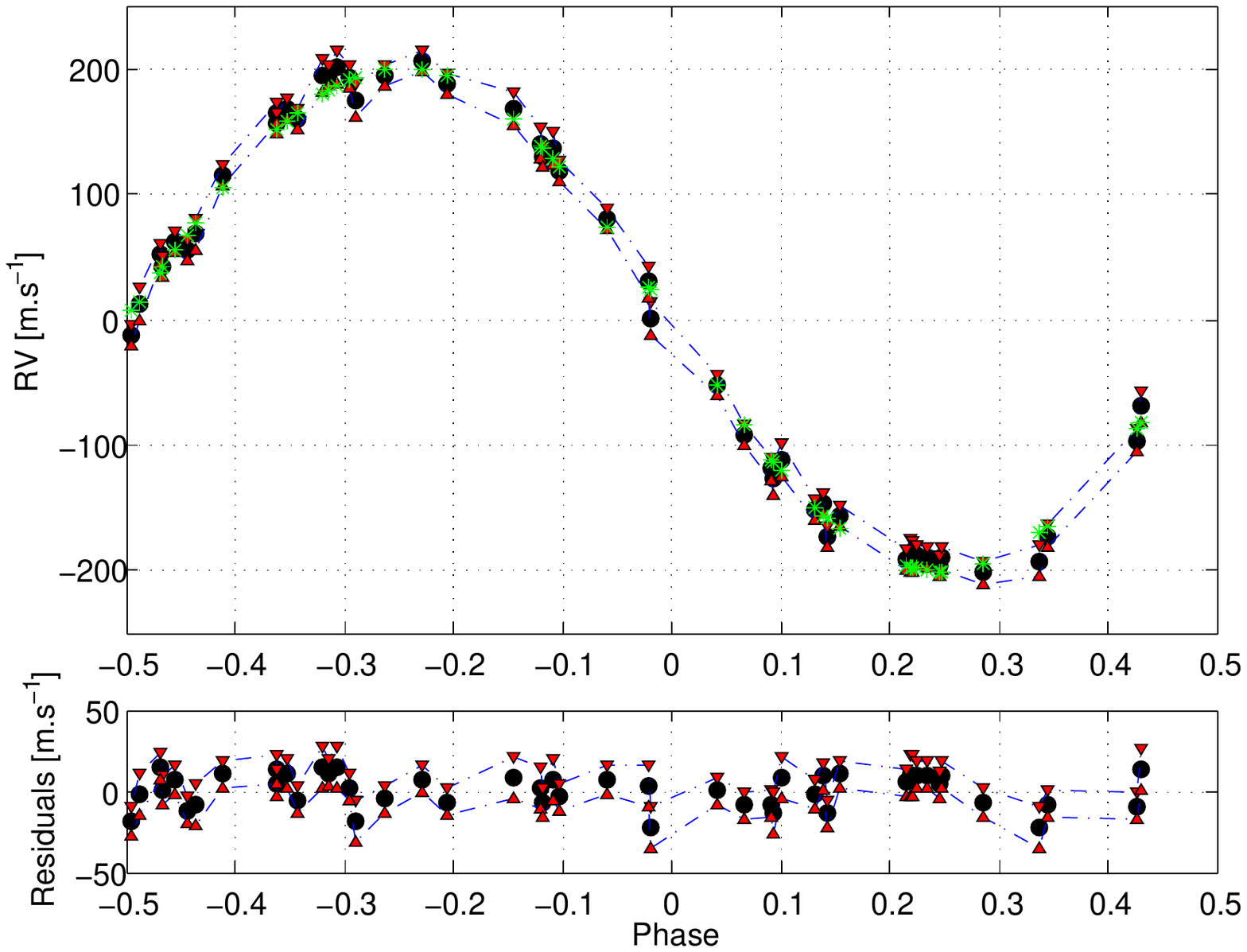}}
  ~ 
  \subfloat[]{\label{fig:ddp_e_unif}\includegraphics[trim = 30mm 71mm 25mm 69mm,clip,width=!,height=70mm]{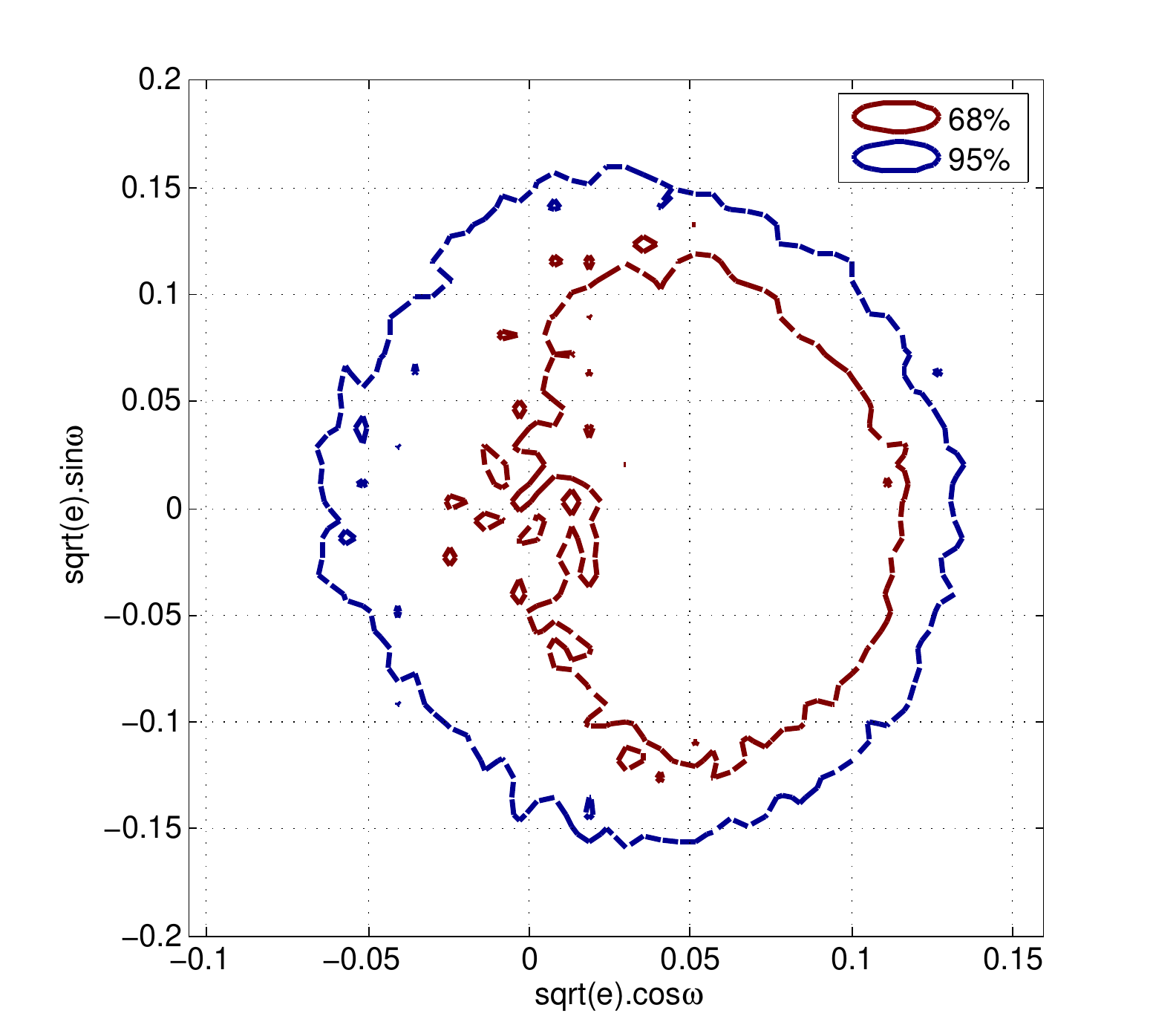}}

  \caption{Incidence of HD\,189733's RV measurements on inferences obtained from the \textit{Spitzer}/IRAC 8-$\mu$m photometry. \protect\subref{fig:RVfit} Overall Keplerian fit (green) of the out-of-transit RV data (black dots) with their 1\,$\sigma$ error bars (red triangles) from \citet{Winn2006} and \citet{Boisse2009}. \protect\subref{fig:ddp_e_unif} Marginal PPD (68\%- and 95\%-confidence intervals) of $\sqrt{e}\cos\omega$ and $\sqrt{e}\sin\omega$ obtained solely from the RV data. This shows that the RV data do not constrain further HD\,189733b's eccentricity than the \textit{Spitzer}/IRAC 8-$\mu$m photometry for low complexity brightness models (see Figs.\,\ref{fig:uniform_e_ddp}, \ref{fig:101_e_ddp} and \ref{fig:102_e_ddp}). However, for more complex models that favor a localized hot spot and larger $\sqrt{e}\sin\omega$, HD\,189733's RV measurements may affect our inferences by rejecting the solutions involving $\sqrt{e}\sin\omega$ $ \gtrsim $ 0.15---especially in the context of the $ e $-$ b $-$\rho_{\star}$-BD correlation (see Sect.\,\ref{sec:unipolar}).}
  \label{fig:RV_inputs}
\end{figure*}

\clearpage

\begin{figure*}[!h]
  \centering
  
  \subfloat[]{\hspace{+00mm}\label{fig:103_e_ddp_RV}\includegraphics[trim = 35mm 85mm 35mm 92mm,clip,width=0.5\textwidth,height=!]{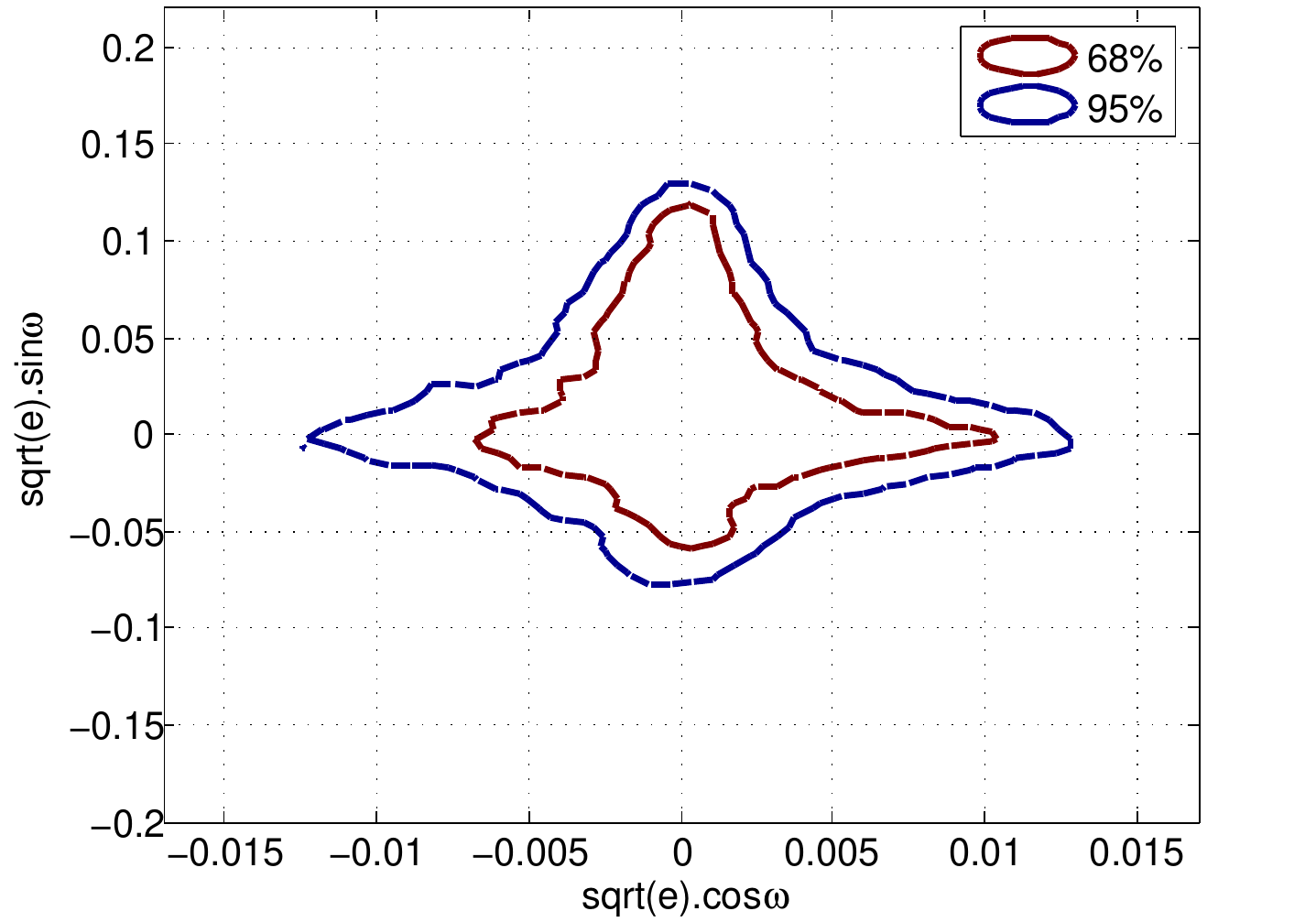}}
  ~ 
  \subfloat[]{\label{fig:3_e_ddp_RV}\includegraphics[trim = 35mm 85mm 35mm 92mm,clip,width=0.5\textwidth,height=!]{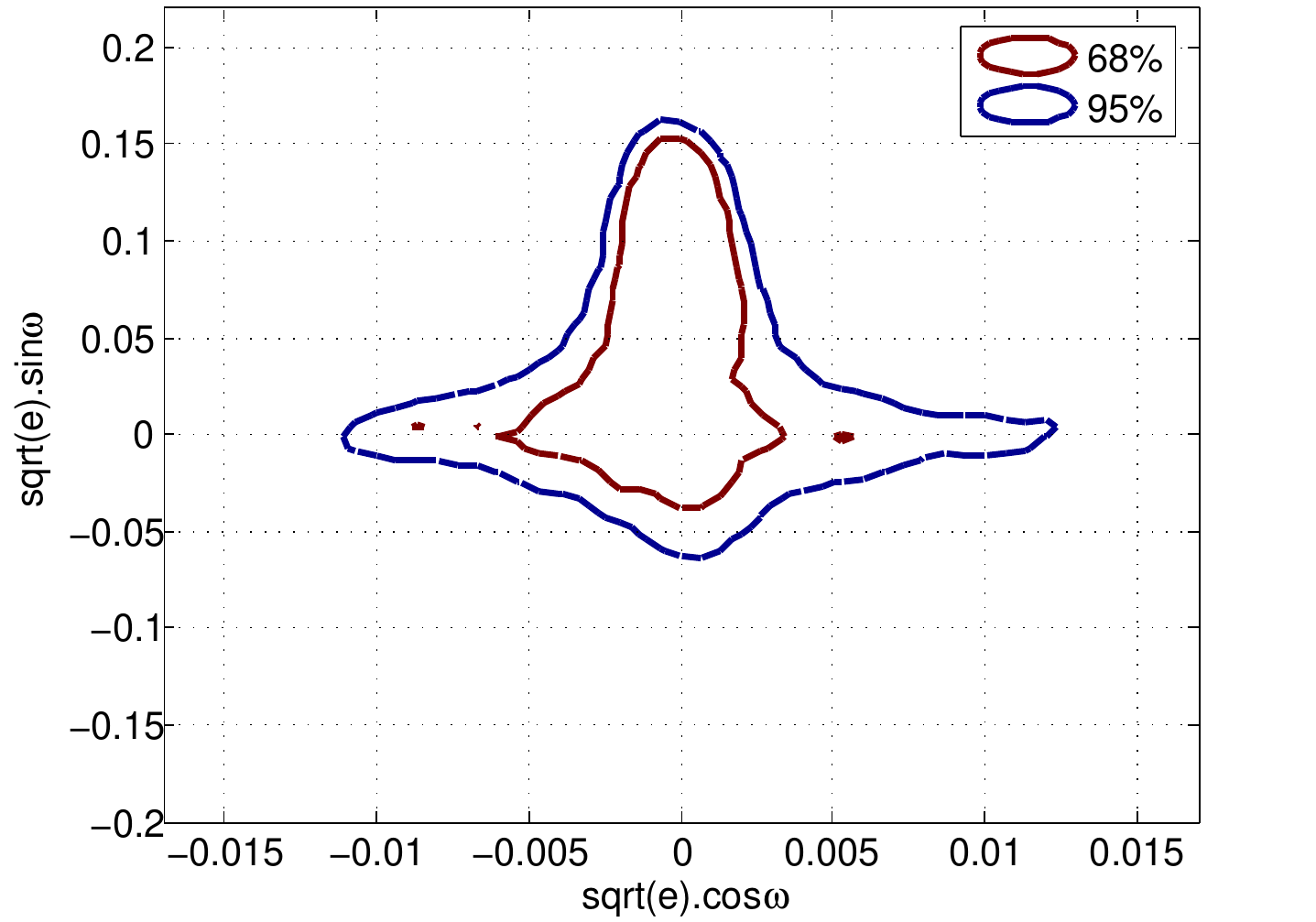}}

   \subfloat[]{\hspace{+00mm}\label{fig:103_erho_ddps_RV}\includegraphics[trim = 35mm 85mm 35mm 92mm,clip,width=0.5\textwidth,height=!]{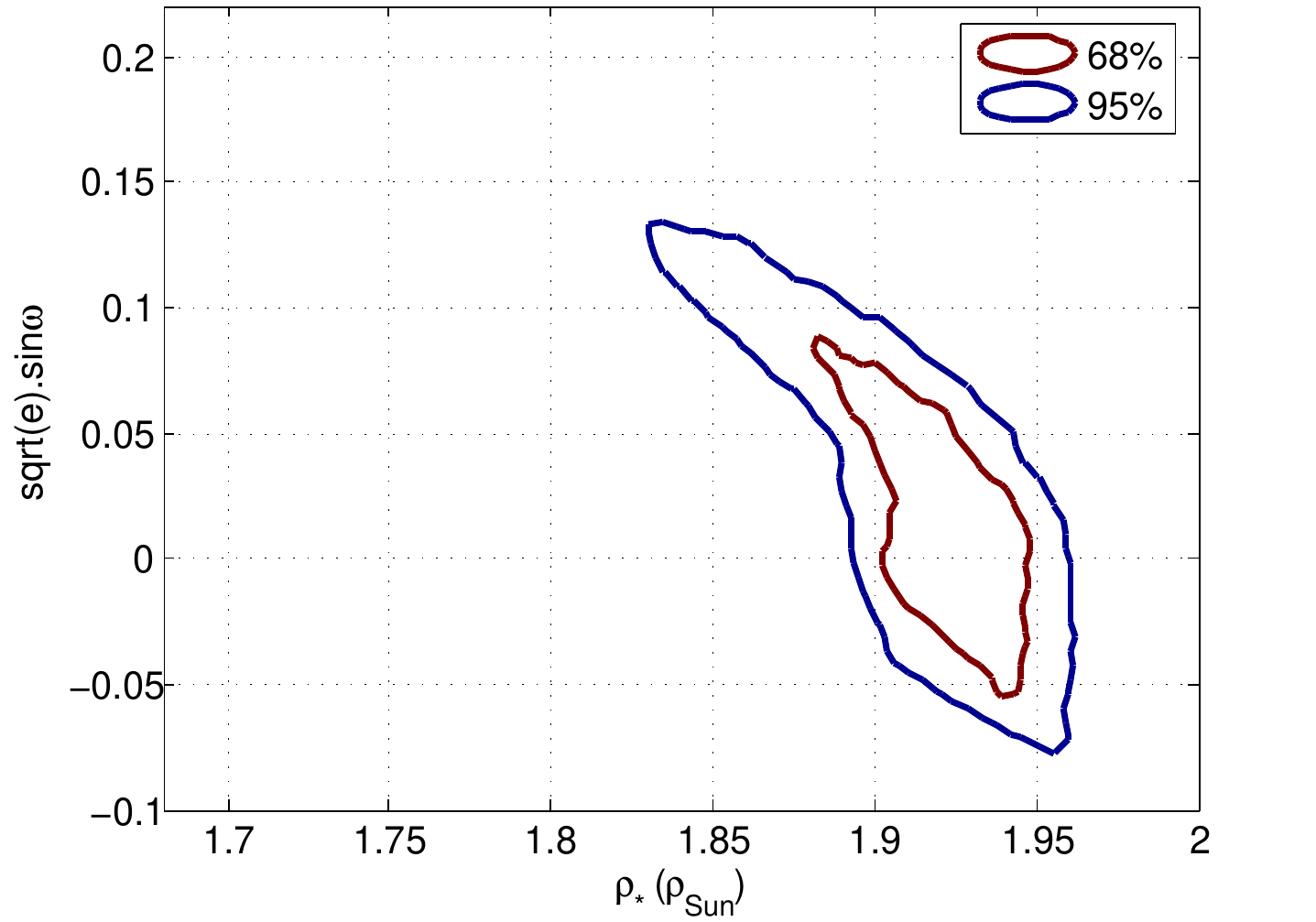}}
  ~  
  \subfloat[]{\label{fig:3_erho_ddps_RV}\includegraphics[trim = 35mm 85mm 35mm 92mm,clip,width=0.5\textwidth,height=!]{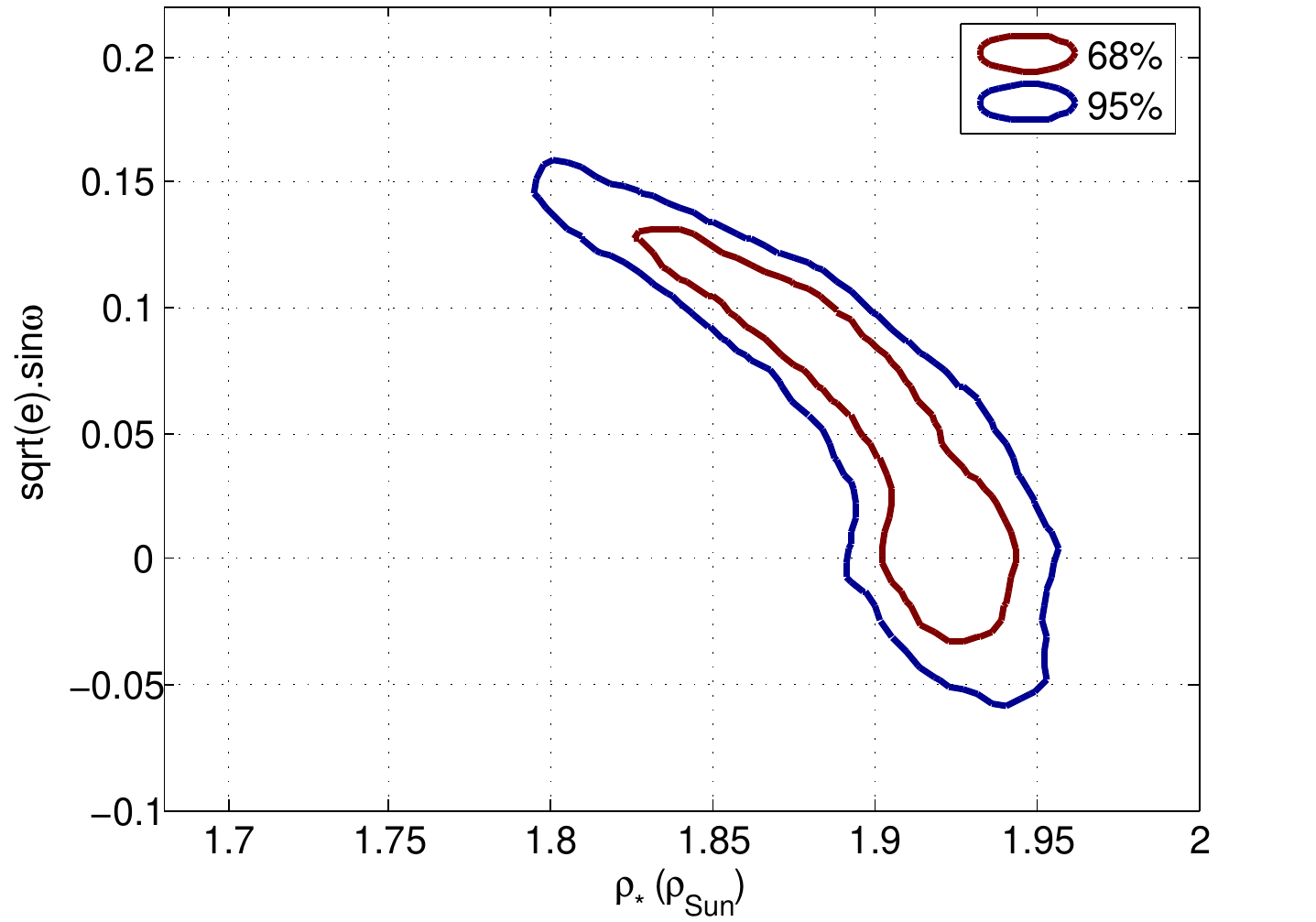}}

  \caption{Incidence of the RV measurements on the system-parameter PPD estimated using complex brightness models that suggest a non-zero eccentricity from the \textit{Spitzer}/IRAC 8-$\mu$m photometry (see Figs.\,\ref{fig:103_e_ddp} and \,\ref{fig:3_e_ddp}). \protect\subref{fig:103_e_ddp_RV} \&  \protect\subref{fig:3_e_ddp_RV} Marginal PPDs (68\%- and 95\%-confidence intervals) of $\sqrt{e}\cos\omega$ and $\sqrt{e}\sin\omega$ for the $\Gamma_{SH,3}$ and the $\Gamma_2$ brightness models, respectively. Conceptually these correspond to the marginalized product of the PPDs estimated using solely the photometry (see Figs.\,\ref{fig:103_e_ddp} and \,\ref{fig:3_e_ddp}) and using solely the RV data (Fig.\,\ref{fig:RVfit}). In particular, these highlight that the solutions involving $\sqrt{e}\sin\omega$ $ \gtrsim $ 0.15 are strongly rejected by HD\,189733's RV data, leading to a redistribution of the probability density. \protect\subref{fig:103_erho_ddps_RV} \&  \protect\subref{fig:3_erho_ddps_RV} Marginal PPDs of $\rho_{\star}$ and $\sqrt{e}\sin\omega$ for the $\Gamma_{SH,3}$ and $\Gamma_2$ brightness models, respectively. These show the impact of RV data in the context of the $ e $-$ b $-$\rho_{\star}$-BD correlation (see Sect.\,\ref{sec:unipolar}); by rejecting the solutions involving $\sqrt{e}\sin\omega$ $ \gtrsim $ 0.15, HD\,189733's RV data favors solutions consistent with the inferences obtained with less complex brightness models (see Fig\,\ref{fig:correl_mono}). This emphasizes the necessity of global approaches to consistently probe the highly-correlated parameter space of exoplanetary data.}
  \label{fig:correl_multi_RV}
\end{figure*}

\begin{table}
\caption{Parameter estimates obtained from the photometry and the RV measurements. \label{tab:BFP_RV}}
	\centering
	\setlength{\extrarowheight}{3pt}
	\setlength{\tabcolsep}{8pt}

	\begin{tabular}{c c c c c}
	
	\hline\hline
	\multirow{2}{*}{\textbf{Parameters (units)}} & \multicolumn{2}{c}{\textbf{Unipolar brightness}} & \multicolumn{2}{c}{\textbf{Multipolar brightness}}\\
	  & $\Gamma_{SH,1}$ & $\Gamma_{2}$ & $\Gamma_{SH,2}$ & $\Gamma_{SH,3}$\\
	\hline

			$b (R_\star)$	& $0.6591\pm^{0.0028}_{0.0023}$ & $0.6604\pm^{0.0063}_{0.0031}$ & $0.6590\pm^{0.0029}_{0.0025}$ & $0.6592\pm^{0.0036}_{0.0025}$
			  \\
			$\sqrt{e}\cos\omega$ & $0.0012\pm^{0.0039}_{0.0022}$ & $-0.0002\pm^{0.0027}_{0.0028}$ & $0.0003\pm^{0.0043}_{0.0038}$ & $0.0003\pm^{0.0046}_{0.0038}$
			 \\
			$\sqrt{e}\sin\omega$ &  $0.007\pm^{0.050}_{0.030}$  &  $0.031\pm^{0.079}_{0.043}$  &  $0.006\pm^{0.052}_{0.031}$  & $0.009\pm^{0.056}_{0.034}$
			\\
			$\rho_\star (\rho_\odot)$ &			$1.922\pm^{0.017}_{0.022}$ &			$1.912\pm^{0.023}_{0.051}$ &			$1.922\pm^{0.018}_{0.021}$ &			$1.922\pm^{0.019}_{0.025}$

	\end{tabular}
	
\end{table}

\clearpage

\begin{figure*}
   \centering
   
      \subfloat[]{\hspace{+00mm}\label{fig:103_brightness_ddps_RV}\includegraphics[angle = -90, trim = 70mm 10mm 70mm 10mm,clip,width=\textwidth,height=!]{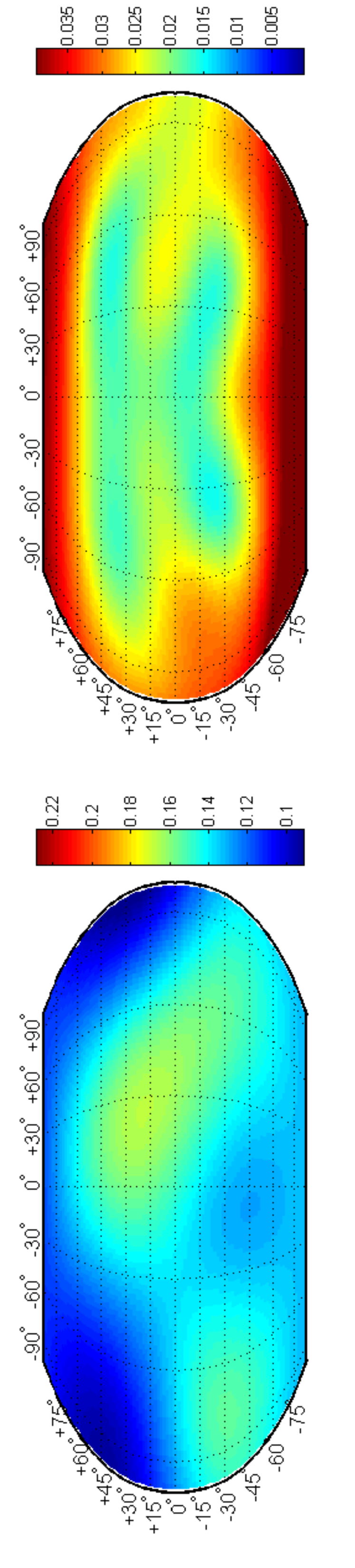}}
      
            \subfloat[]{\hspace{+00mm}\label{fig:3_brightness_ddps_RV}\includegraphics[angle = -90, trim = 70mm 10mm 70mm 10mm,clip,width=\textwidth,height=!]{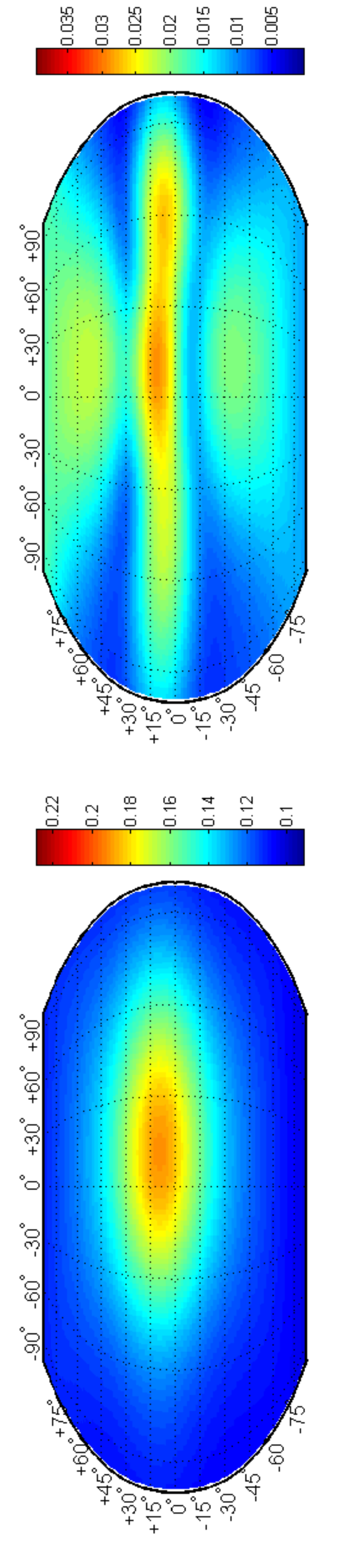}}

  \vspace{+0.5cm}
    \caption{Incidence of HD\,189733's RV measurements on HD\,189733b's dayside brightness distribution, estimated using complex brightness models that suggest non-zero eccentricity from the \textit{Spitzer}/IRAC 8-$\mu$m photometry (see Figs.\,\ref{fig:103_e_ddp} and \,\ref{fig:3_e_ddp}). \textbf{Left:} Relative brightness distribution at HD\,189733b's dayside. \textbf{Right:} Dayside standard deviation. \protect\subref{fig:103_brightness_ddps_RV} Estimate using the $\Gamma_{SH,3}$ brightness model. \protect\subref{fig:3_brightness_ddps_RV} Estimate using the $\Gamma_2$ brightness model. The incidence of HD\,189733's RV data is an extension of the brightness
patterns (compare with Figs.\,\ref{fig:103_brightness_ddps} and \ref{fig:3_brightness_ddps}, respectively) associated with a decrease of the brightness peak---conservation of the hemisphere integrated flux. These show that localized brightness pattern that are favored by the photometry are rejected by the RV data
in the context of the $ e $-$ b $-$\rho_{\star}$-BD correlation; because these are associated with $\sqrt{e}\sin\omega$ $ \gtrsim $ 0.15.}
  \label{fig:complex_brightness_RV}

\end{figure*}

\end{document}